\documentclass{aa_v2}  
\usepackage{graphicx}
\usepackage{multirow}
\usepackage{txfonts}
\usepackage[pdfencoding=auto,psdextra]{hyperref}
\hypersetup{
    colorlinks=true,
    linkcolor=blue,
    filecolor=magenta,      
    urlcolor=blue,
    citecolor=blue
}
\usepackage[utf8]{inputenc}
\newcommand{\orcid}[1]{}

\usepackage[switch, modulo]{lineno}
\renewcommand{\linenumbers}[0]{}

\usepackage{euclid}
\begin{document} 

   \title{An empirical derivation of the average dust attenuation law at $2<z<7$}

   \author{G.~Rodighiero \inst{1, 2},
          G.~E.~Esposito \inst{2, 1},
          D.~Calzetti \inst{3},
          P.~Benotto \inst{2, 1},
          M.~Catone  \inst{1, 2},     
          P.~Cassata \inst{1, 2}, 
          G.~Gandolfi \inst{4}, 
          L.~Bisigello \inst{2}, 
          S.~Carniani\inst{5},
          A.~Renzini \inst{2},
          I.~Shivaei \inst{6},
          B.~Vulcani  \inst{2},
          A.~Puglisi \inst{7},
          A.~Enia \inst{8, 9},
          S.~Álvarez-Córdoba \inst{10}
          }

   \institute{
   $^{1}$ Dipartimento di Fisica e Astronomia “G. Galilei”, Universit\`a di Padova, Vicolo dell'Osservatorio 3, 35131 Padova, Italy \\
   $^{2}$ INAF, Osservatorio Astronomico di Padova, Vicolo dell'Osservatorio 5, 35122 Padova, Italy \\
   $^{3}$ Department of Astronomy, University of Massachusetts Amherst, 710 North Pleasant Street, Amherst, MA 01003, USA\\
   $^{4}$ INAF, Osservatorio Astronomico di Roma, Via Frascati 33, 00078 Monteporzio Catone, Roma, Italy \\
   $^{5}$ Scuola Normale Superiore, Piazza dei Cavalieri 7, I-56126 Pisa, Italy\\
   $^{6}$ Centro de Astrobiolog\'ia (CAB), CSIC-INTA, Ctra. de Ajalvir km 4, Torrej\'on de Ardoz, E-28850, Madrid, Spain \\
   $^{7}$ School of Physics and Astronomy, University of Southampton, Highfield SO17 1BJ, UK\\
   $^{8}$ CEA Saclay, DFR/IRFU, Service d'Astrophysique, Bat. 709, 91191 Gif-sur-Yvette, France\\
   $^{9}$ INAF-Osservatorio di Astrofisica e Scienza dello Spazio di Bologna, Via Piero Gobetti 93/3, 40129 Bologna, Italy\\
   $^{10}$ Instituto de Radioastronomía y Astrofísica, UNAM, Campus Morelia, A.P. 3-72, Morelia 58089, Mexico\\
    }
   
 
 \abstract{
    \textit{Context.} Dust attenuation significantly affects the observed spectral energy distributions (SEDs) of galaxies, introducing significant uncertainties in the derivation of key physical properties such as star formation rates, stellar masses, and metallicities. While attenuation curves have been extensively studied in the Local Universe and at intermediate redshift, direct spectroscopic constraints at earlier cosmic epochs have remained limited prior to the advent of \JWST (JWST).
    
    \textit{Aims.}
    We aim to derive the average dust attenuation law of massive $\log_{10}(M_\star/M_\odot)>9$ star-forming galaxies over the redshift range $2<z<7$ using JWST spectroscopy.
    
    \textit{Methods.} We combined JWST/NIRSpec spectroscopy from the JWST Advanced Deep Extragalactic Survey (JADES) survey with deep multi-wavelength photometry from the ASTRODEEP-JWST catalogues. Using a mass-selected sample of $\sim 120$ galaxies with $\log_{10}(M_\star/M_\odot) > 9$ and reliable Balmer decrement (\ha\,/\hb) measurements, we constructed stacked SEDs in bins of Balmer optical depth and derived the selective attenuation curve following the empirical methodology introduced by \citet{Calzetti_1994}. The wavelength coverage is further extended towards the near-infrared (NIR) using JWST/MIRI photometry.
    
    \textit{Results.} The resulting attenuation curve spans the rest-frame range $0.16\mbox{--}1.14\,\micron$ and is well described by a smooth function. We derived a normalisation factor of $R_V \simeq 3.98$, with $\Delta R_V = 0.16$ providing an estimate of its sensitivity to the  adopted long-wavelength extrapolation, finding that the average  attenuation law is remarkably consistent with the local starburst relation in both slope and normalisation. However, compared to several determinations at intermediate redshift, our curve appears systematically flatter in the ultraviolet (UV). We find no significant evidence to support a $2175\,\AA$ UV bump in the average attenuation curve.
    
    \textit{Conclusions.} 
    Our results provide a new direct empirical constraint on the average dust attenuation law for massive star-forming galaxies at $2<z<7$ based on JWST spectroscopy. Despite the diversity of attenuation properties observed in individual systems, the ensemble-average behaviour remains consistent with the local starburst relation, suggesting that the main physical mechanisms regulating dust attenuation were already in place at early cosmic times. The relatively shallow UV slope compared to several previous empirical determinations may reflect differences in dust–star geometry, radiative-transfer effects, and/or effective grain-size distributions among galaxy populations.
    }

   \authorrunning{G. Rodighiero et al.}

    \maketitle
    \nolinenumbers

\section{Introduction}
Dust in the interstellar medium (ISM) plays a central role in shaping the observed properties of galaxies. Its wavelength-dependent attenuation modifies the emergent stellar continuum and nebular emission, introducing substantial uncertainties in measurements of star formation rates (SFRs), stellar masses, metallicities, and the overall energy budget of galaxies. Recovering intrinsic galaxy properties therefore requires an accurate characterisation of the dust attenuation curve, whose shape reflects both the composition and size distribution of dust grains, as well as the relative geometry of stars, gas, and dust \citep[e.g.][]{Calzetti2000,Weingartner01,Witt2000}. Variations in the attenuation curve, most pronounced in the rest-frame ultraviolet (UV), carry key information on the physical conditions of the ISM and on the processes governing dust production, growth, and destruction.

Over the past three decades, significant progress has been made in constraining extinction and attenuation curves in both the Local Universe \citep[e.g.][]{cardelli_1989,Calzetti_1994,Gordon2003,Battisti_2016,Salim_2018,Salim_2020} and at intermediate redshift \citep[e.g.][]{Reddy_2015,Salmon2016,Shivaei_2020,Battisti_2022}. Observational constraints were also obtained at substantially higher redshifts prior to \textit{James Webb} Space Telescope (JWST) using a variety of complementary techniques. In particular, extinction curves along individual lines of sight were constrained from quasar spectra and gamma-ray burst afterglows out to $z \sim 6\mbox{--}8$ \citep[e.g.][]{Maiolino2004,Gallerani2010,Bolmer2018,Zafar18}, revealing both departures from and similarities to extinction curves measured in the Local Universe. Constraints on dust attenuation properties have also been obtained through broadband spectral energy distribution (SED) modelling and combinations of rest-frame UV and infrared (IR) observations (e.g. \citealt{Burgarella2007,Burgarella2013,Kriek2013, Buat2015,Salim2018}; see also the review by \citealt{Salim2020}), revealing substantial diversity in attenuation-curve shapes. At higher redshifts, such diversity appears to persist up to $z\sim5$ (e.g. \citealt{Boquien2022}). Nonetheless, pre-JWST studies of galaxy attenuation at these epochs faced significant limitations. At high redshift, attenuation curves for integrated galaxy populations were generally inferred through broadband SED modelling, comparisons between dusty and low-dust analogues, or rest-frame UV and IR diagnostics, all of which depend to varying degrees on assumptions about the intrinsic stellar populations and both dust absorption and emission. Spectroscopic studies based on nebular reddening, such as those relying on the Balmer decrement, were instead largely restricted to $z\lesssim2.5$ by the difficulty of detecting rest-frame optical emission lines from the ground \citep[e.g.][]{Reddy_2015}. As a result, assembling sizeable and homogeneous galaxy samples with simultaneous spectroscopic constraints on nebular reddening and broad rest-frame UV-to-optical continuum coverage remained challenging prior to JWST.

The advent of the JWST has fundamentally transformed this landscape. Its unprecedented sensitivity and continuous wavelength coverage from $0.7\mbox{--}5.0\,\micron$ allow rest-UV and optical light to be probed at epochs previously inaccessible, while medium-resolution NIRSpec spectroscopy delivers robust spectroscopic redshifts, nebular line measurements, and continuum constraints for typical star-forming galaxies well into the reionisation era. 

Recent JWST studies have started to investigate the physical origin of the observed diversity of attenuation curves at high redshift. In particular, several galaxies appear to exhibit significantly flatter attenuation curves than those commonly observed in the Local Universe. One possible interpretation is that these grey attenuation curves reflect dust populations dominated by relatively large grains, produced by massive stellar sources and only weakly reprocessed within the ISM owing to the young ages of these systems \citep[e.g.][]{Markov2025}. An alternative interpretation emphasises the role of the relative geometry between stars and dust, showing that radiative transfer effects, the spatial distribution of dust, and the fraction of unobscured young stars can substantially modify the observed attenuation curve even for a fixed intrinsic dust extinction law \citep[e.g.][]{SeonDraine2016,Fisher2025}. These scenarios should not be regarded as mutually exclusive. Indeed, both theoretical models and numerical simulations indicate that attenuation curves are shaped by the combined effects of dust grain composition and size distribution, dust--star geometry, optical depth, and the evolutionary state of the ISM \citep[e.g.][]{McKinnon2018,Narayanan2018,Hirashita2020,Sommovigo26a}. Disentangling the relative importance of these different physical mechanisms remains one of the major challenges in interpreting dust attenuation measurements at high redshift. In this context, robust empirical determinations of the average attenuation law provide an important observational benchmark against which different physical models can be tested.

Another point of consideration is the UV-bump feature. The $2175\,\AA$ UV bump is one of the most prominent features of dust attenuation curves and provides a powerful diagnostic of dust composition and processing in galaxies.  It is commonly attributed to small carbonaceous dust grains, including polycyclic aromatic hydrocarbons \citep[PAHs;][]{Gordon2024}.  JWST has considerably expanded the body of evidence to support the presence of this feature at high redshift. A pronounced UV bump was first identified spectroscopically in the $z\simeq6.7$ galaxy JADES-GS-z6-0, supporting that chemically evolved carbonaceous dust may already be present within the first $\sim 800\,\mathrm{Myr}$ of cosmic history \citep{Witstok2023}. Subsequent studies have reported additional detections and constraints on the UV bump in galaxies across the reionisation era and Cosmic Dawn \citep[e.g.][]{Markov_2024,Markov2025,Fisher2025,Ormerod2025}, revealing substantial diversity in both the presence and strength of the feature. Together with the growing evidence of a wide range of attenuation-curve slopes, these results suggest that dust attenuation properties at high redshift are more diverse than previously assumed and may differ significantly from local empirical prescriptions. These findings further motivate the adoption of flexible attenuation parametrisations when modelling JWST observations and highlight the importance of deriving empirical attenuation laws directly from spectroscopic observations. Indeed, \cite{Shivaei2025} adopt a flexible attenuation curve formalism to fit NIRCam grism spectra and photometry for $\sim 3300$ galaxies at $1<z<9$, finding systematic flattening of the attenuation curve towards earlier epochs and higher $A_V$. These studies emphasise the importance of not imposing a fixed attenuation prescription when interpreting JWST data; specifically, they showed that allowing the curve shape to vary instead is essential to capture the physical diversity of high-redshift galaxies. These results provide important context and motivation for the detailed spectroscopic analysis presented in this paper.

In this work, we exploit the spectroscopic capabilities of JWST to derive an empirical determination of the average dust attenuation law for massive star-forming galaxies with $\log_{10}(M_\star/M_\odot)>9$ over the redshift range $2 < z < 7$. By combining NIRSpec spectroscopy from the JADES survey with deep multi-wavelength photometry from the ASTRODEEP-JWST catalogues \citep{Merlin_2024}, we applied an empirical approach based on the relation between nebular reddening and stellar continuum reddening originally introduced by \cite{Calzetti_1994} attenuation curve.\footnote{By selective attenuation, we refer to a curve that describes the differential attenuation as a function of wavelength, independent of the total dust content in galaxies. This curve is derived from ratios between average spectral templates constructed from galaxy samples grouped into bins of Balmer optical depth. Consequently, the zero point of the relation is arbitrary.} This method provides an empirical spectroscopic constraint on the average attenuation law of star-forming galaxies at early cosmic epochs, complementing previous observationally constrained studies based on different empirical and modelling approaches \citep{Reddy_2015,Battisti_2016,Battisti_2022,Shivaei_2020,Markov2025,Fisher2025}.

This paper is organised as follows. In Sect.\,\ref{sec:data}, we describe the photometric and spectroscopic datasets used in this work. In Sect.\,\ref{sec:meth} we present the sample selection and the methodology adopted to derive the attenuation curve. Section\,\ref{sec:sed_fitting} describes the SED fitting procedure and the derivation of the physical properties of the galaxy sample. In Sect.\,\ref{sec:dustatt}, we present the derivation of the selective and total attenuation curves. In Sect.\,\ref{sec:Disc}, we discuss the implications of our results in the context of recent JWST studies of dust attenuation at high redshift. Finally, Sect.\,\ref{sec:conclusions} summarises our main conclusions. Throughout this paper, we adopt a flat $\Lambda$CDM cosmology with $H_0 = 70\,\kmsMpc$ and $\Omega_{\Lambda} = 0.7$. We assume a \citet{Kroupa} initial mass function (IMF) and AB magnitudes \citep{1983ApJ...266..713O}.

\section{Data sets}\label{sec:data}

\subsection{ASTRODEEP NIRCAM photometry}

For the purposes of this work, we exploited the recently released \textsc{ASTRODEEP-JWST} photometric catalogues \citep{Merlin_2024}, which provide a homogeneous, deep multi-wavelength dataset optimised for the study of galaxy populations in the high-redshift Universe. The catalogues combine NIRCam imaging from eight JWST programs: GLASS-JWST \citep{Treu2022}, UNCOVER \citep{Bezanson2024}, DDT2756, GO3990, CEERS \citep{Finkelstein2022}, JADES \citep{Eisenstein2023}, NGDEEP \citep{Bagley2024}, and PRIMER (GO 1837; PI: Dunlop), spanning six well-studied extragalactic fields.

The \textsc{ASTRODEEP} pipeline integrates JWST/NIRCam mosaics with up to 16 HST bands, with a wavelength coverage ranging from $0.44$ to $4.4\,\micron$. All images are homogenised onto the NIRCam pixel grid, aligned with Gaia-DR3 astrometry \citep{GaiaDR3}, and scaled to \si{\uJy} units, allowing for consistent colour measurements across fields. Source detections were performed on a weighted stack of the F356W and F444W mosaics, optimised for the identification of faint (and dusty) high-redshift systems. The adopted detection strategy corresponds to an effective signal-to-noise ratio $\mathrm{S/N}\!\approx\!2$ threshold, ensuring high completeness for galaxies at $z\gtrsim4$.

Photometry is extracted using the \texttt{A-PHOT} package \citep{2019A&A...622A.169M}, operating on point spread function-matched (PSF-matched) images. Colours are measured in fixed circular apertures and scaled to total fluxes via Kron-like elliptical apertures defined on the detection stack. Empirical PSF, built from high-S/N stars across all fields, are used to homogenise the resolution, improving the reliability of fluxes and colour gradients. Each catalogue includes both global and locally background-subtracted measurements; in this work we adopt the latter, which provides more stable estimates in crowded regions and near bright cluster members.

The dataset reaches typical $5\sigma$ limiting magnitudes of $\mathrm{AB}\approx 29 \mbox{--} 30$ in the short-wavelength NIRCam bands and $\mathrm{AB}\approx 28.5 \mbox{--} 30.5$ in the long-wavelength filters, depending on the field depth. Detection completeness exceeds 90\% at $\mathrm{AB}\approx 28.5 \mbox{--} 29.5$ and $50\%$ at $\mathrm{AB}\approx 30 \mbox{--} 30.5$, with NGDEEP and JADES providing the deepest coverage. Each source is assigned a detailed quality flag that describes blending, saturation, missing coverage, and morphological classification (point-like vs. extended). Spurious detections are identified via a Principal Component Analysis-based diagnostic that incorporates SExtractor \citep{Bertin1996} morphology parameters; all flagged spurious or saturated sources are excluded from our analysis.

\subsection{JADES: NIRSpec spectra and emission line catalogue}\label{jades}

We also made use of imaging and spectroscopic data from JADES, a large GTO programme designed to obtain some of the deepest JWST observations in the GOODS-South and GOODS-North legacy fields \citep{Deugenio2025,CurtisLake2025}. JADES combines broad-band NIRCam imaging with medium- and low-resolution NIRSpec spectroscopy, providing an unparalleled view of galaxies in the early Universe across $0.6\mbox{--}5.3\,\micron$. Our analysis relies heavily on the JADES NIRSpec spectroscopic campaign, which includes observations obtained with the \textsc{PRISM}, \textsc{R1000}, and \textsc{R2700} configurations. The spectroscopic footprint consists of multiple NIRSpec MSA configurations in both GOODS fields, targeting galaxies spanning a wide range of redshifts, luminosities, and stellar masses.

Importantly, the JADES public release includes fully calibrated  nebular emission-line flux measurements for all detected lines, together with  their uncertainties, continuum levels, and quality flags. These homogeneous  measurements, derived from the dedicated JADES extraction and fitting  pipeline \citep{Deugenio2025}, provide fluxes for key transitions such as \ha, \hb, [\ion{O}{III}], [\ion{O}{II}], [\ion{N}{II}], [\ion{S}{II}], and numerous rest-UV emission lines at $z\gtrsim4$.

All NIRSpec spectra are processed with the dedicated JADES reduction pipeline \citep{Deugenio2025} which supplements the standard JWST calibration steps with customised routines optimised for faint-source extraction, background subtraction, bad-pixel identification, spectral registration, and noise modelling. The high data quality permits secure detection of rest-frame optical and UV nebular lines out to $z>10$, enabling detailed physical characterisation of galaxies in the early Universe.

\subsection{MIRI photometry}\label{MIRI}
For the JWST/MIRI images of GOODS-S, we used the mosaics provided by the Data Release 1 of the Systematic Mid-infrared Instrument Legacy Extragalactic Survey \citep[SMILES,][]{2024ApJ...975...83R, 2024ApJ...976..224A}, which were converted to \si{\uJy} and matched to the pixel scale of the JADES NIRCam images using \texttt{SWarp} \citep{2002ASPC..281..228B}. Then, similarly to the procedure outlined in \cite{Merlin_2024}, the RMS maps, which include the Poisson noise, were rescaled by a multiplicative factor to make them consistent with the dispersion of fluxes measured in empty regions of the scientific images within $300$ random circular apertures with diameter equal to the FWHM of the filter relative to the measurement image. Finally, we performed template-fitting photometry using \texttt{T-PHOT} v2.0 \citep{2016A&A...595A..97M} using synthetic JWST/NIRCam and JWST/MIRI PSFs generated with STPSF \citep{Perrin2014S}. The overall JWST/MIRI catalogue will be presented in Catone et al. (in prep.). 

\section{Methodology and sample selection}\label{sec:meth}
To derive the wavelength-dependent dust attenuation curve, we adopted the empirical methodology introduced by \citet{Calzetti_1994} and refined by \citet{Battisti_2016}. This methodology relies on the relation between the reddening of the ionised gas (probed by the Balmer decrement) and the reddening of the stellar continuum. We adapted this approach to the combination of deep JWST/NIRSpec spectroscopy from JADES and broad-wavelength photometry from the ASTRODEEP-JWST catalogue. In Sect.\,\ref{sec:sed_fitting}, we describe how we used it to derive the physical properties of the spectroscopic sample. We note, however, that the relation between nebular and stellar attenuation exhibits substantial intrinsic scatter and could depend on galaxy properties and redshift \citep[e.g.][]{Battisti_2022}. Therefore, in the present analysis, the Balmer optical depth is used as a statistical tracer of the average stellar continuum reddening across the galaxy population, rather than as a precise predictor of the continuum attenuation of individual galaxies. We further investigated this relation and its scatter for our sample in Sect.\,\ref{beta-tau-relation}.

To this end, we started with the 4086 targets available in the JADES DR3 release. Our selection required reliable measurements of both \ha\ and \hb\ emission lines to estimate the Balmer optical depth, $\tau_B^l$ (see Sect.\,\ref{sec:balmer_depth}), as well as rest-frame UV photometry to derive the UV-$\beta$ slope (see Sect.\,\ref{sec:beta}). Thus, we first selected targets with available \ha\ and \hb\ fluxes from the medium-resolution grating spectra (able to resolve \ha\ from [\ion{N}{II}]), obtaining 584 sources (after excluding sources flagged by the data-reduction quality flags in the JADES catalogue). By cross-matching this spectroscopic sample with the photometric catalogues (using a positional tolerance of $\ang{;;1}$), we found 504 objects spanning a redshift interval of $z \sim 0.6\mbox{--}6.9$.

Slit loss corrections were applied by comparing the spectra of each source with its Kron photometry obtained from the JADES photometric catalogue \citep{Rieke2023}. Each spectrum was convolved with the NIRCam filter throughput curves,\footnote{\url{https://jwst-docs.stsci.edu/gsc.tab=0}} and the resulting synthetic photometry was fitted to the catalogue values using a first-order polynomial with \texttt{scipy.optimize.curve\_fit} \citep{Virtanen2020}. Both the spectra and their uncertainties were then scaled accordingly. 
We also tested the impact of the uncertainties associated with the wavelength-dependent slit-loss corrections on the spectral continuum shape. For each galaxy, we generated Monte Carlo realisations of the coefficients of the first-order polynomial correction, accounting for the uncertainties and covariance of the fitted slope and intercept. For each realisation, we evaluated the correction relative to its value at the observed wavelength corresponding to $5500\,\AA$ in the rest frame, consistently with the normalisation adopted in the construction of the stacked spectral templates. For the $118$ galaxies entering the final stacks, the median uncertainty induced on the relative continuum shape is typically $\sim1\mbox{--}3\%$ over most of the wavelength range. Repeating the test separately in the four Balmer-optical-depth bins used to construct the attenuation templates, we find no systematic dependence on $\tau_{\mathrm{B}}^{l}$. The median wavelength-dependent perturbation within each $\tau_{\mathrm{B}}^{l}$ bin remains below $\sim1\%$ over nearly the entire wavelength range. We therefore conclude that uncertainties in the slit-loss corrections have a negligible impact on the relative shapes of the stacked spectra and, hence, on the derived average attenuation curve. Fourteen sources lacking photometric counterparts were excluded from subsequent analysis.

\subsection{Balmer optical depth}\label{sec:balmer_depth}
Dust attenuation in galaxies can be quantified through wavelength-dependent optical depth, $\tau(\lambda)$, which relates the intrinsic and observed flux densities as
\begin{equation}
F_{\lambda}^{\rm obs} = F_{\lambda}^{\rm int} e^{-\tau(\lambda)}.
\end{equation}
As anticipated, in this work we use the Balmer decrement, $F(\mathrm{H}\alpha)/F(\mathrm{H}\beta)$, to trace  the  dust attenuation in the ionised gas. We define the Balmer optical depth as
\begin{equation}
    \tau_{\rm B}^l = \ln\left[\frac{F({\rm H\alpha})/F({\rm H\beta})}{2.75}\right],
\end{equation}
where the superscript \textit{l} indicates that $\tau_B$ has been obtained from the emission line and should be distinguished from optical depths associated with the stellar continuum. We adopted an intrinsic Case B recombination ratio of $\mathrm{H}\alpha/\mathrm{H}\beta=2.75$, appropriate for $T_{\rm e}=2\times10^4\,\mathrm{K}$ and $n_{\rm e}=100\,\mathrm{cm}^{-3}$ \citep{Storey1995}. This choice reflects the higher electron temperatures expected for the low-metallicity star-forming regions
typically observed at high redshift, compared to the canonical $T_{\rm e}=10^4\,\mathrm{K}$ assumption underlying the commonly adopted
intrinsic ratio of $2.86$ used in local and intermediate redshift studies \citep[e.g.][]{Calzetti_1994,Battisti_2016,Reddy_2015}.

\subsection{Stellar continuum reddening}\label{sec:beta}
The stellar continuum reddening is traced using the rest-UV continuum slope $\beta$, defined as $F_\lambda \propto \lambda^{\beta}$. For each galaxy, we computed $\beta$ using NIRCam photometry as
\begin{equation}
\beta_{\mathrm{UV}} = 
\frac{\log_{10}\!\left[ F_{\lambda}(\mathrm{FUV}) / F_{\lambda}(\mathrm{NUV}) \right]}
{\log_{10}\!\left( \lambda_{\mathrm{FUV}} / \lambda_{\mathrm{NUV}} \right)} \,, 
\end{equation}
where the flux density is in $\mathrm{erg\,s^{-1}\,cm^{-2}\,\AA^{-1}}$.

We linearly interpolated the fluxes measured in the bands closest to the rest-frame far-UV (FUV) and near-UV (NUV) bands. We adopted $\lambda_{\mathrm{FUV}} = 1516\,\AA$ and $\lambda_{\mathrm{NUV}} = 2267\,\AA$, as in \citet{Battisti_2016}. To account for some bad photometric data, we discarded rest-frame UV bands for which the flux ratio between adjacent photometric measurements differed by more than an order of magnitude, or where the uncertainty on the flux was larger than the measured flux itself. The associated uncertainties were propagated through a Monte Carlo approach, repeatedly perturbing the data within their observational errors and deriving confidence intervals from the percentiles of the resulting distribution of regression parameters. 

\subsection{Sample selection and cleaning criteria}\label{sample}
Starting from the sample of $504$ sources with \ha\ and \hb\ emission-line measurements and photometric counterparts, we applied a series of selection criteria during the estimation of $\tau_B^l$ and $\beta$, as well as during the spectral analysis, to ensure the quality of the final sample.

Flux measurements in the NUV and FUV rest-frame bands, centred on $\lambda_\mathrm{NUV} = 2267\,\AA$ and $\lambda_\mathrm{FUV} = 1516\,\AA$, respectively, are required to compute the UV-$\beta$ slope. Therefore, our analysis is limited to targets at redshifts $z > 1.86$, ensuring that these rest-frame wavelengths fall within the observed photometric coverage. We further performed a photometric quality assessment by removing bands affected by spurious flux measurements, including catalogue-flagged values.
 
Afterwards, we removed targets with negative (non-physical) $\tau^l_B$, while keeping those that were consistent with zero within their uncertainties. Moreover, three sources with $\tau^l_B>0.9$ were excluded because this high-$\tau^l_B$ regime is sparsely populated and cannot be robustly sampled in our analysis. We then excluded active galactic nuclei (AGNs) from the sample, relying on the classifications provided by \cite{Maiolino_AGN} and \cite{Scholtz_AGN}, so that we could be sure to retain only star-formation-dominated systems and avoid contamination from non-stellar ionising sources.

Finally, the spectral quality of the remaining sources was assessed through a combination of visual inspection S/N analysis. By visual inspection, we identified a systematic, shutter-related instrumental feature in some spectra, appearing as an artificial excess around $\sim 4000\,\AA$. These sources were excluded from the analysis. An example of this peculiar feature is shown in Fig.\,\ref{fig:bump_plot}. We also removed targets that  exhibit an unusually prominent Balmer break (i.e. $\mathrm{D_n}4000>1.5$) compared to the bulk of the sample to avoid biases in the derivation of the attenuation curve. Such spectra are likely dominated by more evolved stellar populations and are therefore not representative of the actively star-forming systems that we aim to characterise in this study.

We examined the continuum S/N across the entire spectral range. Since the most informative portion of the attenuation curve lies in the UV $-$  which is also the noisiest region due to the limited resolving power of the prism $-$  we retained only sources with $\mathrm{S/N} > 2$ in the rest-frame interval $1500 < \lambda/\AA < 4000$ to better constrain the shape of the selective attenuation curve. This requirement may in principle introduce selection biases, as faint, high-redshift, or heavily dust-obscured galaxies are expected to exhibit lower UV fluxes and thus lower S/N in this wavelength range. However, the median redshift of the sample changes negligibly (changes by only $\sim1.4\%$) before and after the cut, indicating that no significant redshift-dependent bias is introduced. Moreover, all excluded targets show relatively low Balmer optical depths ($\tau^l_B \lesssim 0.5$), suggesting that they are not significantly dust-obscured. Therefore, this selection does not bias the sample against dusty galaxies.

As discussed in Sect.\,\ref{stellar_mass_estimate}, we restricted our analysis to galaxies with $\log_{10}(M_\star/M_\odot) > 9$. This conservative mass selection facilitates a more homogeneous comparison with previous studies of relatively massive star-forming galaxies at early universe \citep[e.g.][]{Simmonds2024}. Consequently, the attenuation curve derived in this work should be regarded as representative of the massive star-forming galaxy population at $2<z<7$, rather than of the full galaxy population over the same redshift interval. We note that this mass cut does not introduce any systematics in the redshift distribution of the sample. Indeed, as shown in Fig.\,\ref{z_tau_sSFR}, the average redshifts of the sources in each $\tau^l_B$ bin are compatible within $1\sigma$ with the full $\log_{10}(M_\star/M_\odot) > 9$ sample (i.e. $z\sim3.45$). The application of the selection criteria, summarised in Table\,\ref{sample_selection_table}, results in a clean sample of 118 galaxies, where the number of sources discarded at each step is also indicated.

\begin{table}[t] 
    \centering
    \caption{Summary of the selection sample and cleaning steps.} \label{sample_selection_table} 
    \renewcommand{\arraystretch}{1.2}
        \begin{tabular}{lcc}
        \hline
        \hline
        \textbf{Filtering step} & \textbf{Removed\,targets} & \textbf{Sample size} \\ 
        \hline \noalign{\vskip 1pt}
        JADES DR3 spectra & --- & 4086 \\
        \ha\ and \hb\ emission lines & 3502 & \phantom{0}584 \\
        ASTRODEEP photometry & \phantom{0}\phantom{0}80 & \phantom{0}504 \\
        $\beta$ computation & \phantom{0}\phantom{0}81 & \phantom{0}423\\
        $\tau^l_B$ computation & \phantom{0}\phantom{0}33 & \phantom{0}390 \\
        AGN  & \phantom{0}\phantom{0}21  & \phantom{0}369 \\
        Spectral visual inspection & \phantom{0}\phantom{0}15 & \phantom{0}354 \\
        $S/N < 2\,(1500\mbox{--}4000\,\AA)$ & \phantom{0}\phantom{0}25 & \phantom{0}329 \\
        $M_{\star} > 10^9\,M_\odot $ & \phantom{0}211 & \phantom{0}118 \\
        \hline
        \hline
        \end{tabular}
\end{table}
\begin{table}[t]
    \centering
    \caption{Priors used in the \textsc{Bagpipes} configuration for this work.} \label{priors_bagpipes} 
    \begin{tabular}{ccc}
    \hline \hline \noalign{\vskip 1pt}
     & \textbf{Parameters} & \textbf{Range} \\ 
    \hline \noalign{\vskip 1pt}
    \textbf{Dust} & $\mathrm{A}_V\,[\mathrm{mag}]$  & [0, 6] \\
    \noalign{\vskip 2pt}
    \textbf{Nebular Emission} & $\log_{10}(U)$       & [-4, -1] \\
    \noalign{\vskip 4pt}
    \multirow{4}*{\textbf{Delayed-\textbf{$\tau$} model}} & $T_0$ [Gyr] &  [0.001, 15] \\
                & $\tau$ [Gyr] & [0.01, 10] \\
                & $Z/Z_{\odot}$ & [0, 2.5] \\
                & $\log_{10} (M_{*}/M_{\odot})$ & [6, 12.5] \\
    \hline \hline
    \end{tabular}
    \tablefoot{The prior is assumed uniform within the allowed ranges.}
\end{table}

\section{SED fitting and physical properties}\label{sec:sed_fitting}

To derive the physical properties of our galaxy sample, we performed a photometric SED fitting using the \textsc{Bagpipes} code \citep{Carnall_2018}. The SED fitting was performed using the full set of 16 HST and JWST/NIRCam photometric bands available from the ASTRODEEP-JWST catalogue, spanning $0.44\mbox{--}4.44\,\mu{\rm m}$. These comprise the HST/ACS F435W, F606W, F775W, and F814W filters; the HST/WFC3 F105W, F125W, F140W, and F160W filters; and the JWST/NIRCam F090W, F115W, F150W, F200W, F277W, F356W, F410M, and F444W
filters. MIRI photometry was not included in the SED fitting. \texttt{Bagpipes} is a Bayesian framework designed to model galaxy spectra from the far-UV to microwave wavelengths by incorporating stellar population synthesis, nebular emission, and dust attenuation. The posterior distributions of the model parameters were estimated using the \textsc{MultiNest} nested sampling algorithm \citep{multinest_1,multinest_2}. Our specific model configuration includes a delayed-$\tau$ model with $\mathrm{SFR(t)} \propto (t-T_0)\,\mathrm{e}^{-(t-T_0)/\tau}$ for the star formation history (SFH), where $T_0$ is the time since star formation began and $\tau$ the timescale of its decline. Stellar populations models come from \cite{Bruzual_Charlot}, based on the MILES library for the UV-optical region. The assumed IMF is the \citet{Kroupa}, while nebular emission lines and continuum were modelled using the \texttt{Cloudy} photoionisation code \citep{ferland2023cloudy}, with the ionisation parameter $\log_{10}(U)$ treated as a free parameter. For the dust attenuation we adopted the attenuation relation presented \citep[][hereafter C00]{Calzetti2000}. We report in Table\,\ref{priors_bagpipes} the specific priors adopted for each input parameter of the SED fitting procedure. The redshift is fixed to that of the spectroscopic measurement provided by JADES.
\begin{figure}
    \centering
    \includegraphics[width=1\linewidth]{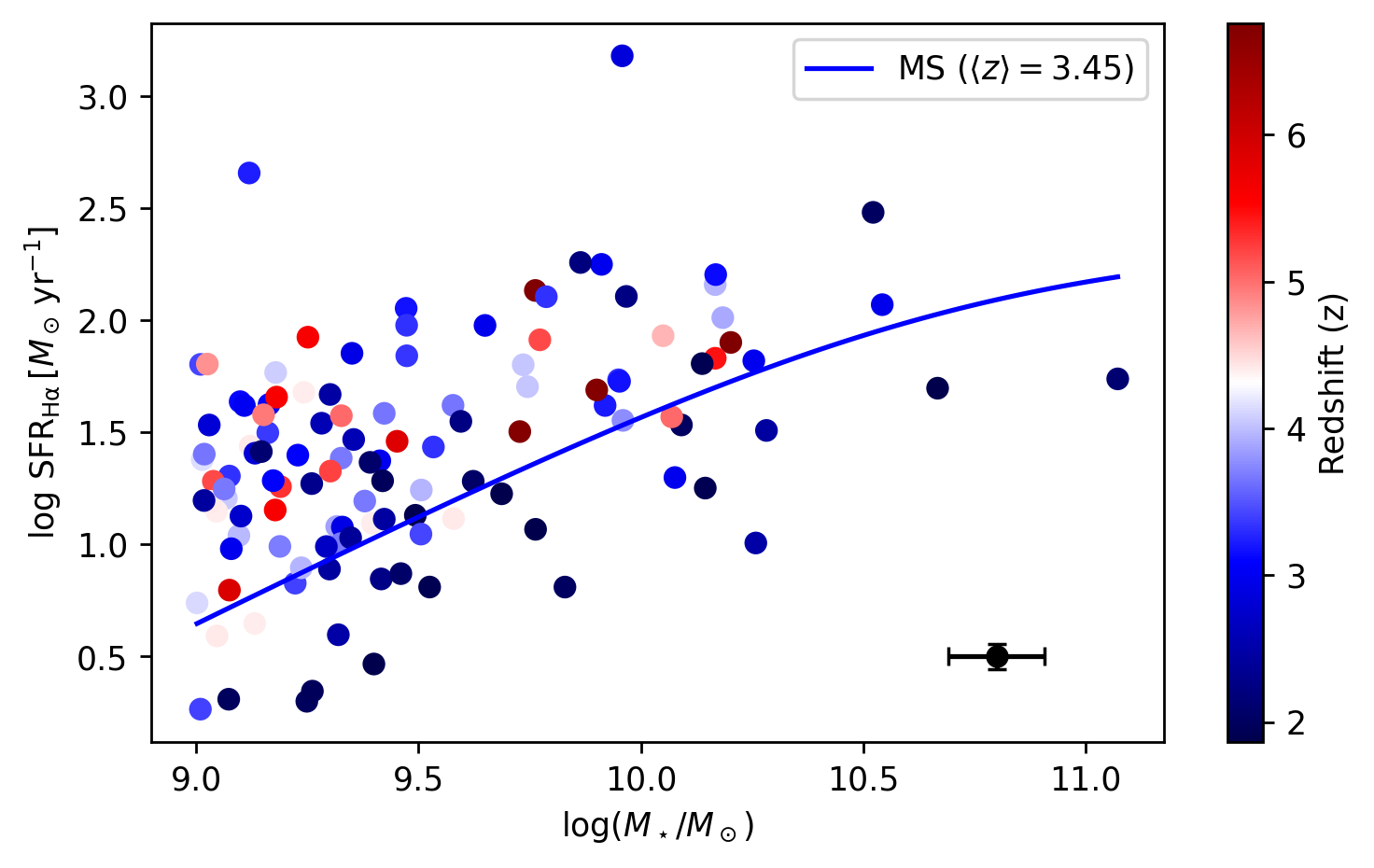}
    \caption{Distribution of SFR(\ha) as function of $M_\star$ for our sample, colour-coded by redshift; the blue line shows the star-forming main sequence (MS) at the mean redshift of the sample from \citet{MS_Popesso}, the error bar in the lower-right corner indicates the typical uncertainty on the two quantities.}
    \label{fig:MS}
\end{figure}
\begin{figure}
    \centering
    \includegraphics[width=1\linewidth]{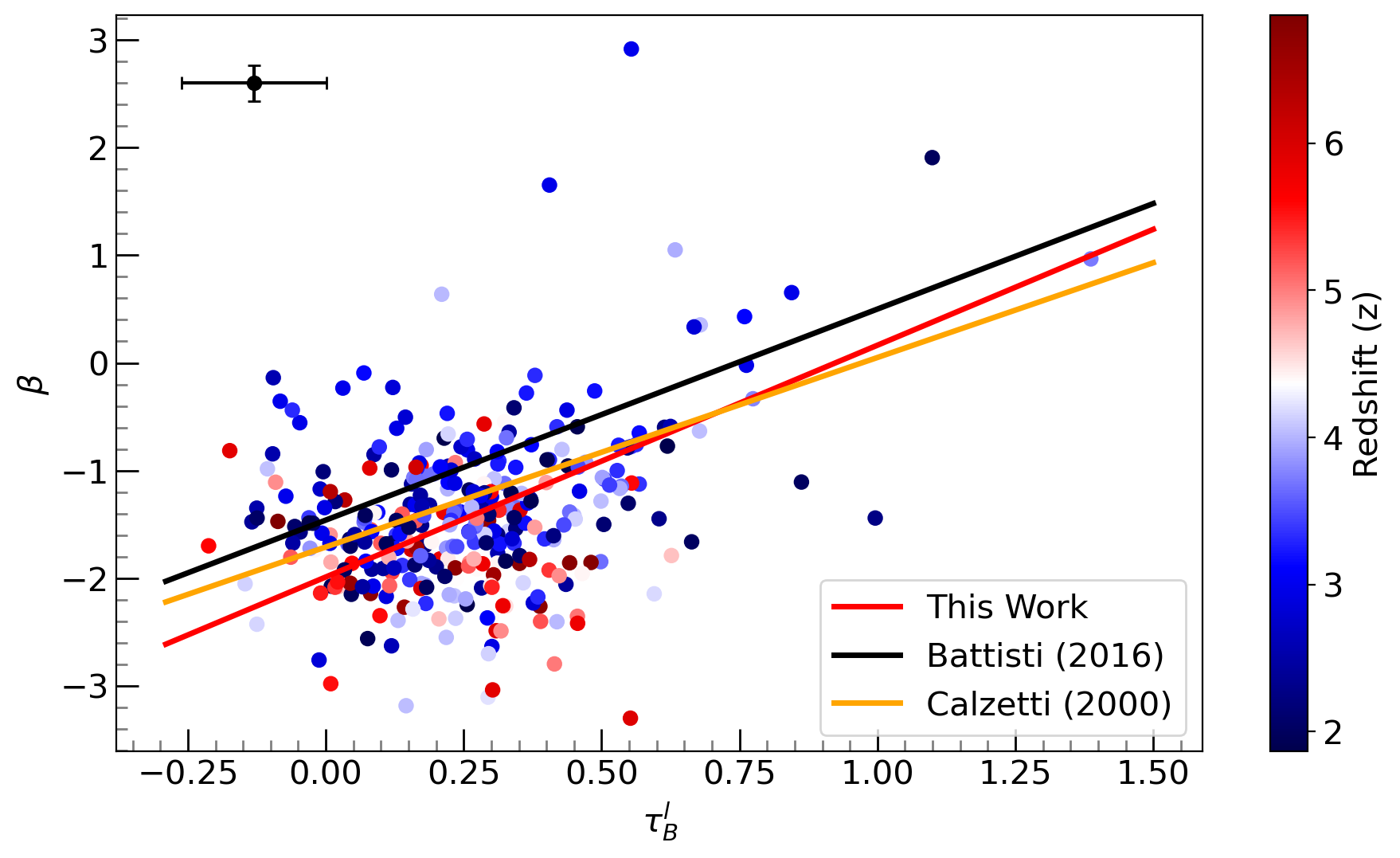}
    \caption{UV $\beta$ slope as a function of the Balmer optical depth, $\tau_B^l$, colour-coded with the redshift $z$ of each target. The black error-bar is the mean error on the parameters $\beta$ and $\tau^l_B$. The red line is our best linear fit (see details in the main text).}
    \label{fig:beta-tau}
\end{figure}
\begin{figure*}[h]
    \sidecaption
    \centering
    \includegraphics[angle=0,width=12.9cm]{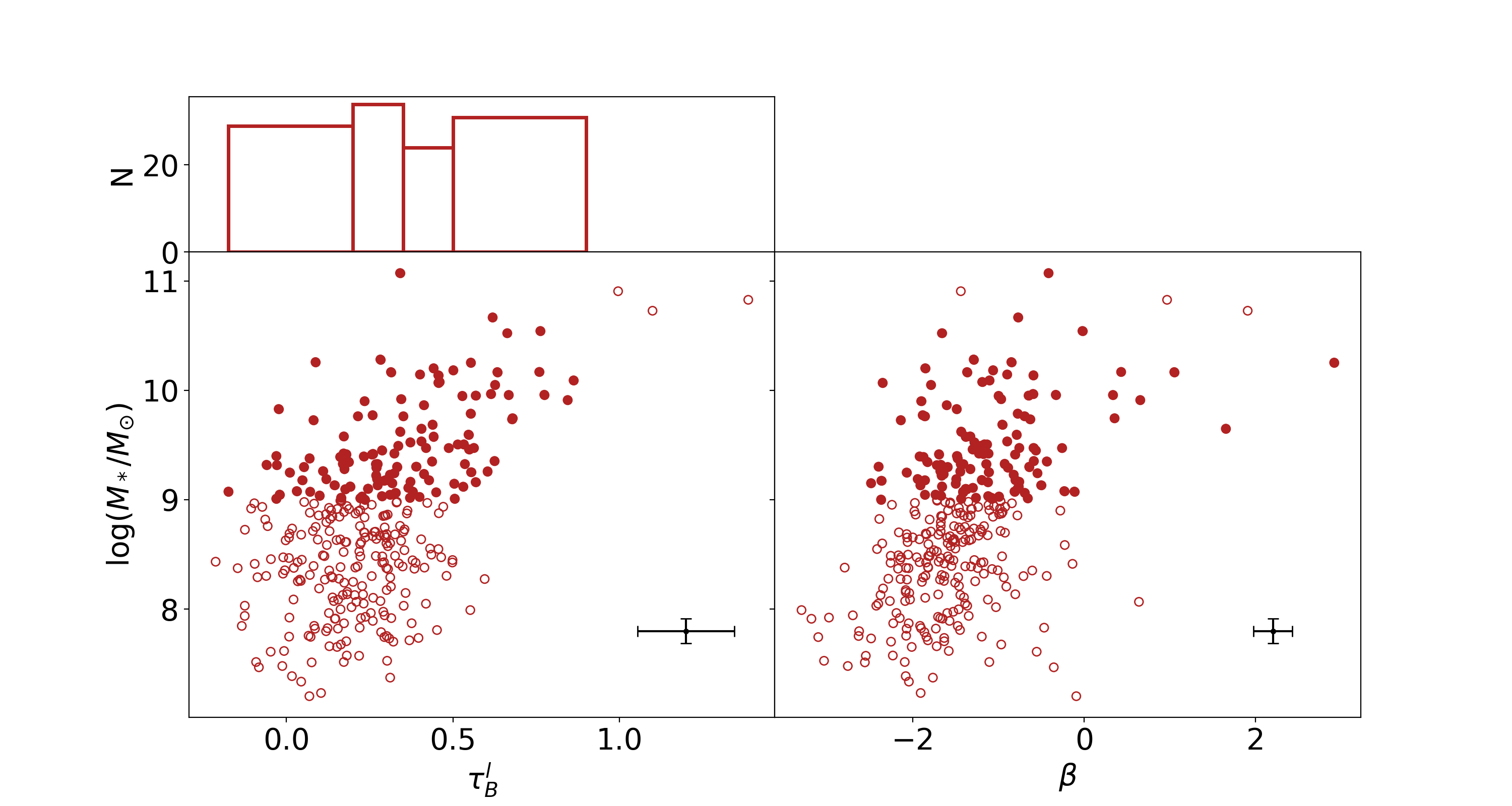}
    \caption{Dependence of the stellar masses on Balmer optical depth, $\tau_B^l$ (left panel), and on the UV continuum slope, $\beta$ (right panel), for the sample of $301$ galaxies at $1.86 < z < 7$. The filled brown circles mark the sources included in the derivation of the attenuation law. The histogram on top of the left panel indicates the $\tau_B^l$ bins used to compute the corresponding average templates.}\label{fig:mass_dust}
\end{figure*}

We used the \cite{Kennicutt1998} relation to convert the \ha\ luminosities into SFRs. The resulting observed SFRs were then corrected for dust attenuation by adopting the Galactic extinction curve $k(\lambda)$ of \cite{cardelli_1989}, which was also used to derive the corresponding nebular colour excess $E(B-V)_{\rm gas}$. We then estimated specific star formation rates (sSFR) from SFR(\ha) and the $M_\star$ derived from the SED-fitting. In Fig.\,\ref{fig:MS}, we show the distribution of SFR(\ha) as a function of $M_\star$.

We note that the stellar masses adopted in this work are model-dependent quantities and therefore inherit the assumptions of the adopted SED-fitting framework, including the assumed SFH, stellar population synthesis models, IMF, nebular emission, and dust attenuation prescription. Previous studies have shown that different modelling assumptions generally lead to systematic offsets in stellar-mass estimates of the order of $\sim 0.1\,\mathrm{dex}$ \citep{Carnall2019}. These uncertainties can increase to $\sim 0.4\,\mathrm{dex}$ when adopting different dust attenuation prescriptions \citep{Markov23}, while even larger discrepancies (in extreme cases approaching an order of magnitude) have been reported for highly bursty galaxies when adopting fundamentally different SFH parameterisations \citep{Topping2022}. Overall, recent studies suggest that the adopted SFH is among the primary drivers of systematic differences in stellar-mass estimates, while the attenuation prescription represents one of several additional modelling assumptions contributing to the inferred values \citep[e.g.][]{Ormerod2025}. Accordingly, the stellar masses used in this work should not be regarded as absolute measurements, but rather as internally consistent estimates within a well-defined SED-fitting framework. In the present work, these SED-derived stellar masses are used exclusively to define an approximately mass-complete galaxy sample, following the standard approach adopted in recent JWST studies \citep[e.g.][]{Simmonds2025}. The empirical attenuation curve is subsequently derived directly from the observed spectra and Balmer optical depths, independently of the best-fitting SED solution. The SED-derived stellar masses are therefore used exclusively for sample selection and do not enter the empirical derivation of the attenuation curve itself.

\section{The average dust attenuation in star-forming galaxies at $2<z<7$}\label{sec:dustatt}

\subsection{$\beta\mbox{--}\tau^l_B$ Relation}
\label{beta-tau-relation}

We can characterise dust attenuation in star-forming galaxies by comparing the UV continuum reddening, measured through $\beta_{\rm UV}$, with the optical reddening of the ionised gas traced by the Balmer optical depth, $\tau_B^l$. Since both quantities probe dust attenuation, a positive correlation is expected. Figure~\ref{fig:beta-tau} shows the relation between the two quantities. We fit the $\beta\mbox{--}\tau_B^l$ relation using the Bayesian linear-regression method implemented in the \texttt{linmix} package \citep{Kelly2007}, which accounts for measurement uncertainties in both variables and intrinsic scatter.

The best-fit relation is
\begin{equation}
\beta=(2.15\pm0.26)\,\tau_B^l-(1.99\pm0.08).
\end{equation}
We recovered the expected positive correlation between $\beta$ and $\tau_B^l$, consistent with a foreground-like dust geometry in which the reddening of the stellar continuum is correlated with that of the ionised gas. A Pearson correlation test yields $r=0.434$ with a $p$-value of $9.0\times10^{-7}$, indicating a statistically significant, moderately positive linear correlation between $\beta$ and $\tau_B^l$. The relation nevertheless exhibits substantial scatter, indicating that dust attenuation is not the sole parameter regulating the observed UV continuum slope. If the dispersion were driven primarily by variations in dust geometry, we might expect the scatter to decrease towards low $\tau_B$, where the dust column becomes small and geometric effects are expected to be less pronounced. Instead, we observe a comparable level of scatter over the full range of $\tau_B$, suggesting that dust geometry alone is unlikely to account for the observed dispersion. This interpretation is consistent with previous studies showing that variations in stellar populations, metallicity, and SFH also contribute to the intrinsic scatter of the $\beta\mbox{--}\tau_B$ relation \citep[e.g.][]{Battisti_2016,Reddy_2015,Salim_Nar2020}.

As discussed by \citet{Battisti_2022}, the $\beta\mbox{--}\tau_B^l$ relation displays significant diversity in both slope and normalisation across different galaxy samples. Our best-fit relation is consistent in slope with those derived by \citet{Calzetti_1994} and \citet{Battisti_2016}, while exhibiting a bluer intercept (i.e. the value of $\beta$ at $\tau_B^l=0$). The similar slope indicates that the average relation between UV continuum reddening and Balmer optical depth is broadly consistent with previous empirical determinations, whereas the bluer intercept is consistent with a bluer intrinsic UV continuum. This may reflect differences in stellar populations, metallicity, or other galaxy properties, although the present data do not allow for these effects to be disentangled. We therefore refrain from attributing the observed differences to any single physical mechanism.

\begin{figure}
    \centering
    \includegraphics[width=0.85\linewidth]{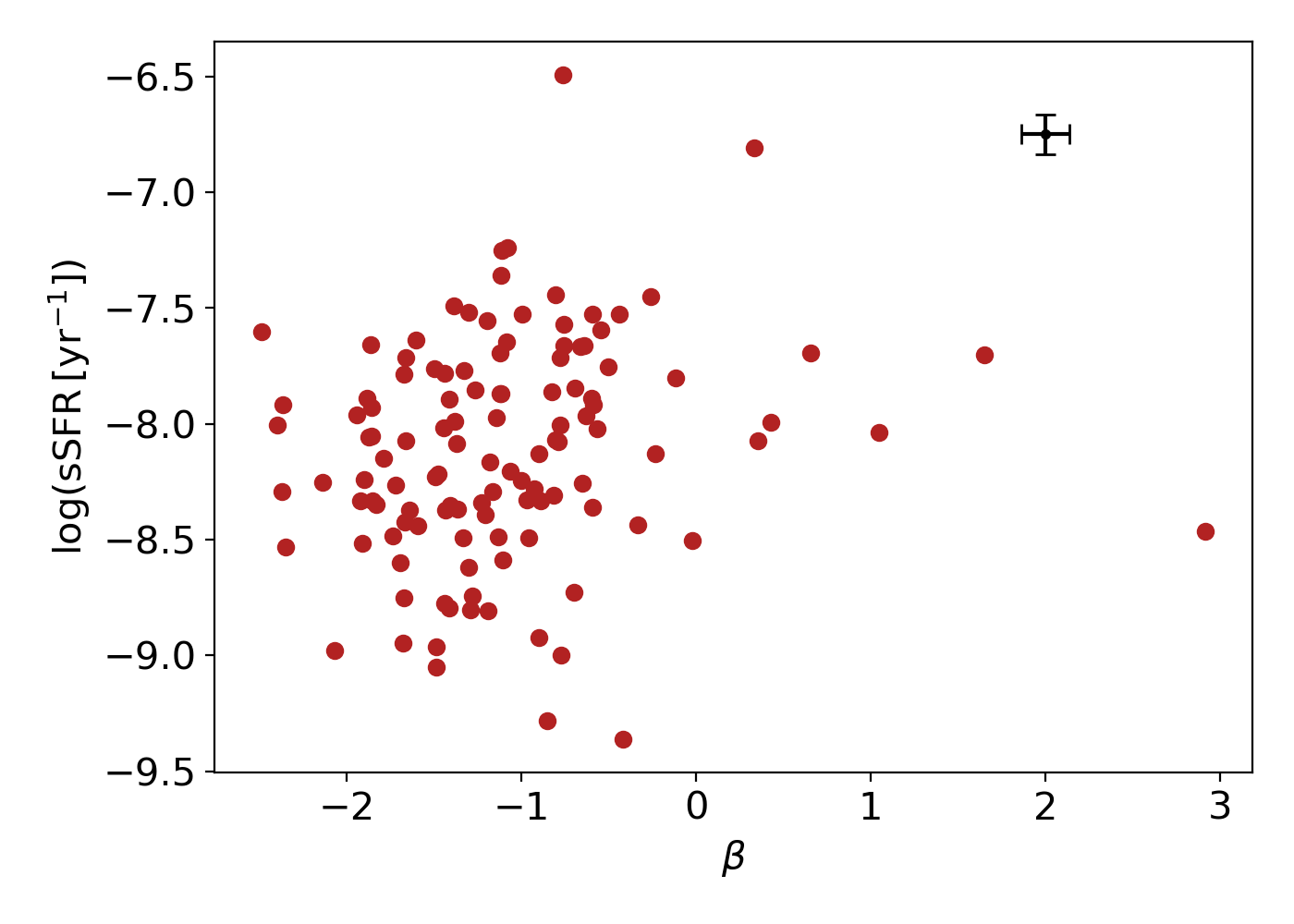}
    \includegraphics[width=0.85\linewidth]{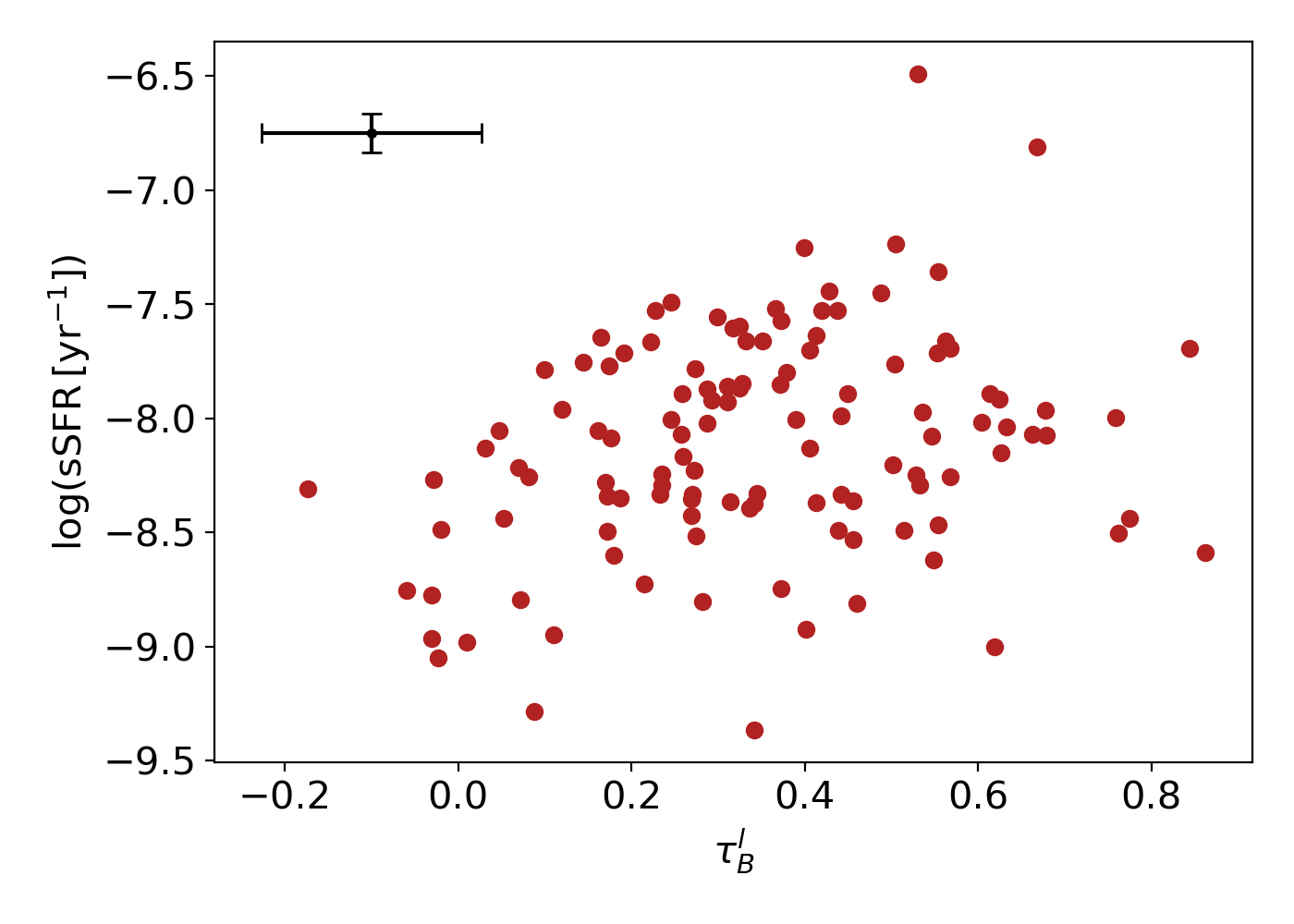}
    \caption{Specific star formation rate (sSFR), derived from the dust-corrected \ha\ luminosity, as a function of the UV continuum slope $\beta$ (top panel) and the Balmer optical depth $\tau_B^{\,l}$ (bottom panel) for the galaxies in our sample. Each point represents an individual galaxy, while the error bars in the upper-left corners indicate the typical uncertainties on the plotted quantities.} 
    \label{fig:sSFR}
\end{figure}

We do not find any clear evidence of a redshift dependence of the relation over the explored range, as indicated by the colour coding (see colourbar) in Fig.\,\ref{fig:beta-tau}. If it had been included in the compilation shown in Fig.\,10 of \cite{Battisti_2022}, our slope would not align with the reported trend, thereby reducing the evidence to support a redshift evolution for the $\beta\mbox{-}\tau^l_B$ slope. Although higher-redshift galaxies tend to exhibit slightly bluer UV slopes \citep{Cullen2023,Cullen2024,Dottorini2025}, this trend could be driven by selection effects rather than intrinsic evolution \citep[e.g.][]{Ikki25,Rodighiero2026}. For this reason, we analysed the full sample without binning in redshift, for the moment.

\begin{figure}
    \centering
    \includegraphics[width=1.\linewidth]{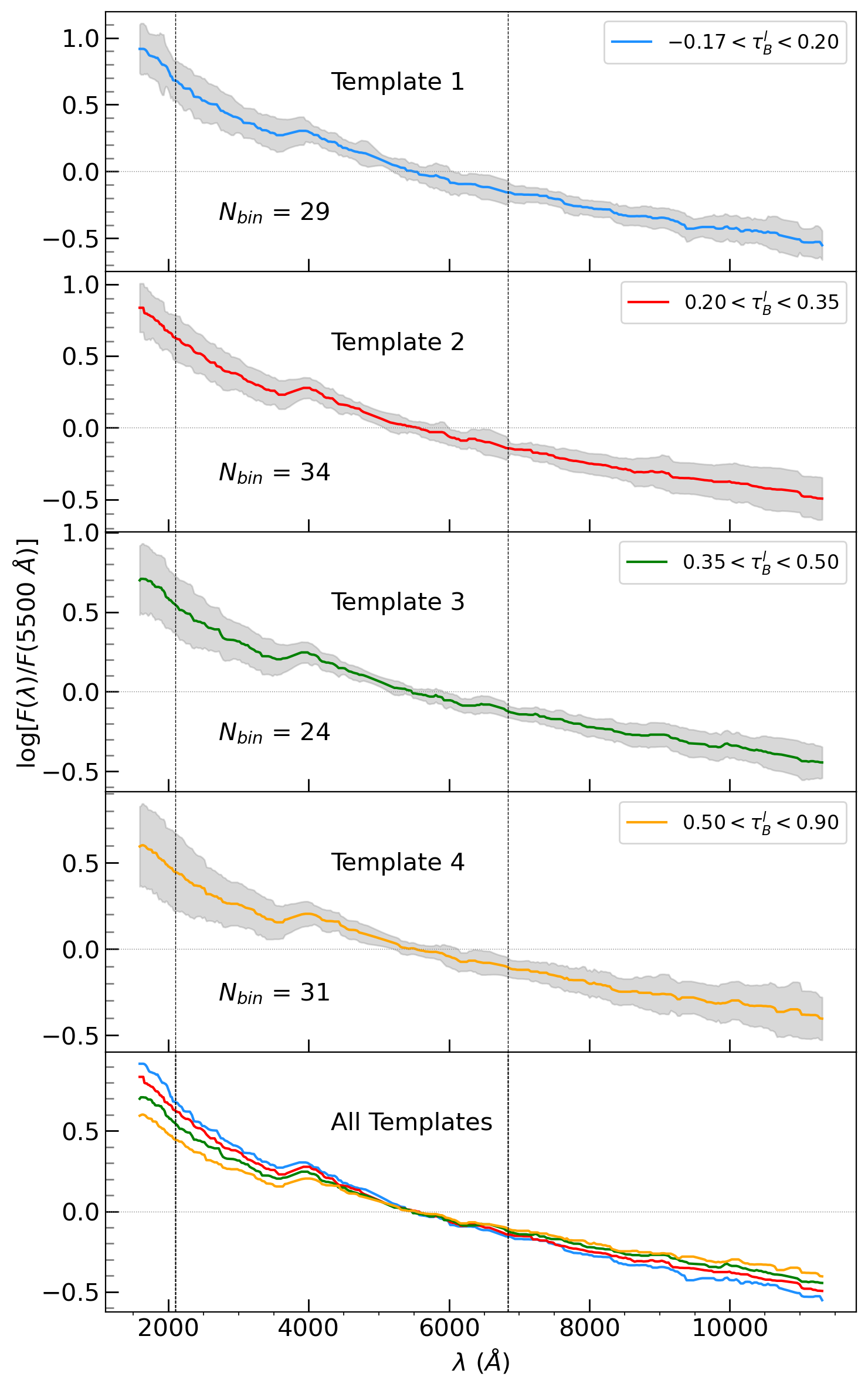}
    \caption{Average SEDs in $\tau^l_B$ bins. The gray region shows the $1\sigma$ dispersion of the sample spectra in each bin. The vertical dotted lines delimit the central wavelength range that is common to all targets in the sample; the number of spectra contributing to each bin within this region is indicated. The outer regions correspond to wavelengths covered by more than $50\%$ of the sources. The bottom panel overlap on the same scale the four templates to highlight the reddening of the UV continuum at increasing $\tau^l_B$. The shaded regions enclose approximately $68\%$ of the individual spectra contributing to each template at a given wavelength and therefore represent the object-to-object dispersion within each $\tau_{\mathrm{B}}^{l}$ bin, rather than the uncertainty on the mean template.}
    \label{fig:templates}
\end{figure}
\begin{figure}
    \centering
    \includegraphics[width=1.\linewidth]{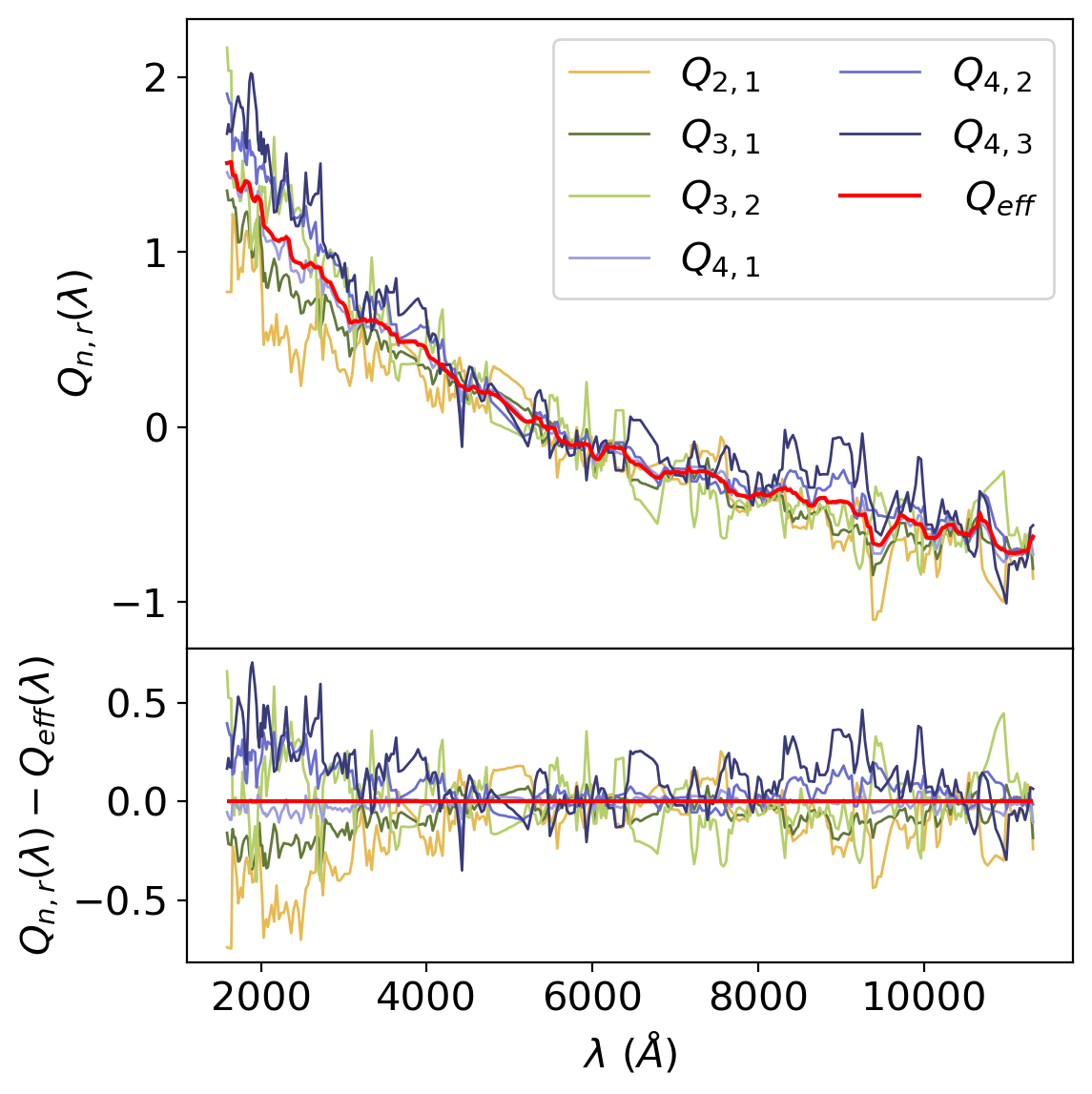}
    \caption{Selective attenuation curves derived from the stacked spectra in different $\tau_l^B$ bins. The top panel report the individual relative curves $Q_{n,r}(\lambda)$ obtained by comparing higher-dust bins to the lowest-$\tau_l^B$ reference bin (coloured lines), together with the effective average curve $Q_{\rm eff}(\lambda)$ (red line). The lower panel shows the residuals with respect to $Q_{\rm eff}(\lambda)$, demonstrating the consistency of the wavelength dependence across bins. }
    \label{fig:Qnr}
\end{figure}
\begin{figure}
    \centering
    \includegraphics[width=1.\linewidth]{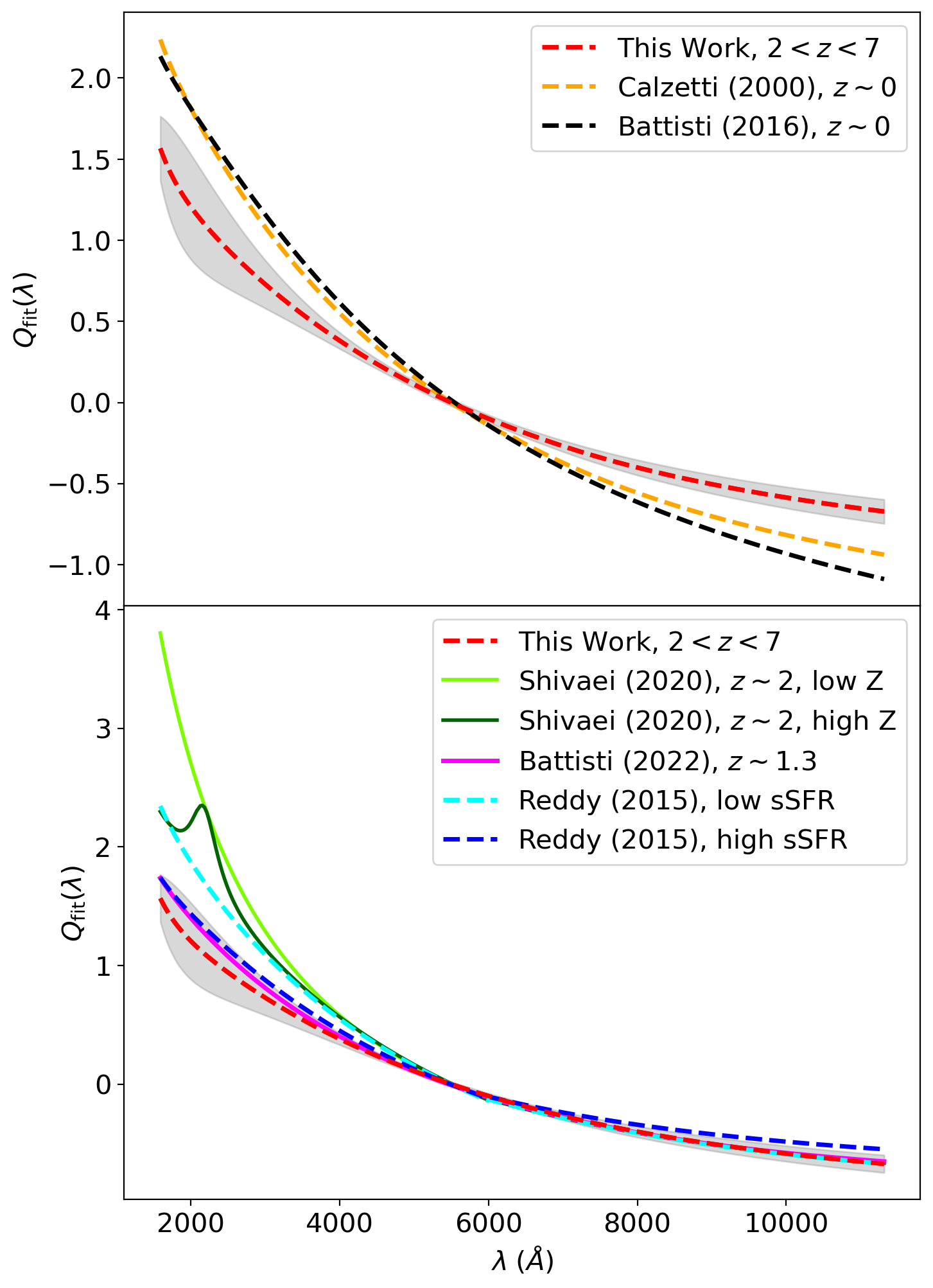}
    \caption{Effective selective attenuation curve $Q_{\rm eff}(\lambda)$ derived in this work for galaxies at $2<z<7$ (red dashed line) compared with previous empirical determinations. The grey shaded region represents the dispersion among the fits obtained from the individual $Q_{n,r}(\lambda)$ determinations and provides an empirical indication of the sensitivity of the derived selective attenuation curve to the choice of template pair; it should not be interpreted as a formal statistical uncertainty on $Q_{\rm eff}(\lambda)$. The upper panel shows the comparison with Local Universe relations from C00 and \citet{Battisti_2016}, highlighting that our curve is systematically flatter in the UV. The lower panel compares our result with attenuation curves derived at intermediate redshift \citep{Reddy_2015, Shivaei_2020, Battisti_2022}.}
    \label{fig:qeff}
\end{figure}

\subsection{Stellar mass dependencies}\label{stellar_mass_estimate}

We investigated the dependence of the dust attenuation tracers on stellar mass to assess whether the observed $\beta\mbox{-}\tau_B^l$ relation may be primarily driven by underlying mass trends. Figure \ref{fig:mass_dust} shows the UV continuum slope, $\beta$, and the Balmer optical depth, $\tau_B^l$, as a function of stellar mass for the final sample.

A clear positive correlation is observed between $\tau_B^l$ and $M_\star$, indicating that more massive galaxies experience stronger nebular attenuation. This trend is consistent with the well-established mass--dust relation observed at lower redshift \citep[e.g.][]{McLure2018}, where galaxies with higher stellar masses tend to host larger dust reservoirs. The tendency of more massive galaxies to be more metal-rich is also well established through the mass--metallicity relation \citep[e.g.][]{Tremonti2004}, which has now been traced to high redshift with JWST \citep[e.g.][]{Lam2026}. Recent JWST observations also indicate a close connection between nebular reddening and metallicity at high redshift \citep[e.g.][]{Karthikeyan2026}. The UV continuum slope $\beta$ also becomes systematically redder towards increasing stellar mass, consistent with the increasing dust attenuation in more massive systems.

We further compare the scatter of the two relations by fitting the $\log(M_\star/M_\odot)$--$\tau_B^l$ and $\log(M_\star/M_\odot)$--$\beta$ distributions. We measure a scatter of $\sigma(\tau_B^l)\simeq0.13$, compared to a substantially larger dispersion of $\sigma(\beta)\simeq0.55$. The tighter relation between $\tau_B^l$ and stellar mass suggests that the nebular attenuation traced by the Balmer decrement varies more uniformly across the galaxy population than the UV continuum slope. In contrast, the larger dispersion observed in $\beta$ likely reflects the combined influence of several physical properties, including stellar population age, metallicity, SFH, and dust attenuation, all of which contribute to shaping the UV continuum. Since $\beta$ and $\tau_B^l$ probe different attenuation components, we do not interpret the different scatter as implying that $\tau_B^l$ is intrinsically a superior tracer of dust attenuation.

To further investigate the role of star-formation activity, Fig.\,\ref{fig:sSFR} shows the relations between both attenuation indicators and the specific star-formation rate derived from the dust-corrected \ha\ luminosity. We quantify the strength of these trends using Pearson correlation coefficients. We find a weak but statistically significant positive correlation between $\tau_B^l$ and sSFR ($r=0.275$, $p=0.0026$). In contrast, the $\beta$--sSFR relation is weaker ($r=0.164$) and, given its $p$-value ($p=0.076$), does not provide statistically significant evidence of a linear correlation. This suggests that star-formation activity may contribute to the variation in nebular attenuation, although the modest correlation coefficient and substantial scatter indicate that it is unlikely to be the sole driver. The weaker $\beta$--sSFR trend is consistent with the additional sensitivity of the UV continuum slope to stellar population properties, metallicity, SFH, and dust attenuation. These results are consistent with the expected connection between sSFR and the average properties and evolutionary state of the underlying stellar populations \citep[e.g.][]{Langeroodi24}. However, the large scatter in the observed relations suggests that their distributions most likely reflect the combined influence of dust content, stellar populations, metallicity, and SFH.

\subsection{Construction of attenuation templates}\label{sec:templates}
Building on the correlations identified in the previous sections, we went on to exploit the measured values, $\tau^l_B$, to construct average spectral templates from which the selective attenuation curve can be empirically derived. Following \citet{Battisti_2016}, we divide the mass-selected sample into four bins of Balmer optical depth $\tau_B^l$ (see histogram in Fig.\,\ref{fig:mass_dust} and Fig.\,\ref{appB}, top left panel), and construct an average continuum SED for each bin by combining the NIRSpec spectra. For each bin, we constructed an average continuum SED by combining the NIRSpec spectra according to the following steps: 
1) the spectra were converted to rest-frame wavelengths using the spectroscopic redshifts;
2) the most prominent emission lines were removed; 3) the spectra were regridded to the same resolution; and 4) normalised at $5500\,\AA$ (rest-frame). Finally, the spectra were averaged.

To extend the wavelength coverage of the attenuation curve, this procedure was applied separately to three distinct rest-frame spectral regions, which were then combined into a single, continuous average template. The results of this procedure are presented in Fig.\,\ref{fig:templates}, where the four average spectral templates are reported, covering the wavelength range $0.16 \lesssim \lambda/\micron \lesssim 1.14$. In particular, the central interval, $0.21 \lesssim \lambda/\micron \lesssim 0.68$, corresponds to the rest-frame range common to all targets and therefore provides the highest statistical reliability. The number of objects that contribute to this interval in each $\tau^l_B$ bin is shown in Fig.\,\ref{fig:templates}. The outer regions of the templates are also well sampled, each including more than $50\%$ of the total sample. Following \cite{Battisti_2016}, we characterise the intrinsic dispersion of each template through the central 68\% range of the individual normalised spectra. We stress that this quantity represents the object-to-object dispersion within each $\tau_{\mathrm{B}}^{l}$ bin, rather than the uncertainty on the mean template.

After constructing the average spectral templates, we apply a median filter with a kernel size of seven spectral elements to suppress small-scale fluctuations while preserving the broad continuum shape. Because the spectral sampling differs among the three rest-frame wavelength regions, the corresponding smoothing window is not constant in wavelength and is approximately $102$, $177$, and $193\,\AA$ in the three regions, respectively. These smoothing scales are much narrower than the width of broad attenuation features such as the $2175\,\AA$ UV bump, and are therefore not expected to significantly alter such features. More generally, however, the full template-construction procedure, including rebinning and averaging spectra within each $\tau_B^l$ bin, may dilute features whose strength or central wavelength varies among individual galaxies. The attenuation curve derived from the composite spectra should therefore be interpreted as primarily tracing the average broad-band attenuation properties of the galaxy population rather than preserving all spectral features present in individual objects.

The composite spectra in Fig.\,\ref{fig:templates} show that a clear systematic suppression of the UV continuum is observed with increasing $\tau^l_B$, accompanied by a progressive attenuation of the spectral slope. The optical continuum shows a comparatively milder variation, consistent with a wavelength-dependent attenuation curve that is steeper in the UV regime. The smooth and monotonic behaviour of the stacked spectra indicates that $\tau_B^l$ effectively traces the overall dust attenuation of the systems. The high S/N of the composites allows us to robustly extract the selective attenuation curve in the subsequent analysis, while reducing the impact of stochastic variations among individual galaxy spectra. Possible systematic differences in the underlying galaxy populations across the $\tau_B^l$ bins are discussed in Appendix\,\ref{appB}.

\subsection{Derivation of the selective attenuation curve}\label{sec:selective_curve}
Using the four stacked spectral templates constructed in Sect.\,\ref{sec:templates}, we derived the wavelength-dependent selective attenuation curve. The underlying assumption of this differential approach is that galaxies grouped by increasing $\tau_B^l$ are, on average, sufficiently comparable in their intrinsic properties that the dominant systematic differences among their composite spectra can be attributed to differential dust attenuation. We assess this assumption in Appendix\,\ref{appB} by comparing the redshift, sSFR, and stellar-mass distributions of the galaxies in the different $\tau_B^l$ bins.

For each attenuation bin $n$, we compute the relative optical depth relative to a lower-dust reference bin $r$ ($r<n$) via
\begin{equation}
\tau_{n,r}(\lambda) = -\ln\left[\frac{F_n(\lambda)}{F_r(\lambda)}\right],
\end{equation}
where $F_n(\lambda)$ and $F_r(\lambda)$ are the average flux densities of the corresponding templates. The selective attenuation curve is then defined as
\begin{equation}
Q_{n,r}(\lambda) = \frac{\tau_{n,r}(\lambda)}{\Delta \tau_l^B},
\end{equation}
with $\Delta \tau_l^B = \tau_{l,B,n} - \tau_{l,B,r}$.

We consider all ordered template combinations ($Q_{2,1}$, $Q_{3,1}$, $Q_{4,1}$, $Q_{3,2}$, $Q_{4,2}$, $Q_{4,3}$), normalising each by the median $\tau_l^B$ difference of the corresponding bins. Because the templates $Q_{n,r}(\lambda)$ are obtained as ratios of average spectral templates, they constrain only the selective attenuation, losing information about the total dust content; its zero-point is therefore arbitrary. Following \citet{Calzetti_1994} and \citet{Battisti_2016}, we imposed $Q_{n,r}(5500\,\AA) = 0$. However, due to the noise present in the templates $Q_{n,r}(\lambda)$, we normalised each template by subtracting the average value measured over the wavelength range $4500\mbox{--}6500\,\AA$.
\begin{table}
    \centering
    \caption{Best-fit coefficients and associated $1\sigma$ uncertainties for the parametrisations of the selective attenuation curve,
    $Q_{\rm fit}(\lambda)$ in Eq.(\,\ref{qfit}), and the total attenuation curve $k(\lambda)$ in Eq.\,(\ref{k_equation}).}\label{coeff_qfit}
    \label{tab:att_coeff}
    \begin{tabular}{ccc}
    \hline\hline\noalign{\vskip 1pt}
    Equation & Coefficient & Value \\
    \hline\noalign{\vskip 1pt}
    Eq.~7 & $a_0$ & $-1.647 \pm 0.060$ \\
          & $a_1$ & $\phantom{+}1.217 \pm 0.082$ \\
          & $a_2$ & $-0.195 \pm 0.033$ \\
          & $a_3$ & $\phantom{+}0.013 \pm 0.004$ \\
          & $b_0$ & $-1.326 \pm 0.092$ \\
          & $b_1$ & $\phantom{+}0.740 \pm 0.126$ \\
    \hline\noalign{\vskip 1pt}
    Eq.~10 & $c_0$ & $-1.601 \pm 0.030$ \\
           & $c_1$ & $\phantom{+}1.160 \pm 0.048$ \\
           & $c_2$ & $-0.177 \pm 0.021$ \\
           & $c_3$ & $\phantom{+}0.011 \pm 0.003$ \\
           & $d_0$ & $-1.307 \pm 0.055$ \\
           & $d_1$ & $\phantom{+}0.726 \pm 0.079$ \\
    \hline\noalign{\vskip 1pt}
           & $f$ & $\phantom{+}3.778 \pm 0.240$ \\
           & $R_V$ & $3.976$ \\
    \hline
    \end{tabular}
\end{table}
\begin{figure}[h]
    \centering
    \includegraphics[width=1\linewidth]{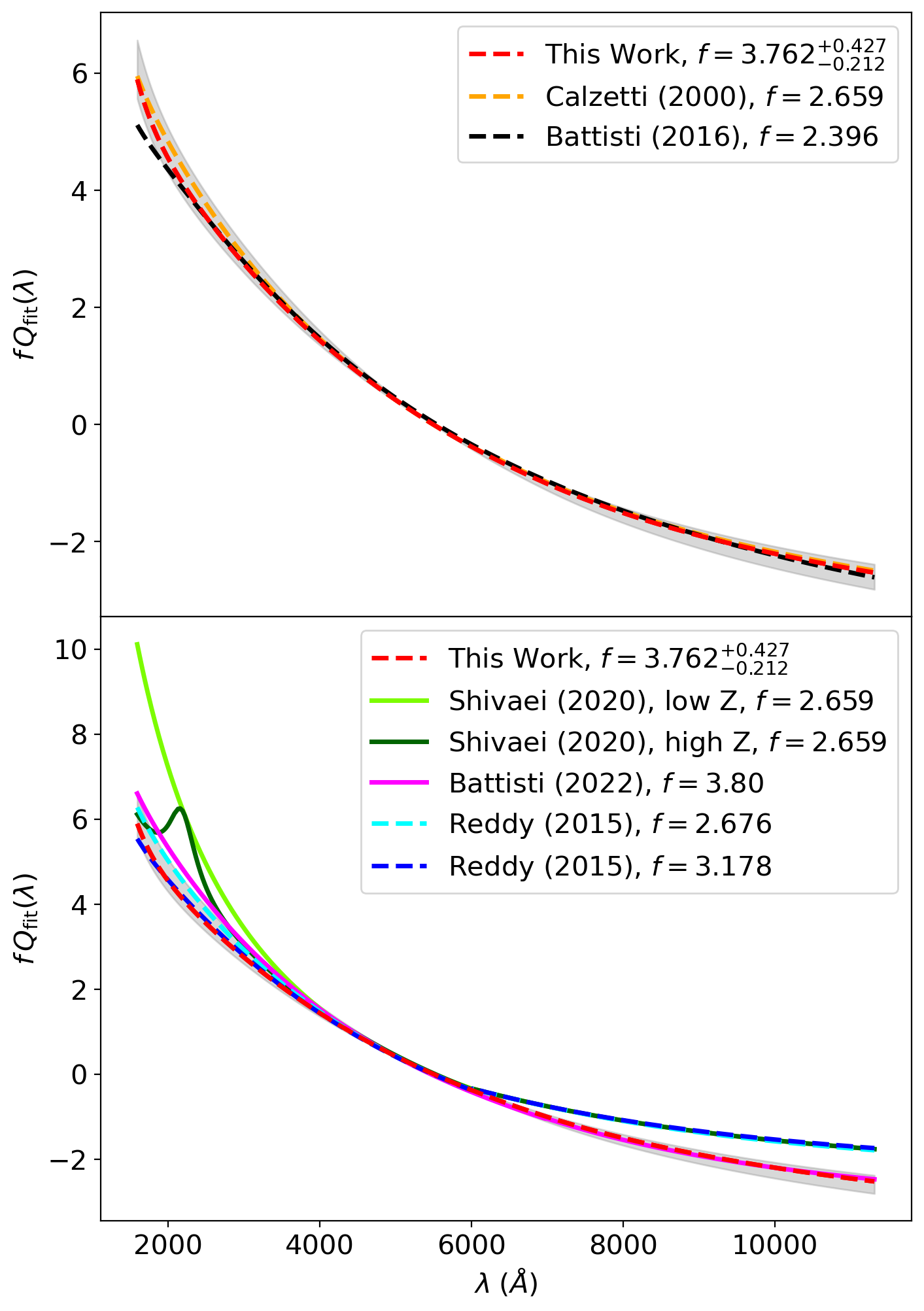}
    \caption{Total attenuation curve obtained by scaling the effective selective attenuation curve $Q_{\rm eff}(\lambda)$ by the normalisation factor, $f$, in Eq.\,(\ref{eq:ffactor}). The red dashed line shows the attenuation law derived in this work for galaxies at $2<z<7$, with the grey shaded region indicating the envelope of the $fQ_{\rm eff}$ relation obtained varying the value of f derived from fitting each $Q_{n,r}$ template. This envelope therefore reflects the sensitivity to the range of $f$ values derived from the individual $Q_{n,r}(\lambda)$ determinations, rather than a formal propagation of the statistical uncertainties of the individual spectra. \emph{Top panel}: comparison with the local attenuation curves of \citet{Calzetti2000} and \citet{Battisti_2016}. \emph{Bottom panel}: comparison with attenuation curves derived at intermediate redshift \citep{Reddy_2015, Shivaei_2020, Battisti_2017}. The value of the normalisation factor $f$ adopted for each relation is indicated in the legend.}
    \label{fig:fQeff}
\end{figure}

The individual curves and their average are shown in Fig.\,\ref{fig:Qnr}. Despite minor small-scale fluctuations, the $Q_{n,r}(\lambda)$ shapes are highly consistent. Their mean defines the effective selective attenuation curve, $Q_{\rm eff}(\lambda)$, which is smooth and monotonic, characterised by a steep UV rise and progressive flattening towards longer wavelengths. Following  \citet{Battisti_2016}, the dispersion among the individual $Q_{n,r}(\lambda)$ determinations provides an empirical indication of the sensitivity of the derived selective attenuation curve to the particular combination of reference and attenuated templates. We can parameterise $Q_{\rm eff}(\lambda)$  with a polynomial function of $\lambda\,(\mu$m$)$. The best fit relations are
\begin{equation}
Q_{\rm fit}(\lambda) =
\begin{cases}
a_0 + \dfrac{a_1}{\lambda}
+ \dfrac{a_2}{\lambda^2}
+ \dfrac{a_3}{\lambda^3},
& 0.16 \leq \lambda < 0.70~\mu{\rm m}, \\[6pt]
b_0 + \dfrac{b_1}{\lambda},
& 0.70 \leq \lambda < 1.14~\mu{\rm m}.
\end{cases}\label{qfit}
\end{equation}
The best-fit coefficients and their associated $1\sigma$ uncertainties are reported in Table\,\ref{coeff_qfit}. The continuity in $\lambda_0 = 0.70\,\micron$ is guaranteed by conditions $Q_1(\lambda_0) = Q_2(\lambda_0)$ and $\left.\frac{dQ_1}{d\lambda}\right|_{\lambda_0} = \left.\frac{dQ_2}{d\lambda}\right|_{\lambda_0}$. This analytical form provides a convenient representation of the empirically derived attenuation law at $2 < z < 7$.

Figure\,\ref{fig:qeff} compares the effective selective attenuation curve derived in this work with previous empirical determinations. The upper panel shows the comparison with local relations by C00 and \citet{Battisti_2016}: while the overall shape remains broadly similar, our $Q_{\rm eff}(\lambda)$ is systematically flatter than both curves. The lower panel extends the comparison to higher-redshift studies. Our selective attenuation curve in the blue region ($\lambda < 5000\,\AA$) is consistent with the results of \citet{Reddy_2015} at $z \sim 2$ for galaxies in the high sSFR bin, and \citet{Battisti_2022} at $z\sim1.3$. We note that the UV region of the selective attenuation curve $Q_{eff}(\lambda)$, shows a dependence on sSFR, with higher sSFR values corresponding to a flatter $Q_{eff}(\lambda)$. Indeed, \cite{Reddy_2015} and \cite{Battisti_2022} works show samples with $\langle\mathrm{sSFR}\rangle$ similar to ours as reported in Table\,\ref{tab:ssfr_f}. where the scaling factor f is also reported (see Sect.\,\ref{sec:normalization}). In contrast, samples with lower $\langle\mathrm{sSFR}\rangle$ exhibit steeper selective extinction curves in the UV.

As an additional test, we repeated the analysis by dividing the sample into two broad redshift bins and deriving the selective attenuation curve independently for each subsample. The resulting curves are consistent within the uncertainties over the entire wavelength range, indicating no statistically significant evolution in the average attenuation curve across the redshift interval explored in this work. We therefore proceed by adopting the attenuation curve derived from the full sample.

\subsection{Normalisation and recovery of the total attenuation curve}\label{sec:normalization}
To convert $Q_{\rm eff}(\lambda)$ into a total attenuation curve $k(\lambda)=A_\lambda/E(B-V)$, we again follow the procedure of \citet{Battisti_2016}. The scaling factor $f$, which takes into account the differential attenuation between ionised gas and stellar continuum, is defined so that $k(B)-k(V)=1$ via
\begin{equation}\label{eq:ffactor}
    f = \left[ Q_{\rm fit}(B) - Q_{\rm fit}(V) \right]^{-1},
\end{equation}
where the $B$ and $V$ bands are assumed to be $4400\,\AA$ and $5500\,\AA$, respectively. We computed $Q(B)$ and $Q(V)$ from Eq.\,(\ref{qfit}). The value of $f$ for our average selective attenuation curve, $Q_{\rm fit}(\lambda)$, is $f = 3.762_{-0.212}^{+0.427}$, where the uncertainty reflects the spread between the maximum and minimum values obtained from fits using individual $Q_{n,r}(\lambda)$.

The resulting $fQ_{\rm fit}(\lambda)$ curve, together with previous works results, is shown in Fig.\,\ref{fig:fQeff}. When introducing the factor $f$, the slopes of most attenuation curves become strikingly similar. Our curve is consistent with most of them, except for the curves by \citet{Reddy_2015}, which deviates at $\lambda > 6000\,\AA$ and the curve at low metallicity by \citet{Shivaei_2020}, which is still steeper in the UV range.

The samples with higher sSFR show higher values of $f$, as shown in Table\,\ref{tab:ssfr_f}. This suggests that in systems with higher sSFR the gas is more attenuated than the stars, and the scaling factor $f$ used to extract the attenuation on the stellar continuum is higher.
\begin{table}
\centering
\caption{Comparison of $f$ as a function of sSFR for different works.}\label{tab:ssfr_f}
    \begin{tabular}{cccc}
    \hline
    \hline \noalign{\vskip 1pt}
    $\langle\log_{10}(\mathrm{sSFR})\rangle$ & $f$ & $z$ & \textbf{Works} \\ 
    \hline \noalign{\vskip 2pt}
    $-9.8\phantom{0}$  & $2.396^{+0.33}_{-0.29}$ & $z \sim 0\phantom{.0}$ & \cite{Battisti_2016} \\ 
    \noalign{\vskip 2pt}
    $-9.36$ & $2.659$ & $z \sim 0\phantom{.0}$ & \cite{Calzetti2000} \\ 
    \noalign{\vskip 2pt}
    $-9.2\phantom{0}$  & $2.659$ & $z \sim 2\phantom{.0} $& \cite{Reddy_2015} \\ 
    \noalign{\vskip 2pt}
    $-8.9\phantom{0}$  & $2.659$ & $z \sim 2\phantom{.0}$ & \cite{Shivaei_2020} \\
    \noalign{\vskip 2pt}
    $-8.5\phantom{0}$  & $3.178$ & $z \sim 2\phantom{.0}$ & \cite{Reddy_2015} \\ 
    \noalign{\vskip 2pt}
    $-8.4\phantom{0}$  & $3.80^{+1.12}_{-1.12}$  &$ z \sim 1.3$ & \cite{Battisti_2022} \\ 
    \noalign{\vskip 2pt}
    $-8.11$ & $3.762^{+0.427}_{-0.212}$ & $2 < z < 7$ & This work \\ 
    \noalign{\vskip 1pt}
    \hline
    \hline
    \end{tabular}
\end{table}
The total attenuation curve can finally be derived as
\begin{equation}\label{eq:totattcurve}
    k(\lambda) = f\,Q_{\rm eff}(\lambda) + R_V,
\end{equation}
where $R_V$ is the normalisation constrained either from the long-wavelength behaviour of the derived curve (assuming $k\rightarrow0$ in the NIR) or from an energy-balance argument when rest-frame IR data are available.
\begin{figure}[h]
    \centering
    \includegraphics[width=1\linewidth]{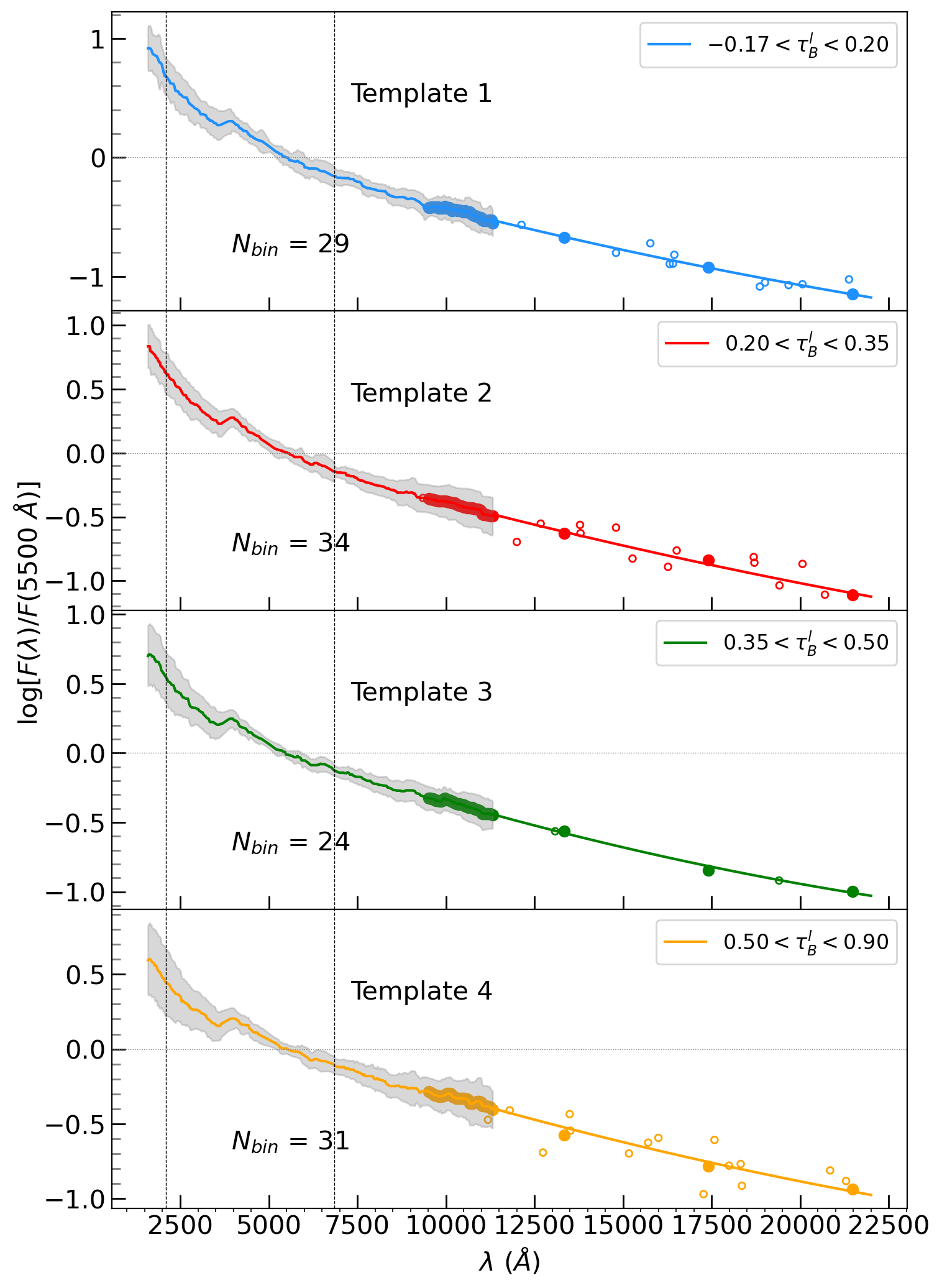}
    \caption{Average spectral templates in the four $\tau_B^{l}$ bins after extending the wavelength coverage using JWST/MIRI photometry. Coloured points show the observed MIRI fluxes associated with the galaxies contributing to each bin, while the solid lines represent the stacked NIRSpec spectra normalised at $5500\,\AA$. The inclusion of MIRI data allows for the average SEDs to be traced up to $\sim2.2\,\micron$, reducing the extrapolation required to estimate $R_V$ from the long-wavelength behaviour of the attenuation curve. Open circles show the MIRI photometric measurements of individual galaxies, while filled circles indicate the corresponding average  fluxes computed in wavelength bins. The numbers of galaxies with available MIRI observations in the four $\tau_B^l$ bins are $N=8$, $7$, $2$, and $8$, respectively.}
    \label{sed_MIRI}
\end{figure}

Since we lacked far-IR (FIR) constraints, to determine $R_V$, we evaluated the curve at $\lambda_n = 2.85\,\micron$ and imposed the condition $k(\lambda_n)=0$. This wavelength corresponds to the regime where the extinction curves of the Small and the Large Magellanic Clouds (SMC and LMC, respectively) and the Milky Way approach negligible attenuation \citep{Gordon2003,Reddy_2015}. Applying this procedure to our best-fit relation in Eq.\,(\ref{qfit}) yields $R_V = 3.82$.

\begin{figure}
    \centering
    \includegraphics[width=1\linewidth]{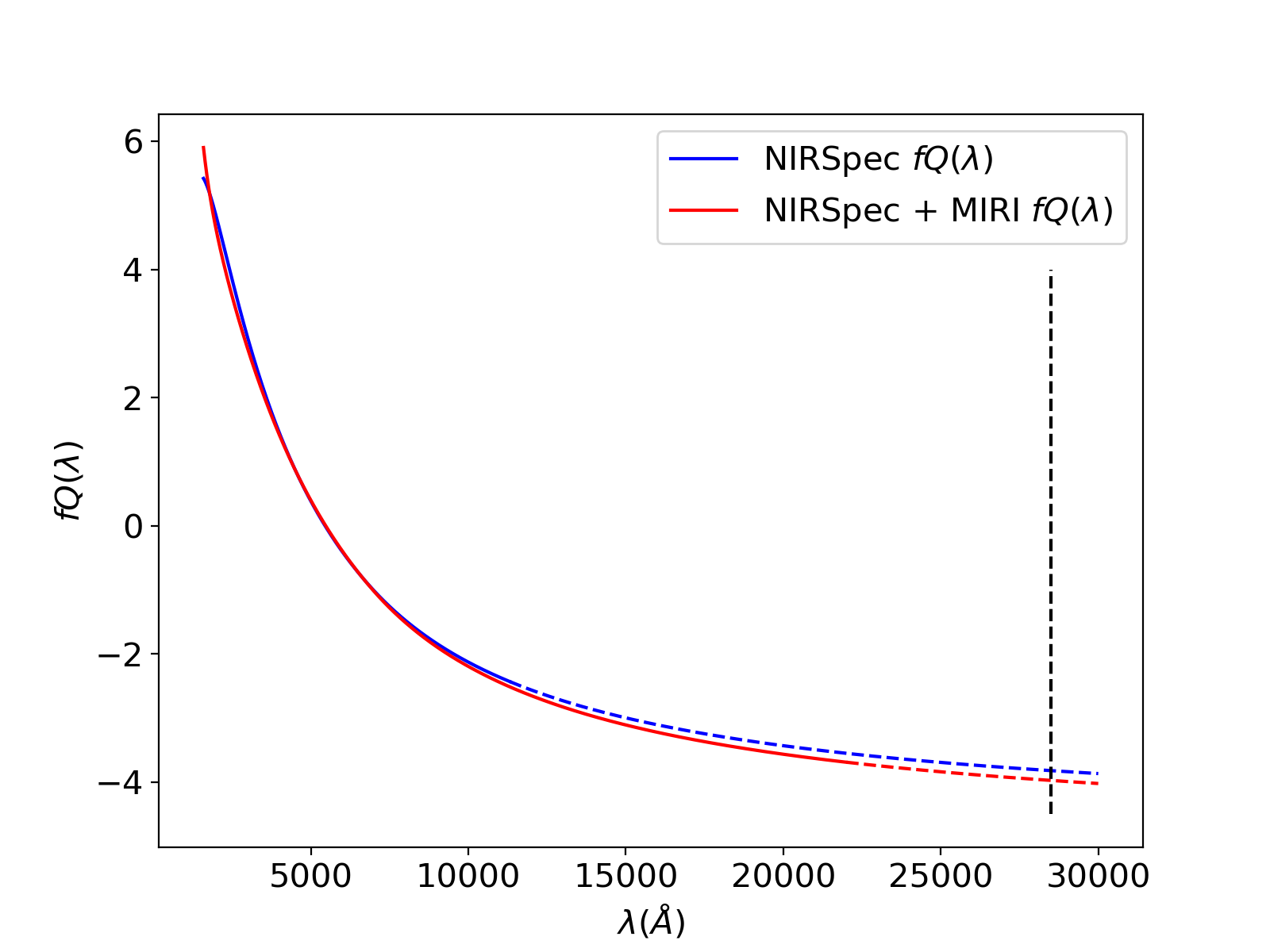}
    \caption{Determination of $R_V$ from the long-wavelength extrapolation of the attenuation curve. The blue line shows the polynomial fit to $fQ_{\rm eff}(\lambda)$ obtained using only the NIRSpec-based attenuation curve, while the red line includes the extension provided by the MIRI photometry. The vertical dashed line marks $\lambda = 2.85\,\micron$, where the condition $k(\lambda)=0$ is imposed to derive $R_V$.The comparison illustrates how the inclusion of MIRI data reduces the extrapolation required to estimate the normalisation of the total attenuation curve.}
    \label{extrapolation}
\end{figure}

However, since our spectral coverage is limited to $\sim1.2\,\micron$, a non negligible extrapolation is required. To mitigate this issue, we included in the analysis the observed MIRI photometry for the JADES sources with a counterpart in the catalogue provided by Catone et al. (in prep) and described in Sect.\,\ref{MIRI}. The analysis allows us to extend the average templates to $\sim2.2\,\micron$ (see Fig.\,\ref{sed_MIRI}), thereby reducing the extrapolation to $\sim 2.85\,\micron$. Since spectral template 3 contains only a few targets in the MIRI wavelength range, we excluded this bin from the computation of $Q_{\rm eff}(\lambda)$. Combining NIRSpec and MIRI, we obtain $R_V =3.98$. Since the inclusion of the MIRI photometry extends the wavelength range over which $Q_{\rm eff}(\lambda)$ is constrained, we refit the selective attenuation curve and recompute the corresponding normalisation factor, obtaining $f=3.778$. This value is slightly different from the NIRSpec-only value, $f=3.762$, reported above. Figure\,\ref{extrapolation} highlights the differences in this procedure, showing the best-fit polynomial functions to $f\,Q_{\rm eff}(\lambda)$ when adopting only NIRSpec (blue line) or the same in combination with MIRI to predict $R_V$. We adopt the value obtained including MIRI ($R_V=3.98$; red curve in Fig.\,\ref{extrapolation}) to derive the total attenuation curve. We use the difference between the NIRSpec-only and NIRSpec+MIRI extrapolated  values, $\Delta R_V=0.16$, as an estimate of the sensitivity of  $R_V$ to the long-wavelength extrapolation. This should not be interpreted as a formal statistical $1\sigma$ uncertainty and is likely a lower limit to the total uncertainty, since a full propagation of the uncertainties associated with the extrapolation procedure is not included. Moreover, a robust estimate of $R_V$ would require a full treatment of the dust energy balance. The final fit of the total attenuation curve is given via
\begin{equation}
k(\lambda) =
\begin{cases}
f\left(c_0 + \dfrac{c_1}{\lambda}
+ \dfrac{c_2}{\lambda^2}
+ \dfrac{c_3}{\lambda^3}\right) + R_V,
& 0.16 \leq \lambda < 0.70~\mu{\rm m}, \\[6pt]
f\left(d_0 + \dfrac{d_1}{\lambda}\right) + R_V,
& 0.70 \leq \lambda < 2.20~\mu{\rm m}.
\end{cases}
\label{k_equation}
\end{equation}
The corresponding best-fit coefficients with their associated 1$\sigma$ uncertainties are also reported in Table\,\ref{tab:att_coeff}.

\begin{figure*}
    \includegraphics[width=1.0\linewidth]{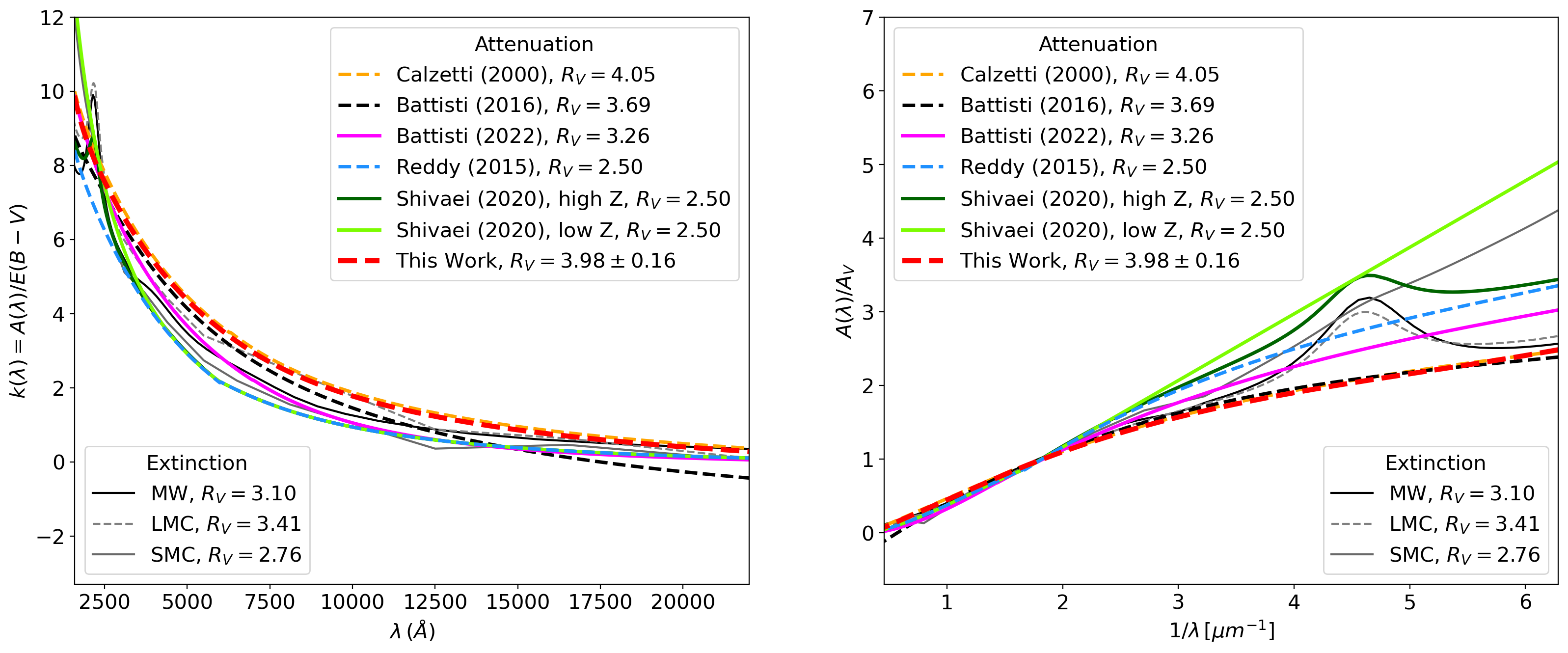}  
    \caption{Comparison between the total attenuation curve derived in this work and previous attenuation and extinction relations. \emph{Left panel}: Attenuation curves expressed as $k(\lambda)=A_\lambda/E(B-V)$ as a function of wavelength. \emph{Right panel}: Same curves shown as $A(\lambda)/A(V)$ versus $1/\lambda$. Our result for galaxies at $2<z<7$ (red dashed line) is compared with attenuation curves from C00, \citet{Battisti_2016}, \citet{Battisti_2022}, \citet{Reddy_2015}, and \citet{Shivaei_2020}, as well as with the extinction curves of the Milky Way, LMC, and SMC. The normalisation $R_V$ adopted for each relation is indicated in the legend.}
    \label{total_attenuation}
\end{figure*}

In Fig.\,\ref{total_attenuation} we compare the total attenuation curve derived in this work with a compilation of commonly adopted attenuation and extinction relations. The comparison is shown in terms of both $k(\lambda)=A_{\lambda}/E(B-V)$ (left panel) and in normalised form $A(\lambda)/A(V)$ as a function of $1/\lambda$ (right panel), enabling a direct assessment of both the general normalisation and the shape of the curve. 

Our attenuation curve for galaxies at $2<z<7$ is remarkably consistent with the local C00 relation, both in its UV-optical slope and in its normalisation. In particular, in UV ($\lambda \lesssim 3000\,\AA$) our curve closely follows the C00 curve, indicating that a C00-like attenuation behaviour provides an adequate description of the average dust attenuation properties of our sample.

In contrast, compared to other empirical determinations at intermediate redshift, our curve appears to be systematically flatter, most noticeably in the UV part of the spectrum. Several of the intermediate-redshift relations shown in Fig.\,\ref{total_attenuation} \citep[e.g.][]{Reddy_2015, Shivaei_2020,Battisti_2022} rise more steeply towards short wavelengths, implying stronger UV attenuation per unit $E(B-V)$ than inferred here. Again, the flatter UV behaviour of our curve therefore suggests a reduced wavelength dependence of the attenuation at high redshifts, which may reflect differences in dust geometry, star-dust mixing, and/or the effective grain-size distribution in our mass-selected galaxy sample.

At optical and NIR wavelengths, the various attenuation curves tend to converge, consistent with the weaker wavelength dependence of attenuation beyond the Balmer break. Finally, the comparison with the Milky Way, LMC, and SMC extinction curves highlights the expected distinction between extinction and attenuation laws: while extinction curves trace the intrinsic dust properties along individual lines of sight, attenuation curves encode the combined effect of dust and the relative spatial distribution of stars and dust within galaxies.

\subsection{Stellar to nebular colour excess}
We calculated the ratio $E(B-V)_{\rm stars}/E(B-V)_{\rm gas}$ using the relation in \citet{Calzetti_1994} via
\begin{equation}
    \frac{E(B-V)_{\rm star}}{E(B-V)_{\rm gas}} = 1/f,
\end{equation}
where $f$ is the $f$-factor of our total attenuation.\footnote{We note that \cite{Battisti_2016} and \cite{Battisti_2022} adopt a slightly different definition of the stellar to nebular colour excess:
\begin{equation}
\frac{E(B-V)_{\rm stars}}{E(B-V)_{\rm gas}} = \frac{k(\hb)-k(\ha)}{f}
\end{equation}
where $k(\lambda)$ is the Cardelli extinction curve \citep{cardelli_1989}. However, the factor ${k(\hb)-k(\ha)}$ is $\sim1.06$, implying variations for the ratio  within the uncertainties of our measurements.} The result is $E(B-V)_{\rm stars}/E(B-V)_{\rm gas} = 0.27 \pm 0.02$, which is less than the C00 relation $E(B-V)_{\rm stars}/E(B-V)_{\rm gas} = 0.44 \pm 0.03$. This implies that, in our sample, the stellar emission is less attenuated relative to the gas compared to the sample of C00. 

The low $E(B-V)_{\rm stars}/E(B-V)_{\rm gas}$ ratio indicates significant differential attenuation between the stellar continuum and the nebular emission, with the ionised gas experiencing systematically stronger reddening. Such differential attenuation has been widely observed in star-forming galaxies both locally and at high redshift \citep[e.g.][]{Calzetti2000,Reddy_2015,Shivaei_2020}. A natural interpretation arises from the different stellar populations traced by nebular emission and the UV continuum. Ionising photons, and hence the Balmer emission lines, are produced by the youngest stellar populations ($\lesssim 7\mbox{--}10\,\mathrm{Myr}$), which are still preferentially associated with their dusty natal environments. In contrast, the UV continuum is produced by stellar populations extending to ages of $\sim100\,\mathrm{Myr}$. As stars progressively disperse from their birth clouds on timescales of order $\sim10\,\mathrm{Myr}$ \citep[e.g.][]{Grasha2018,Pedrini2026}, a significant fraction of the UV continuum can emerge from less obscured regions, while the nebular emission remains preferentially associated with more heavily attenuated star-forming regions. This age-dependent geometry is consistent with the two-component dust model of \citet{CharlotFall2000}, in which young stars and their associated nebular emission experience additional attenuation from their natal birth clouds, while the broader stellar continuum is predominantly affected by the diffuse ISM. The relative importance of these components may further depend on galaxy properties and star-formation activity \citep[e.g.][]{Price_2014,Shivaei_2020}. Our measurements therefore primarily indicate significant differential attenuation between the stellar continuum and the ionised gas, rather than uniquely constraining the detailed geometry of the dust distribution.

\section{Discussion}\label{sec:Disc}
The empirical attenuation curve derived in this work provides a new constraint on the average dust attenuation properties of massive star-forming galaxies over the redshift range $2<z<7$ with $\log_{10}(M_\star/M_\odot) > 9$. Accordingly, our conclusions primarily apply to this mass-selected galaxy population and should not be directly extrapolated to lower-mass galaxies at the same redshifts. These results can then be interpreted in the broader context of recent JWST studies that have begun to characterise the diversity and evolution of dust attenuation curves at high redshift.

\subsection{Comparison with recent JWST studies}
Our results provide an empirical measurement of the average attenuation curve in $2<z<7$ derived using spectroscopic Balmer decrements and rest-frame UV photometry. Interestingly, the resulting curve is consistent with the local starburst relation of C00 in both slope and normalisation, while being flatter than several attenuation curves previously derived for intermediate-redshift galaxy samples, particularly in the UV regime. This behaviour is also consistent with the trends reported by \citet{Shivaei2025} and \citet{Markov_2024}, where the UV–optical slope of the attenuation curve depends strongly on dust content and galaxy properties.

The comparison highlights an important difference in methodology. While studies such as \citet{Shivaei2025} and \citet{Markov_2024} derived attenuation curves on a galaxy-by-galaxy basis through SED fitting with flexible dust prescriptions, our approach follows the empirical method originally introduced by \citet{Calzetti_1994}, relying on the relation between nebular reddening and stellar continuum reddening. As a result, our curve represents an ensemble-average attenuation law rather than the full distribution of slopes observed among individual galaxies. We note that the empirical approach, which infers the attenuation curve from the correlation between nebular reddening and stellar-continuum reddening across a galaxy population, can yield different results from approaches that infer attenuation curves by fitting individual SEDs with flexible dust prescriptions, because the former compares galaxies as a population while the latter treats galaxies individually. Thus, the empirical approach carries robustness in the results but loses the ability to distinguish subtleties (e.g. derive separate curves for high $A_V$ and low $A_V$ cases),  while the hybrid approach can identify such subtleties, under the strong assumption that the complex star formation histories of galaxies are reproduced faithfully. Both approaches thus carry strengths and weaknesses.

The agreement between the average curve obtained here and the C00 relation suggests that despite the significant diversity of attenuation curves among individual systems, the population-averaged attenuation behaviour of the selected massive star-forming galaxies over $2<z<7$ remains remarkably similar to that observed in local starburst galaxies. This is also consistent with recent JWST studies finding that the correlation between Balmer decrement and stellar mass remains largely unchanged up to $z\sim7$ \citep{Woodrum2025,Fisher2025,Karthikeyan2026}. However, the Balmer decrement specifically traces attenuation towards the ionised gas and does not directly constrain the shape of the stellar attenuation curve, which additionally depends on factors such as dust grain properties and the relative geometry of stars and dust. The observed stability of the Balmer decrement--stellar mass relation should therefore not be interpreted as evidence that the physical mechanisms regulating dust production and distribution are already fully established at these epochs.

\subsection{Implications for the evolution of dust properties}
The physical interpretation of attenuation curve variations has recently been explored both observationally and theoretically. \citet{Markov2025} investigated the redshift evolution of dust attenuation curves using JWST/NIRSpec spectroscopy of galaxies at $2<z<12$. As anticipated in the Introduction, their analysis revealed a systematic flattening of the attenuation curve with increasing redshift, accompanied by a progressive decrease in the strength of the $2175\,\AA$ UV bump. These trends were interpreted consistent with dust in early galaxies is dominated by large grains formed in the ejecta of core-collapse supernovae, which subsequently undergo processing in the interstellar medium on timescales of the order of $\sim0.5\mbox{--}1\,\mathrm{Gyr}$. \citeauthor{Markov2025} therefore proposed that the flattening of attenuation curves at high redshift reflects the early stage of dust evolution, when grain processing and shattering in the ISM have not yet produced the smaller grains responsible for steeper UV extinction slopes. This interpretation may also be consistent with the relatively young ages of high-redshift galaxies, for which the contribution of asymptotic giant branch (AGB) stars to the overall dust budget may be reduced compared with more evolved systems, given the longer evolutionary timescales required for low- and intermediate-mass stars to enter the AGB phase \citep{Ferrarotti2006}. 

The average attenuation curve derived in this work is broadly consistent with this picture and with the average curve of \citet{Markov_2024} across $2<z<12$ (see the blue curve in their Fig.\,1). Although our curve remains similar to the C00 relation, it appears flatter in the UV compared to several higher-redshift empirical curves reported in the literature. This behaviour may reflect the combined effects of grain-size distribution and dust-star geometry in galaxies at these epochs. As emphasised by \citet{Markov2025}, a dust population weighted towards larger grains can contribute to shallower wavelength dependence, although the observed attenuation slope also depends strongly on dust–star geometry and radiative-transfer effects.

Another key factor is the role of dust-star geometry and radiative transfer effects. Increasing dust column densities or more complex spatial distributions of stars and dust can flatten the observed attenuation curve through scattering and differential obscuration. Indeed, theoretical studies and observations have shown that the attenuation slope correlates with $A_V$, with dustier galaxies generally exhibiting shallower curves. \citet{Shivaei2025} demonstrate that this dependence on attenuation is a dominant driver of the diversity in attenuation slopes observed across galaxy populations. 

The comparison presented in Figs.\,\ref{fig:qeff}--\ref{total_attenuation} highlights that the main differences among attenuation curves emerge in the UV regime, where the wavelength dependence of dust attenuation is most sensitive to the physical properties of dust grains and to the relative geometry between stars and dust. In addition to variations in the overall UV slope, another key diagnostic feature is the presence of the broad absorption feature centred at $2175\,\AA$, commonly referred to as the UV bump. The strength of this feature varies significantly among galaxies and provides important insight into the composition and evolution of dust grains. In the following, we discuss the implications of the absence of this feature in the attenuation curve derived in this work.

\subsection{Absence of the $2175\,\AA$ UV bump}
An important aspect of our derived attenuation curve is that we do not find significant evidence of a UV bump at $2175\,\AA$.
Within the uncertainties, the attenuation law appears smooth across the UV range, lacking the characteristic feature commonly observed in the Milky Way extinction curve and in a fraction of galaxies at intermediate redshift.

The presence and strength of the $2175\,\AA$ feature are generally attributed to small carbonaceous grains or PAHs. Its absence in our average attenuation curve is therefore consistent with scenarios in which such grains are either not yet abundant or are efficiently destroyed in the interstellar medium of high-redshift galaxies. This interpretation is supported by recent JWST-based studies. For example, \citet{Markov2025} report that the strength of the UV bump decreases with increasing redshift, becoming progressively weaker beyond $z \sim 5$. Their analysis suggests that dust in early galaxies is dominated by large grains produced in core-collapse supernova ejecta, while the smaller carbonaceous grains responsible for the UV bump emerge later through grain processing and dust production in AGB stars.

At intermediate redshift, observational results indicate a more complex picture. \citet{Kashino2022} reported strong evidence of a prominent $2175\,\AA$ feature in star-forming galaxies at $1.3 \lesssim z \lesssim 1.8$, consistent with earlier spectroscopic studies that reported a significant fraction of galaxies exhibiting a UV bump \citep[e.g.][]{Noll2009}. These results suggest that the carriers of the feature are already in place in relatively evolved systems at these epochs.

On the other hand, \citet{Shivaei_2020} showed that attenuation curves depend strongly on galaxy properties. In particular, high-metallicity systems exhibit shallow, C00-like attenuation curves with a significant UV bump, whereas lower-metallicity galaxies tend to display steeper, SMC-like curves. The dependence of attenuation-curve shape on metallicity reported by \citet{Shivaei_2020} highlights the importance of galaxy properties when interpreting an ensemble-average curve. However, given the heterogeneous metallicity distribution of our sample, our measurements do not allow us to determine whether metallicity plays a significant role in the absence of the UV bump.
 
These results point towards a redshift-dependent evolution of the UV bump strength. While the feature is commonly observed in a fraction of galaxies at intermediate redshift, it appears to weaken or disappear on average at higher redshift, in agreement with our findings. This trend suggests that the conditions required to produce a prominent UV bump (whether related to dust composition or dust-star geometry) are less prevalent in the early Universe.

Finally, we caution that the lack of a detectable $2175\,\AA$ feature in our average attenuation curve does not necessarily imply that the UV bump is absent from individual galaxies in our sample. If the feature is present only in a subset of galaxies (as in the source detected by \citealt{Witstok2023} at $z\sim7$), or if its strength and central wavelength vary among individual systems, the stacking procedure may dilute its signature in the ensemble-average curve. In addition, the spectral processing involved in constructing the average templates, including rebinning and median filtering, may further reduce the contrast of a weak feature. We therefore interpret our result as lacking any significant evidence to support a prominent UV bump in the ensemble-average attenuation curve, rather than as evidence suggesting the absence of the feature in individual galaxies. Alternatively, the weak average signature may reflect a dominant dust population with a relatively small contribution from the carbonaceous grains associated with the UV bump, although the present analysis does not allow these possibilities to be distinguished.

\subsection{The origin of the average attenuation law}
Taken together, the emerging picture from JWST studies is that attenuation curves exhibit both intrinsic diversity and systematic evolution with galaxy properties and cosmic time. Within this context, the attenuation law derived here can be interpreted as the effective average curve resulting from the combination of galaxies spanning a range of dust contents, geometries, and evolutionary stages.

The similarity of our average curve over $2<z<7$ to both the C00 relation and the average curve reported by \citet{Markov_2024} over $2<z<12$ suggests that an approximately C00-like population-averaged attenuation law may persist over a broad range of cosmic epochs.
However, the flatter UV behaviour we observe relative to some higher-redshift studies may indicate that the dust properties and grain-size distributions in these early systems differ from those in the Local Universe.

Future JWST observations will enable this issue to be explored further by combining spectroscopic constraints on nebular attenuation with spatially resolved imaging and FIR measurements of dust emission. Such datasets will help disentangle the relative roles of dust composition, grain-size evolution, and star–dust geometry in shaping the attenuation curves of galaxies across cosmic time.

\section{Summary and conclusions}\label{sec:conclusions}

In this work, we present a new direct empirical determination of the average dust attenuation law for massive star-forming galaxies over the redshift range $2<z<7$ using JWST spectroscopy. By combining NIRSpec spectra from the JADES survey with deep multi-wavelength photometry from the ASTRODEEP-JWST catalogues, we were able to derive the attenuation curve following the empirical methodology originally developed by \cite{Calzetti_1994} and later applied by \cite{Battisti_2016}. Our analysis exploits the relation between the reddening of the stellar continuum and the reddening of the ionised gas traced by the Balmer decrement.

Starting from an initial spectroscopic sample, we constructed a clean dataset of $118$ star-forming galaxies with $\log_{10}(M_\star/M_\odot) \geq 9$  and reliable measurements of the \ha/\hb\ ratio and rest-frame UV photometry. Using the Balmer optical depth $\tau^{l}_{B}$ as a tracer of dust attenuation, we grouped galaxies into bins of $\tau^{l}_{B}$ and constructed stacked spectral templates spanning the wavelength range $\sim0.16\mbox{--}1.14\,\micron$. The ratios of these templates allowed us to derive the selective attenuation curve $Q_{\rm eff}(\lambda)$.

Our main results can be summarised as follows:
\begin{itemize}
\item We find a positive correlation between the UV continuum slope, $\beta$, and the Balmer optical depth, $\tau^{l}_{B}$, consistent with the expectation that galaxies with stronger nebular reddening tend to exhibit redder stellar continua. However, the relation shows substantial intrinsic scatter, likely driven by variations in stellar populations, star formation histories, dust attenuation, and dust–star geometry.

\item From stacked spectra in bins of $\tau^{l}_{B}$, we inferred a smooth selective attenuation curve spanning the rest-frame UV to optical range. The effective curve $Q_{\rm eff}(\lambda)$ is well described by a third-order polynomial over $0.16 \leq \lambda/\micron \leq 1.14$. Compared to several empirical determinations in the local universe and at intermediate redshift \citep[e.g. C00,][]{Reddy_2015,Shivaei_2015,Battisti_2017}, our curve total appears to be systematically flatter in the UV. This difference may reflect variations in dust geometry, grain-size distribution, or the relative spatial distribution of stars and dust in early galaxies.

\item The normalisation of the selective attenuation curve, expressed via the scaling factor, $f$, places our $f\,Q_{\rm eff}(\lambda)$ relation in the upper range of literature values. We find $f = 3.778^{+0.411}_{-0.228}$, higher than typical local and intermediate-redshift determinations, consistent with the elevated sSFR of our sample. This trend supports a scenario in which galaxies with higher star formation activity exhibit stronger differential attenuation between nebular and stellar emission, leading to a systematically higher normalisation of the total attenuation curve compared to previous works.

\item By extrapolating the long-wavelength behaviour and imposing $k(\lambda_n)=0$ at $\lambda_n = 2.85\,\micron$, we inferred a normalisation $R_V = 3.98 \pm 0.16$. The resulting attenuation law is remarkably consistent with the local starburst attenuation relation of C00, both in slope and in normalisation. This suggests that despite the large diversity of attenuation curves observed in individual galaxies, the ensemble-average attenuation behaviour of massive star-forming galaxies remains broadly similar from the Local Universe up to $z\sim7$.

\item We find no significant evidence of a UV bump at $2175\,\AA$ in the derived attenuation curve. The absence of this feature may reflect a lower contribution from the carbonaceous dust component associated with the UV bump, more efficient processing or destruction of its carriers, or dilution of the feature during the averaging and smoothing procedures used to construct the ensemble-average attenuation curve. In addition, the dependence of attenuation-curve shape on metallicity reported by \citet{Shivaei_2020} highlights the importance of galaxy properties when interpreting an ensemble-average curve. However, our current analysis does not allow us to determine whether metallicity plays a significant role in the absence of the UV bump. The dependence of the attenuation curve on metallicity will be investigated in detail in a future work.
\end{itemize}

Overall, our results provide a direct empirical constraint on the average attenuation law of massive star-forming galaxies across $2<z<7$ based on JWST spectroscopy. Although individual galaxies exhibit significant diversity in attenuation properties, the population-averaged attenuation law is broadly consistent with the local starburst relation, while exhibiting a somewhat flatter UV slope. 

Future JWST observations will allow these results to be refined by increasing the sample size, probing wider ranges of stellar mass and dust content, and extending the wavelength coverage towards the FIR. Such datasets will be crucial for disentangling the roles of dust composition, grain-size evolution, and star-dust geometry in shaping the attenuation curves of galaxies across cosmic time. Moreover, facilities such as the proposed PRIMA mission \citep{Glenn2025}, in synergy with ALMA, will provide direct measurements of dust emission in the FIR for large samples of galaxies, enabling a more complete and physically grounded characterisation of dust attenuation and its connection to galaxy evolution across cosmic time.

\begin{acknowledgements}
We thank the anonymous referee for their constructive comments and suggestions,
which helped improve the quality and clarity of the manuscript.
G.R., P.C., and B.V are supported  by the European Union -- NextGeneration EU RFF M4C2 1.1 PRIN 2022 project 2022ZSL4BL INSIGHT. G.R. is also supported by ASI/INAF agreement for the PRIMA mission n. 2025-6-HH.0.  This work is also supported by INAF
RF2024 Large Grant "UNDUST: UNveiling the Dawn of the Universe with
JWST" (P.I. P. Santini). 
We thank the JADES collaboration for making available the high-quality JWST/NIRSpec spectroscopic data used in this work.
\end{acknowledgements}

\bibliographystyle{aa}
\bibliography{Bibliografia.bib}

@ARTICLE{Treu2022,
       author = {{Treu}, T. and {Roberts-Borsani}, G. and {Brada{\v c}}, M. and
                 et al.},
        title = "{The GLASS-JWST Early Release Science Program. I. Survey Design and Release Plans}",
      journal = {\apj},
         year = 2022,
       volume = {935},
          eid = {110},
        pages = {110},
          doi = {10.3847/1538-4357/ac8158}
}

@ARTICLE{Bezanson2024,
       author = {{Bezanson}, R. and {Labb{\'e}}, I. and {Whitaker}, K. E. and
                 et al.},
        title = "{The JWST UNCOVER Treasury Survey: Ultradeep NIRSpec and NIRCam Observations before the Epoch of Reionization}",
      journal = {\apj},
         year = 2024,
       volume = {974},
          eid = {92},
        pages = {92},
          doi = {10.3847/1538-4357/ad66cf}
}

@ARTICLE{Finkelstein2022,
       author = {{Finkelstein}, S. L. and {Bagley}, M. B. and
                 {Arrabal Haro}, P. and et al.},
        title = "{A Long Time Ago in a Galaxy Far, Far Away: A Candidate z $\sim$ 12 Galaxy in Early JWST CEERS Imaging}",
      journal = {\apjl},
         year = 2022,
       volume = {940},
          eid = {L55},
        pages = {L55}
}

@ARTICLE{Eisenstein2023,
       author = {{Eisenstein}, Daniel J. and {Willott}, Chris and {Alberts}, Stacey and {Arribas}, Santiago and {Bonaventura}, Nina and {Bunker}, Andrew J. and {Cameron}, Alex J. and {Carniani}, Stefano and {Charlot}, Stephane and {Curtis-Lake}, Emma and {D'Eugenio}, Francesco and {Ferruit}, Pierre and {Giardino}, Giovanna and {Hainline}, Kevin and {Hausen}, Ryan and {Jakobsen}, Peter and {Johnson}, Benjamin D. and {Maiolino}, Roberto and {Rauscher}, Bernard J. and {Rieke}, Marcia and {Rieke}, George and {Rix}, Hans-Walter and {Robertson}, Brant and {Stark}, Daniel P. and {Tacchella}, Sandro and {Williams}, Christina C. and {Willmer}, Christopher N.~A. and {Baker}, William M. and {Baum}, Stefi and {Bhatawdekar}, Rachana and {Boyett}, Kristan and {Chen}, Zuyi and {Chevallard}, Jacopo and {Circosta}, Chiara and {Curti}, Mirko and {Danhaive}, A. Lola and {DeCoursey}, Christa and {Endsley}, Ryan and {de Graaff}, Anna and {Dressler}, Alan and {Egami}, Eiichi and {Helton}, Jakob M. and {Hviding}, Raphael E. and {Ji}, Zhiyuan and {Jones}, Gareth C. and {Kumari}, Nimisha and {L{\"u}tzgendorf}, Nora and {Laseter}, Isaac and {Looser}, Tobias J. and {Lyu}, Jianwei and {Maseda}, Michael V. and {Nelson}, Erica and {Parlanti}, Eleonora and {Perna}, Michele and {Pusk{\'a}s}, D{\'a}vid and {Rawle}, Tim and {Rodr{\'\i}guez Del Pino}, Bruno and {Rujopakarn}, Wiphu and {Sandles}, Lester and {Saxena}, Aayush and {Scholtz}, Jan and {Sharpe}, Katherine and {Shivaei}, Irene and {Silcock}, Maddie S. and {Simmonds}, Charlotte and {Skarbinski}, Maya and {Smit}, Renske and {Stone}, Meredith and {Suess}, Katherine A. and {Sun}, Fengwu and {Tang}, Mengtao and {Topping}, Michael W. and {{\"U}bler}, Hannah and {Villanueva}, Natalia C. and {Wallace}, Imaan E.~B. and {Whitler}, Lily and {Witstok}, Joris and {Woodrum}, Charity},
        title = "{Overview of the JWST Advanced Deep Extragalactic Survey (JADES)}",
      journal = {\apjs},
         year = 2026,
        month = mar,
       volume = {283},
       number = {1},
          eid = {6},
        pages = {6},
          doi = {10.3847/1538-4365/ae3163},
archivePrefix = {arXiv},
       eprint = {2306.02465},
 primaryClass = {astro-ph.GA},
       adsurl = {https://ui.adsabs.harvard.edu/abs/2026ApJS..283....6E}
}

@ARTICLE{Bagley2024,
       author = {{Bagley}, M. B. and {Pirzkal}, N. and {Finkelstein}, S. L. and
                 et al.},
        title = "{The Next Generation Deep Extragalactic Exploratory Public (NGDEEP) Survey}",
      journal = {\apjl},
         year = 2024,
       volume = {965},
          eid = {L6},
        pages = {L6},
          doi = {10.3847/2041-8213/ad2f31}
}

@ARTICLE{Bertin1996,
       author = {{Bertin}, E. and {Arnouts}, S.},
        title = "{SExtractor: Software for source extraction}",
      journal = {\aaps},
         year = 1996,
        month = jun,
       volume = {117},
        pages = {393--404},
          doi = {10.1051/aas:1996164}
}

@ARTICLE{GaiaDR3,
       author = {{Gaia Collaboration} and {Vallenari}, A. and {Brown}, A. G. A. and
                 {Prusti}, T. and {de Bruijne}, J. H. J. and {Arenou}, F. and
                 {Babusiaux}, C. and {Biermann}, M. and {Creevey}, O. L. and
                 {Ducourant}, C. and et al.},
        title = "{Gaia Data Release 3. Summary of the content and survey properties}",
      journal = {\aap},
         year = 2023,
        month = jun,
       volume = {674},
          eid = {A1},
        pages = {A1},
          doi = {10.1051/0004-6361/202243940}
}

@ARTICLE{Virtanen2020,
  author  = {Virtanen, Pauli and Gommers, Ralf and Oliphant, Travis E. and
             Haberland, Matt and Reddy, Tyler and Cournapeau, David and
             Burovski, Evgeni and Peterson, Pearu and Weckesser, Warren and
             Bright, Jonathan and {van der Walt}, St{\'e}fan J. and
             Brett, Matthew and Wilson, Joshua and Millman, K. Jarrod and
             Mayorov, Nikolay and Nelson, Andrew R. J. and Jones, Eric and
             Kern, Robert and Larson, Eric and Carey, C. J. and
             Polat, {\.I}lhan and Feng, Yu and Moore, Eric W. and
             {VanderPlas}, Jake and Laxalde, Denis and Perktold, Josef and
             Cimrman, Robert and Henriksen, Ian and Quintero, E. A. and
             Harris, Charles R. and Archibald, Anne M. and
             Ribeiro, Ant{\^o}nio H. and Pedregosa, Fabian and
             {van Mulbregt}, Paul and {SciPy 1.0 Contributors}},
  title   = {{SciPy} 1.0: Fundamental Algorithms for Scientific Computing in Python},
  journal = {Nature Methods},
  year    = {2020},
  volume  = {17},
  pages   = {261--272},
  doi     = {10.1038/s41592-019-0686-2}
}

@ARTICLE{Kelly2007,
       author = {{Kelly}, Brandon C.},
        title = "{Some Aspects of Measurement Error in Linear Regression of Astronomical Data}",
      journal = {\apj},
         year = 2007,
        month = aug,
       volume = {665},
       number = {2},
        pages = {1489--1506},
          doi = {10.1086/519947},
archivePrefix = {arXiv},
       eprint = {0705.2774},
 primaryClass = {astro-ph}
}

@ARTICLE{Grasha2018,
       author = {{Grasha}, K. and {Calzetti}, D. and {Bittle}, L. and
                 {Johnson}, K.~E. and {Donovan Meyer}, J. and
                 {Kennicutt}, R.~C. and {Elmegreen}, B.~G. and
                 {Adamo}, A. and {Krumholz}, M.~R. and {Fumagalli}, M.
                 and et al.},
        title = "{Connecting young star clusters to CO molecular gas in NGC 7793 with ALMA-LEGUS}",
      journal = {\mnras},
         year = 2018,
        month = nov,
       volume = {481},
       number = {1},
        pages = {1016--1027},
          doi = {10.1093/mnras/sty2154}
}

@ARTICLE{Pedrini2026,
       author = {{Pedrini}, A. and {Adamo}, A. and {Calzetti}, D. and
                 {Bik}, A. and {Haworth}, T.~J. and {Elmegreen}, B.~G.
                 and {Krumholz}, M.~R. and {Linden}, S.~T. and
                 {Gregg}, B. and {Faustino Vieira}, H. and {Bajaj}, V.
                 and {Ryon}, J.~E. and {Ali}, A.~A. and {Andersson}, E.~P.
                 and {Bortolini}, G. and {Cignoni}, M. and
                 {Duarte-Cabral}, A. and {Grasha}, K. and {Lahén}, N.
                 and {Lai}, T.~S.-Y. and {Lapeer}, D. and {Messa}, M.
                 and {Östlin}, G. and {Sabbi}, E. and {Smith}, L.~J.
                 and {Tosi}, M.},
        title = "{The emerging timescale of young star clusters regulated by cluster stellar mass}",
      journal = {Nature Astronomy},
         year = 2026,
        month = may,
          doi = {10.1038/s41550-026-02857-y}
}

@ARTICLE{Karthikeyan2026,
       author = {{Karthikeyan}, Shreya and {Clarke}, Leonardo and {Shapley}, Alice E. and {Lam}, Natalie and {Sanders}, Ryan L. and {Reddy}, Naveen A. and {Topping}, Michael W. and {Brammer}, Gabriel B.},
        title = "{Balmer Decrements and Nebular and Stellar Reddening in JADES Galaxies at 2.7 < z < 7}",
      journal = {\apj},
         year = 2026,
        month = sep,
       volume = {1009},
       number = {1},
          eid = {83},
        pages = {83},
          doi = {10.3847/1538-4357/ae9a7f},
archivePrefix = {arXiv},
       eprint = {2603.11338},
 primaryClass = {astro-ph.GA},
       adsurl = {https://ui.adsabs.harvard.edu/abs/2026ApJ..1009...83K}
}

@ARTICLE{Lam2026,
       author = {{Lam}, Natalie and {Clarke}, Leonardo and {Shapley}, Alice E. and {Sanders}, Ryan L. and {Topping}, Michael W. and {Brammer}, Gabriel B. and {Reddy}, Naveen A. and {Karthikeyan}, Shreya},
        title = "{The JADES Mass-Metallicity and Fundamental Metallicity Relations at $zrsim2$ Using New High-Redshift Metallicity Calibrations}",
         year = 2026,
        month = may,
          eid = {arXiv:2605.30513},
        pages = {arXiv:2605.30513},
          doi = {10.48550/arXiv.2605.30513},
archivePrefix = {arXiv},
       eprint = {2605.30513},
 primaryClass = {astro-ph.GA},
       adsurl = {https://ui.adsabs.harvard.edu/abs/2026arXiv260530513L}
}

@ARTICLE{Tremonti2004,
       author = {{Tremonti}, Christy A. and {Heckman}, Timothy M.
                 and {Kauffmann}, Guinevere and {Brinchmann}, Jarle
                 and {Charlot}, St{\'e}phane and {White}, Simon D.~M.
                 and {Seibert}, Mark and {Peng}, Eric W.
                 and {Schlegel}, David J. and {Uomoto}, Alan
                 and {Fukugita}, Masataka and {Brinkmann}, Jon},
        title = "{The Origin of the Mass--Metallicity Relation:
                  Insights from 53,000 Star-forming Galaxies
                  in the Sloan Digital Sky Survey}",
      journal = {\apj},
         year = 2004,
       volume = {613},
       number = {2},
        pages = {898-913},
          doi = {10.1086/423264}
}

@ARTICLE{Gordon2024,
       author = {{Gordon}, Karl D. and {Fitzpatrick}, E. L.
                 and {Massa}, Derck and {Bohlin}, Ralph
                 and {Chastenet}, Jérémy and {Murray}, Claire E.
                 and {Clayton}, Geoffrey C. and {Lennon}, Daniel J.
                 and {Misselt}, Karl A. and {Sandstrom}, Karin},
        title = "{Expanded Sample of Small Magellanic Cloud Ultraviolet Dust
                  Extinction Curves: Correlations between the 2175 {\AA} Bump,
                  qPAH, Ultraviolet Extinction Shape, and N(H I)/A(V)}",
      journal = {\apj},
         year = 2024,
        month = jul,
       volume = {970},
       number = {1},
          eid = {51},
        pages = {51},
          doi = {10.3847/1538-4357/ad4be1}
}

@ARTICLE{SeonDraine2016,
       author = {{Seon}, Kwang-Il and {Draine}, Bruce T.},
        title = "{Radiative Transfer Model of Dust Attenuation Curves
                  in Clumpy, Galactic Environments}",
      journal = {\apj},
         year = 2016,
        month = dec,
       volume = {833},
       number = {2},
          eid = {201},
        pages = {201},
          doi = {10.3847/1538-4357/833/2/201}
}

@ARTICLE{Burgarella2007,
       author = {{Burgarella}, D. and {Le Floc'h}, E. and {Takeuchi}, T.~T.
                 and {Huang}, J.~S. and {Buat}, V. and {Rieke}, G.~H.
                 and {Tyler}, K.~D.},
        title = "{Lyman break galaxies at z~1 and the evolution of dust attenuation in star-forming galaxies with redshift}",
      journal = {\mnras},
         year = 2007,
       volume = {380},
       number = {3},
        pages = {986-998},
          doi = {10.1111/j.1365-2966.2007.12063.x}
}

@ARTICLE{Burgarella2013,
       author = {{Burgarella}, D. and {Buat}, V. and {Gruppioni}, C.
                 and {Heinis}, S. and {Magdis}, G. and {Le Floc'h}, E.
                 and {Takeuchi}, T.~T. and {Berta}, S.
                 and {Roussel}, H. and {et al.}},
        title = "{Herschel reveals a dustier Universe: the cosmic evolution of dust attenuation in galaxies}",
      journal = {\aap},
         year = 2013,
       volume = {554},
          eid = {A70},
        pages = {A70},
          doi = {10.1051/0004-6361/201321651}
}

@ARTICLE{Buat2015,
       author = {{Buat}, V. and {Heinis}, S. and {Burgarella}, D.
                 and {De Looze}, I. and {Boquien}, M. and {Ciesla}, L.
                 and {Giovannoli}, E. and {Pannella}, M.
                 and {Gruppioni}, C. and {et al.}},
        title = "{Dust attenuation properties in the local Universe: Calzetti law versus CF00 prescription}",
      journal = {\aap},
         year = 2015,
       volume = {577},
          eid = {A141},
        pages = {A141}
}

@ARTICLE{Kriek2013,
       author = {{Kriek}, M. and {Conroy}, C.},
        title = "{The Dust Attenuation Law in Distant Galaxies: Evidence for Variation with Spectral Type}",
      journal = {\apjl},
         year = 2013,
       volume = {775},
          eid = {L16},
        pages = {L16},
          doi = {10.1088/2041-8205/775/1/L16}
}

@ARTICLE{Salim2018,
       author = {{Salim}, S. and {Boquien}, M. and {Lee}, J.~C.},
        title = "{Dust Attenuation Curves in the Local Universe:
                  Demographics and New Laws for Star-forming Galaxies
                  and High-redshift Analogs}",
      journal = {\apj},
         year = 2018,
       volume = {859},
          eid = {11},
        pages = {11}
}

@ARTICLE{Salim2020,
       author = {{Salim}, S. and {Narayanan}, D.},
        title = "{The Dust Attenuation Law in Galaxies}",
      journal = {\araa},
         year = 2020,
       volume = {58},
        pages = {529-575},
          doi = {10.1146/annurev-astro-032620-021933}
}

@ARTICLE{CharlotFall2000,
       author = {{Charlot}, St{\'e}phane and {Fall}, S. Michael},
        title = "{A Simple Model for the Absorption of Starlight by Dust in Galaxies}",
      journal = {\apj},
         year = 2000,
        month = aug,
       volume = {539},
       number = {2},
        pages = {718-731},
          doi = {10.1086/309250},
archivePrefix = {arXiv},
       eprint = {astro-ph/0003128},
 primaryClass = {astro-ph},
       adsurl = {https://ui.adsabs.harvard.edu/abs/2000ApJ...539..718C}
}

@ARTICLE{Langeroodi24,
       author = {{Langeroodi}, Danial and {Hjorth}, Jens},
        title = "{NIRSpec View of the Appearance and Evolution of Balmer Breaks and the Transition from Bursty to Smooth Star Formation Histories from Deep Within the Epoch of Reionization to Cosmic Noon}",
         year = 2024,
        month = apr,
          eid = {arXiv:2404.13045},
        pages = {arXiv:2404.13045},
          doi = {10.48550/arXiv.2404.13045},
archivePrefix = {arXiv},
       eprint = {2404.13045},
 primaryClass = {astro-ph.GA},
       adsurl = {https://ui.adsabs.harvard.edu/abs/2024arXiv240413045L}
}

@ARTICLE{Markov23,
       author = {{Markov}, V. and {Gallerani}, S. and {Pallottini}, A. and {Sommovigo}, L. and {Carniani}, S. and {Ferrara}, A. and {Parlanti}, E. and {Di Mascia}, F.},
        title = "{Dust attenuation law in JWST galaxies at z {\ensuremath{\sim}} 7-8}",
      journal = {\aap},
         year = 2023,
        month = nov,
       volume = {679},
          eid = {A12},
        pages = {A12},
          doi = {10.1051/0004-6361/202346723},
archivePrefix = {arXiv},
       eprint = {2304.11178},
 primaryClass = {astro-ph.GA},
       adsurl = {https://ui.adsabs.harvard.edu/abs/2023A&A...679A..12M}
}

@ARTICLE{Simmonds2025,
       author = {{Simmonds}, C. and {Tacchella}, S. and {McClymont}, W. and {Curtis-Lake}, E. and {D'Eugenio}, F. and {Hainline}, K. and {Johnson}, B.~D. and {Kravtsov}, A. and {Pusk{\'a}s}, D. and {Robertson}, B. and {Stoffers}, A. and {Willott}, C. and {Baker}, W.~M. and {Belokurov}, V.~A. and {Bhatawdekar}, R. and {Bunker}, A.~J. and {Carniani}, S. and {Chevallard}, J. and {Curti}, M. and {Duan}, Q. and {Helton}, J.~M. and {Ji}, Z. and {Looser}, T.~J. and {Maiolino}, R. and {Maseda}, M.~V. and {Shivaei}, I. and {Williams}, C.~C.},
        title = "{Bursting at the seams: the star-forming main sequence and its scatter at z = 3─9 using NIRCam photometry from JADES}",
      journal = {\mnras},
         year = 2025,
        month = dec,
       volume = {544},
       number = {4},
        pages = {4551-4575},
          doi = {10.1093/mnras/staf1950},
archivePrefix = {arXiv},
       eprint = {2508.04410},
 primaryClass = {astro-ph.GA},
       adsurl = {https://ui.adsabs.harvard.edu/abs/2025MNRAS.544.4551S}
}

@ARTICLE{Carnall2019,
       author = {{Carnall}, Adam C. and {Leja}, Joel and {Johnson}, Benjamin D. and {McLure}, Ross J. and {Dunlop}, James S. and {Conroy}, Charlie},
        title = "{How to Measure Galaxy Star Formation Histories. I. Parametric Models}",
      journal = {\apj},
         year = 2019,
        month = mar,
       volume = {873},
       number = {1},
          eid = {44},
        pages = {44},
          doi = {10.3847/1538-4357/ab04a2},
archivePrefix = {arXiv},
       eprint = {1811.03635},
 primaryClass = {astro-ph.GA},
       adsurl = {https://ui.adsabs.harvard.edu/abs/2019ApJ...873...44C}
}

@ARTICLE{Topping2022,
       author = {{Topping}, Michael W. and {Stark}, Daniel P. and {Endsley}, Ryan and {Bouwens}, Rychard J. and {Schouws}, Sander and {Smit}, Renske and {Stefanon}, Mauro and {Inami}, Hanae and {Bowler}, Rebecca A.~A. and {Oesch}, Pascal and {Gonzalez}, Valentino and {Dayal}, Pratika and {da Cunha}, Elisabete and {Algera}, Hiddo and {van der Werf}, Paul and {Pallottini}, Andrea and {Barrufet}, Laia and {Schneider}, Raffaella and {De Looze}, Ilse and {Sommovigo}, Laura and {Whitler}, Lily and {Graziani}, Luca and {Fudamoto}, Yoshinobu and {Ferrara}, Andrea},
        title = "{The ALMA REBELS Survey: specific star formation rates in the reionization era}",
      journal = {\mnras},
         year = 2022,
        month = oct,
       volume = {516},
       number = {1},
        pages = {975-991},
          doi = {10.1093/mnras/stac2291},
archivePrefix = {arXiv},
       eprint = {2203.07392},
 primaryClass = {astro-ph.GA},
       adsurl = {https://ui.adsabs.harvard.edu/abs/2022MNRAS.516..975T}
}

@ARTICLE{Ormerod2025,
       author = {{Ormerod}, Katherine and {Witstok}, Joris and {Smit}, Renske and {de Graaff}, Anna and {Helton}, Jakob M. and {Maseda}, Michael V. and {Shivaei}, Irene and {Bunker}, Andrew J. and {Carniani}, Stefano and {D'Eugenio}, Francesco and {Bhatawdekar}, Rachana and {Chevallard}, Jacopo and {Franx}, Marijn and {Kumari}, Nimisha and {Maiolino}, Roberto and {Rinaldi}, Pierluigi and {Robertson}, Brant and {Tacchella}, Sandro},
        title = "{Detection of the 2175 {\r{A}} UV bump at z > 7: evidence for rapid dust evolution in a merging reionization-era galaxy}",
      journal = {\mnras},
         year = 2025,
        month = sep,
       volume = {542},
       number = {2},
        pages = {1136-1154},
          doi = {10.1093/mnras/staf1228},
archivePrefix = {arXiv},
       eprint = {2502.21119},
 primaryClass = {astro-ph.GA},
       adsurl = {https://ui.adsabs.harvard.edu/abs/2025MNRAS.542.1136O}
}

@ARTICLE{Witstok2023,
       author = {{Witstok}, Joris and {Shivaei}, Irene and {Smit}, Renske and {Maiolino}, Roberto and {Carniani}, Stefano and {Curtis-Lake}, Emma and {Ferruit}, Pierre and {Arribas}, Santiago and {Bunker}, Andrew J. and {Cameron}, Alex J. and {Charlot}, Stephane and {Chevallard}, Jacopo and {Curti}, Mirko and {de Graaff}, Anna and {D'Eugenio}, Francesco and {Giardino}, Giovanna and {Looser}, Tobias J. and {Rawle}, Tim and {Rodr{\'\i}guez del Pino}, Bruno and {Willott}, Chris and {Alberts}, Stacey and {Baker}, William M. and {Boyett}, Kristan and {Egami}, Eiichi and {Eisenstein}, Daniel J. and {Endsley}, Ryan and {Hainline}, Kevin N. and {Ji}, Zhiyuan and {Johnson}, Benjamin D. and {Kumari}, Nimisha and {Lyu}, Jianwei and {Nelson}, Erica and {Perna}, Michele and {Rieke}, Marcia and {Robertson}, Brant E. and {Sandles}, Lester and {Saxena}, Aayush and {Scholtz}, Jan and {Sun}, Fengwu and {Tacchella}, Sandro and {Williams}, Christina C. and {Willmer}, Christopher N.~A.},
        title = "{Carbonaceous dust grains seen in the first billion years of cosmic time}",
      journal = {\nat},
         year = 2023,
        month = sep,
       volume = {621},
       number = {7978},
        pages = {267-270},
          doi = {10.1038/s41586-023-06413-w},
archivePrefix = {arXiv},
       eprint = {2302.05468},
 primaryClass = {astro-ph.GA},
       adsurl = {https://ui.adsabs.harvard.edu/abs/2023Natur.621..267W}
}

@ARTICLE{Sommovigo26a,
       author = {{Sommovigo}, Laura and {Bartlett}, Deaglan J. and {Cochrane}, Rachel K. and {Ho}, Matthew and {Lovell}, Christopher C. and {Somerville}, Rachel S.},
        title = "{Learning the Universe: The Structure of Dust Attenuation Curves in Galaxy Simulations}",
      journal = {\apj},
         year = 2026,
        month = sep,
       volume = {1008},
       number = {2},
          eid = {218},
        pages = {218},
          doi = {10.3847/1538-4357/ae8e78},
archivePrefix = {arXiv},
       eprint = {2606.10027},
 primaryClass = {astro-ph.GA},
       adsurl = {https://ui.adsabs.harvard.edu/abs/2026ApJ..1008..218S}
}

@ARTICLE{Hirashita2020,
       author = {{Hirashita}, Hiroyuki and {Murga}, Maria S.},
        title = "{Self-consistent modelling of aromatic dust species and extinction curves in galaxy evolution}",
      journal = {\mnras},
         year = 2020,
        month = mar,
       volume = {492},
       number = {3},
        pages = {3779-3793},
          doi = {10.1093/mnras/stz3640},
archivePrefix = {arXiv},
       eprint = {2001.01968},
 primaryClass = {astro-ph.GA},
       adsurl = {https://ui.adsabs.harvard.edu/abs/2020MNRAS.492.3779H}
}

@ARTICLE{Narayanan2018,
       author = {{Narayanan}, Desika and {Conroy}, Charlie and {Dav{\'e}}, Romeel and {Johnson}, Benjamin D. and {Popping}, Gerg{\"o}},
        title = "{A Theory for the Variation of Dust Attenuation Laws in Galaxies}",
      journal = {\apj},
         year = 2018,
        month = dec,
       volume = {869},
       number = {1},
          eid = {70},
        pages = {70},
          doi = {10.3847/1538-4357/aaed25},
archivePrefix = {arXiv},
       eprint = {1805.06905},
 primaryClass = {astro-ph.GA},
       adsurl = {https://ui.adsabs.harvard.edu/abs/2018ApJ...869...70N}
}

@ARTICLE{McKinnon2018,
       author = {{McKinnon}, Ryan and {Vogelsberger}, Mark and {Torrey}, Paul and {Marinacci}, Federico and {Kannan}, Rahul},
        title = "{Simulating galactic dust grain evolution on a moving mesh}",
      journal = {\mnras},
         year = 2018,
        month = aug,
       volume = {478},
       number = {3},
        pages = {2851-2886},
          doi = {10.1093/mnras/sty1248},
archivePrefix = {arXiv},
       eprint = {1805.04521},
 primaryClass = {astro-ph.GA},
       adsurl = {https://ui.adsabs.harvard.edu/abs/2018MNRAS.478.2851M}
}

@ARTICLE{Boquien2022,
       author = {{Boquien}, M{\'e}d{\'e}ric and {Buat}, V{\'e}ronique and {Burgarella}, Denis and {Bardelli}, Sandro and {B{\'e}thermin}, Matthieu and {Faisst}, Andreas and {Ginolfi}, Michele and {Hathi}, Nimish and {Jones}, Gareth and {Koekemoer}, Anton and {Lemaux}, Brian and {Narayanan}, Desika and {Romano}, Michael and {Schaerer}, Daniel and {Vergani}, Daniela and {Zamorani}, Giovanni and {Zucca}, Elena},
        title = "{The ALPINE-ALMA [C II] survey. Dust attenuation curves at z = 4.4-5.5}",
      journal = {\aap},
         year = 2022,
        month = jul,
       volume = {663},
          eid = {A50},
        pages = {A50},
          doi = {10.1051/0004-6361/202142537},
archivePrefix = {arXiv},
       eprint = {2202.11723},
 primaryClass = {astro-ph.GA},
       adsurl = {https://ui.adsabs.harvard.edu/abs/2022A&A...663A..50B}
}

@ARTICLE{Zafar18,
       author = {{Zafar}, T. and {Watson}, D. and {M{\o}ller}, P. and {Selsing}, J. and {Fynbo}, J.~P.~U. and {Schady}, P. and {Wiersema}, K. and {Levan}, A.~J. and {Heintz}, K.~E. and {de Ugarte Postigo}, A. and {D'Elia}, V. and {Jakobsson}, P. and {Bolmer}, J. and {Japelj}, J. and {Covino}, S. and {Gomboc}, A. and {Cano}, Z.},
        title = "{VLT/X-shooter GRBs: Individual extinction curves of star-forming regions}",
      journal = {\mnras},
         year = 2018,
        month = sep,
       volume = {479},
       number = {2},
        pages = {1542-1554},
          doi = {10.1093/mnras/sty1380},
archivePrefix = {arXiv},
       eprint = {1805.07016},
 primaryClass = {astro-ph.GA},
       adsurl = {https://ui.adsabs.harvard.edu/abs/2018MNRAS.479.1542Z}
}

@ARTICLE{Bolmer2018,
       author = {{Bolmer}, J. and {Greiner}, J. and {Kr{\"u}hler}, T. and {Schady}, P. and {Ledoux}, C. and {Tanvir}, N.~R. and {Levan}, A.~J.},
        title = "{Dust reddening and extinction curves toward gamma-ray bursts at z > 4}",
      journal = {\aap},
         year = 2018,
        month = jan,
       volume = {609},
          eid = {A62},
        pages = {A62},
          doi = {10.1051/0004-6361/201731255},
archivePrefix = {arXiv},
       eprint = {1709.06867},
 primaryClass = {astro-ph.GA},
       adsurl = {https://ui.adsabs.harvard.edu/abs/2018A&A...609A..62B}
}

@ARTICLE{Gallerani2010,
       author = {{Gallerani}, S. and {Maiolino}, R. and {Juarez}, Y. and {Nagao}, T. and {Marconi}, A. and {Bianchi}, S. and {Schneider}, R. and {Mannucci}, F. and {Oliva}, T. and {Willott}, C.~J. and {Jiang}, L. and {Fan}, X.},
        title = "{The extinction law at high redshift and its implications}",
      journal = {\aap},
         year = 2010,
        month = nov,
       volume = {523},
          eid = {A85},
        pages = {A85},
          doi = {10.1051/0004-6361/201014721},
archivePrefix = {arXiv},
       eprint = {1006.4463},
 primaryClass = {astro-ph.CO},
       adsurl = {https://ui.adsabs.harvard.edu/abs/2010A&A...523A..85G}
}

@ARTICLE{Maiolino2004,
       author = {{Maiolino}, R. and {Schneider}, R. and {Oliva}, E. and {Bianchi}, S. and {Ferrara}, A. and {Mannucci}, F. and {Pedani}, M. and {Roca Sogorb}, M.},
        title = "{A supernova origin for dust in a high-redshift quasar}",
      journal = {\nat},
         year = 2004,
        month = sep,
       volume = {431},
       number = {7008},
        pages = {533-535},
          doi = {10.1038/nature02930},
archivePrefix = {arXiv},
       eprint = {astro-ph/0409577},
 primaryClass = {astro-ph},
       adsurl = {https://ui.adsabs.harvard.edu/abs/2004Natur.431..533M}
}

@ARTICLE{Glenn2025,
       author = {{Glenn}, Jason and {Meixner}, Margaret and {Bradford}, Charles M. and {Pontoppidan}, Klaus and {Pope}, Alexandra and {Kataria}, Tiffany and {Rocca}, Jennifer and {Luthman}, Elizabeth and {Armus}, Lee and {Baselmans}, Jochem and {Battersby}, Cara and {Bollato}, Alberto and {Burgarella}, Denis and {Chen}, Weibo and {Ciesla}, Laure and {Day}, Peter and {Di Giorgio}, Anna and {Dipirro}, Michael and {Dowell}, Charles Darren and {Echternach}, Pierre and {Essinger-Hileman}, Thomas and {Foote}, Marc and {Gruppioni}, Carlotta and {Hensley}, Brandon and {Henning}, Thomas and {Jellema}, Willem and {Johnson}, Matthew and {Kogut}, Alan and {Krause}, Oliver and {McGuire}, James and {Mills}, Elisabeth and {Moullet}, Arielle and {Rodgers}, Michael and {Sauvage}, Marc and {Smith}, John D. and {Somerville}, Rachel and {Staguhn}, Johannes and {Stevenson}, Thomas and {Tucker}, Carole and {Unwin}, Stephen and {Ziemer}, John and {Cannella}, Matthew and {Dissly}, Richard},
        title = "{PRIMA mission concept}",
      journal = {JATIS},
         year = 2025,
        month = jul,
       volume = {11},
          eid = {031628},
        pages = {031628},
          doi = {10.1117/1.JATIS.11.3.031628},
       adsurl = {https://ui.adsabs.harvard.edu/abs/2025JATIS..11c1628G}
}

@article{noll2009,
  author = {Noll, S. and Burgarella, D. and Giovannoli, E. and Buat, V. and Marcillac, D. and Mu{\~n}oz-Mateos, J. C.},
  title = {Analysis of galaxy spectral energy distributions from far-UV to mid-IR: dust attenuation and stellar populations at $z \sim 2$},
  journal = {\aap},
  year = {2009},
  volume = {507},
  number = {3},
  pages = {1793--1813},
  doi = {10.1051/0004-6361/200912221}
}

@article{kashino2022,
  author = {Kashino, Daichi and Lilly, Simon J. and Renzini, Alvio and Daddi, Emanuele and Zamorani, Giovanni and Silverman, John D. and Ilbert, Olivier and Peng, Yingjie and Mainieri, Vincenzo and Bardelli, Sandro and Zucca, Elena and Kartaltepe, Jeyhan S. and Sanders, David B.},
  title = {The stellar mass versus stellar metallicity relation of star-forming galaxies at $1.6 \le z \le 3.0$ and implications for the evolution of the $\alpha$-enhancement},
  journal = {\apj},
  year = {2022},
  volume = {926},
  number = {2},
  pages = {134},
  doi = {10.3847/1538-4357/ac3a68}
}

@ARTICLE{Perrin2014S,
       author = {{Perrin}, Marshall D. and {Sivaramakrishnan}, Anand and {Lajoie}, Charles-Philippe and {Elliott}, Erin and {Pueyo}, Laurent and {Ravindranath}, Swara and {Albert}, Lo{\"\i}c},
        title = "{Updated point spread function simulations for JWST with WebbPSF}",
      journal = {\procspie},
         year = 2014,
       volume = {9143},
        pages = {91433X},
        month = aug,
          doi = {10.1117/12.2056689},
       adsurl = {https://ui.adsabs.harvard.edu/abs/2014SPIE.9143E..3XP}
}

@ARTICLE{Fisher2025,
       author = {{Fisher}, R. and {Bowler}, R.~A.~A. and {Stefanon}, M. and {Rowland}, L.~E. and {Algera}, H.~S.~B. and {Aravena}, M. and {Bouwens}, R. and {Dayal}, P. and {Ferrara}, A. and {Fudamoto}, Y. and {Gulis}, C. and {Hodge}, J.~A. and {Inami}, H. and {Ormerod}, K. and {Pallottini}, A. and {Phillips}, S.~G. and {Sartorio}, N.~S. and {Schouws}, S. and {Smit}, R. and {Sommovigo}, L. and {Stark}, D.~P. and {van der Werf}, P.~P.},
        title = "{REBELS-IFU: dust attenuation curves of 12 massive galaxies at z ≃ 7}",
      journal = {\mnras},
         year = 2025,
        month = may,
       volume = {539},
       number = {1},
        pages = {109-126},
          doi = {10.1093/mnras/staf485},
archivePrefix = {arXiv},
       eprint = {2501.10541},
 primaryClass = {astro-ph.GA},
       adsurl = {https://ui.adsabs.harvard.edu/abs/2025MNRAS.539..109F}
}

@article{woodrum2025,
       author = {{Woodrum}, Charity and {Shivaei}, Irene and {Witstok}, Joris and {Saxena}, Aayush and {Simmonds}, Charlotte and {Scholtz}, Jan and {Bhatawdekar}, Rachana and {Bunker}, Andrew J. and {Carniani}, St{\'e}fano and {Charlot}, Stephane and {Curti}, Mirko and {Curtis-Lake}, Emma and {Chevallard}, Jacopo and {D'Eugenio}, Francesco and {Hainline}, Kevin and {Helton}, Jakob M. and {Maiolino}, Roberto and {Perna}, Michele and {Rinaldi}, Pierluigi and {Robertson}, Brant and {Straughn}, Amber and {Sun}, Yang and {Tacchella}, Sandro and {Williams}, Christina C. and {Willott}, Chris and {Zhu}, Yongda},
        title = "{JADES: The Star Formation and Dust Attenuation Properties of Galaxies at 3<z<7}",
         year = 2025,
        month = sep,
          eid = {arXiv:2510.00235},
        pages = {arXiv:2510.00235},
          doi = {10.48550/arXiv.2510.00235},
archivePrefix = {arXiv},
       eprint = {2510.00235},
 primaryClass = {astro-ph.GA},
       adsurl = {https://ui.adsabs.harvard.edu/abs/2025arXiv251000235W}
}

@ARTICLE{Simmonds2024,
       author = {{Simmonds}, C. and {Tacchella}, S. and {Hainline}, K. and {Johnson}, B.~D. and {Pusk{\'a}s}, D. and {Robertson}, B. and {Baker}, W.~M. and {Bhatawdekar}, R. and {Boyett}, K. and {Bunker}, A.~J. and {Cargile}, P.~A. and {Carniani}, S. and {Chevallard}, J. and {Curti}, M. and {Curtis-Lake}, E. and {Ji}, Z. and {Jones}, G.~C. and {Kumari}, N. and {Laseter}, I. and {Maiolino}, R. and {Maseda}, M.~V. and {Rinaldi}, P. and {Stoffers}, A. and {{\"U}bler}, H. and {Villanueva}, N.~C. and {Williams}, C.~C. and {Willott}, C. and {Witstok}, J. and {Zhu}, Y.},
        title = "{Ionizing properties of galaxies in JADES for a stellar mass complete sample: resolving the cosmic ionizing photon budget crisis at the Epoch of Reionization}",
      journal = {\mnras},
         year = 2024,
        month = dec,
       volume = {535},
       number = {4},
        pages = {2998-3019},
          doi = {10.1093/mnras/stae2537},
archivePrefix = {arXiv},
       eprint = {2409.01286},
 primaryClass = {astro-ph.GA},
       adsurl = {https://ui.adsabs.harvard.edu/abs/2024MNRAS.535.2998S}
}

@ARTICLE{McLure2018,
       author = {{McLure}, R.~J. and {Dunlop}, J.~S. and {Cullen}, F. and {Bourne}, N. and {Best}, P.~N. and {Khochfar}, S. and {Bowler}, R.~A.~A. and {Biggs}, A.~D. and {Geach}, J.~E. and {Scott}, D. and {Micha{\l}owski}, M.~J. and {Rujopakarn}, W. and {van Kampen}, E. and {Kirkpatrick}, A. and {Pope}, A.},
        title = "{Dust attenuation in 2 < z < 3 star-forming galaxies from deep ALMA observations of the Hubble Ultra Deep Field}",
      journal = {\mnras},
         year = 2018,
        month = may,
       volume = {476},
       number = {3},
        pages = {3991-4006},
          doi = {10.1093/mnras/sty522},
archivePrefix = {arXiv},
       eprint = {1709.06102},
 primaryClass = {astro-ph.GA},
       adsurl = {https://ui.adsabs.harvard.edu/abs/2018MNRAS.476.3991M}
}

@ARTICLE{Kennicutt1998,
       author = {{Kennicutt}, Jr., Robert C.},
        title = "{Star Formation in Galaxies Along the Hubble Sequence}",
      journal = {\araa},
         year = 1998,
        month = jan,
       volume = {36},
        pages = {189-232},
          doi = {10.1146/annurev.astro.36.1.189},
archivePrefix = {arXiv},
       eprint = {astro-ph/9807187},
 primaryClass = {astro-ph},
       adsurl = {https://ui.adsabs.harvard.edu/abs/1998ARA&A..36..189K}
}

@ARTICLE{Ikki25,
       author = {{Mitsuhashi}, Ikki and {Suess}, Katherine A. and {Leja}, Joel and {Dayal}, Pratika and {Feldmann}, Robert and {Fujimoto}, Seiji and {Katz}, Harley and {Nanayakkara}, Themiya and {Narayanan}, Desika and {Price}, Sedona H. and {Weaver}, John R. and {Williams}, Christina C. and {Labbe}, Ivo and {Bezanson}, Rachel and {Atek}, Hakim and {Brammer}, Gabriel and {Cutler}, Sam E. and {Furtak}, Lukas J. and {Pan}, Richard and {Wang}, Bingjie and {Whitaker}, Katherine E.},
        title = "{Discovery of red galaxy candidates at z \raisebox{-0.5ex}\textasciitilde 12: Early dust growth or significant nebular emission with high-temperature stars?}",
         year = 2025,
        month = oct,
          eid = {arXiv:2510.13240},
        pages = {arXiv:2510.13240},
          doi = {10.48550/arXiv.2510.13240},
archivePrefix = {arXiv},
       eprint = {2510.13240},
 primaryClass = {astro-ph.GA},
       adsurl = {https://ui.adsabs.harvard.edu/abs/2025arXiv251013240M}
}

@ARTICLE{Rodighiero2026,
       author = {{Rodighiero}, Giulia and {Ferrara}, Andrea and {Catone}, Michele and {Napolitano}, Lorenzo and {Cassata}, Paolo and {Gandolfi}, Giovanni and {Merlin}, Emiliano and {Grazian}, Andrea and {Renzini}, Alvio and {Bisigello}, Laura and {Castellano}, Marco and {P{\'e}rez-Gonz{\'a}lez}, Pablo G. and {P{\'e}rez-D{\'\i}az}, Borja and {Iani}, Edoardo and {Gruppioni}, Carlotta and {Finkelstein}, Steven L. and {Koekemoer}, Anton M. and {Bianchetti}, Alessandro and {Sinigaglia}, Francesco},
        title = "{EGS-z11-R0: a red, dust-rich galaxy at Cosmic Dawn}",
         year = 2026,
        month = mar,
          eid = {arXiv:2603.15841},
        pages = {arXiv:2603.15841},
          doi = {10.48550/arXiv.2603.15841},
archivePrefix = {arXiv},
       eprint = {2603.15841},
 primaryClass = {astro-ph.GA},
       adsurl = {https://ui.adsabs.harvard.edu/abs/2026arXiv260315841R}
}

@ARTICLE{Dottorini2025,
       author = {{Dottorini}, D. and {Calabr{\`o}}, A. and {Pentericci}, L. and {Mascia}, S. and {Llerena}, M. and {Napolitano}, L. and {Santini}, P. and {Roberts-Borsani}, G. and {Castellano}, M. and {Amorin}, R. and {Dickinson}, M. and {Fontana}, A. and {Hathi}, N. and {Hirschmann}, M. and {Koekemoer}, A.~M. and {Lucas}, R.~A. and {Merlin}, E. and {Morales}, A. and {Pacucci}, F. and {Wilkins}, S. and {Arrabal Haro}, P. and {Bagley}, M. and {Finkelstein}, S.~L. and {Kartaltepe}, J. and {Papovich}, C. and {Pirzkal}, N.},
        title = "{Evolution of the UV slope of galaxies at cosmic morning (z > 4): The properties of extremely blue galaxies}",
      journal = {\aap},
         year = 2025,
        month = jun,
       volume = {698},
          eid = {A234},
        pages = {A234},
          doi = {10.1051/0004-6361/202453267},
archivePrefix = {arXiv},
       eprint = {2412.01623},
 primaryClass = {astro-ph.GA},
       adsurl = {https://ui.adsabs.harvard.edu/abs/2025A&A...698A.234D}
}

@ARTICLE{Cullen2024,
       author = {{Cullen}, F. and {McLeod}, D.~J. and {McLure}, R.~J. and {Dunlop}, J.~S. and {Donnan}, C.~T. and {Carnall}, A.~C. and {Keating}, L.~C. and {Magee}, D. and {Arellano-Cordova}, K.~Z. and {Bowler}, R.~A.~A. and {Begley}, R. and {Flury}, S.~R. and {Hamadouche}, M.~L. and {Stanton}, T.~M.},
        title = "{The ultraviolet continuum slopes of high-redshift galaxies: evidence for the emergence of dust-free stellar populations at z > 10}",
      journal = {\mnras},
         year = 2024,
        month = jun,
       volume = {531},
       number = {1},
        pages = {997-1020},
          doi = {10.1093/mnras/stae1211},
archivePrefix = {arXiv},
       eprint = {2311.06209},
 primaryClass = {astro-ph.GA},
       adsurl = {https://ui.adsabs.harvard.edu/abs/2024MNRAS.531..997C}
}

@ARTICLE{Cullen2023,
       author = {{Cullen}, Fergus and {McLure}, R.~J. and {McLeod}, D.~J. and {Dunlop}, J.~S. and {Donnan}, C.~T. and {Carnall}, A.~C. and {Bowler}, R.~A.~A. and {Begley}, R. and {Hamadouche}, M.~L. and {Stanton}, T.~M.},
        title = "{The ultraviolet continuum slopes ({\ensuremath{\beta}}) of galaxies at z ≃ 8-16 from JWST and ground-based near-infrared imaging}",
      journal = {\mnras},
         year = 2023,
        month = mar,
       volume = {520},
       number = {1},
        pages = {14-23},
          doi = {10.1093/mnras/stad073},
archivePrefix = {arXiv},
       eprint = {2208.04914},
 primaryClass = {astro-ph.GA},
       adsurl = {https://ui.adsabs.harvard.edu/abs/2023MNRAS.520...14C}
}

@ARTICLE{Rieke2023,
       author = {{Rieke}, Marcia J. and {Robertson}, Brant and {Tacchella}, Sandro and {Hainline}, Kevin and {Johnson}, Benjamin D. and {Hausen}, Ryan and {Ji}, Zhiyuan and {Willmer}, Christopher N.~A. and {Eisenstein}, Daniel J. and {Pusk{\'a}s}, D{\'a}vid and {Alberts}, Stacey and {Arribas}, Santiago and {Baker}, William M. and {Baum}, Stefi and {Bhatawdekar}, Rachana and {Bonaventura}, Nina and {Boyett}, Kristan and {Bunker}, Andrew J. and {Cameron}, Alex J. and {Carniani}, Stefano and {Charlot}, Stephane and {Chevallard}, Jacopo and {Chen}, Zuyi and {Curti}, Mirko and {Curtis-Lake}, Emma and {Danhaive}, A. Lola and {DeCoursey}, Christa and {Dressler}, Alan and {Egami}, Eiichi and {Endsley}, Ryan and {Helton}, Jakob M. and {Hviding}, Raphael E. and {Kumari}, Nimisha and {Looser}, Tobias J. and {Lyu}, Jianwei and {Maiolino}, Roberto and {Maseda}, Michael V. and {Nelson}, Erica J. and {Rieke}, George and {Rix}, Hans-Walter and {Sandles}, Lester and {Saxena}, Aayush and {Sharpe}, Katherine and {Shivaei}, Irene and {Skarbinski}, Maya and {Smit}, Renske and {Stark}, Daniel P. and {Stone}, Meredith and {Suess}, Katherine A. and {Sun}, Fengwu and {Topping}, Michael and {{\"U}bler}, Hannah and {Villanueva}, Natalia C. and {Wallace}, Imaan E.~B. and {Williams}, Christina C. and {Willott}, Chris and {Whitler}, Lily and {Witstok}, Joris and {Woodrum}, Charity},
        title = "{JADES Initial Data Release for the Hubble Ultra Deep Field: Revealing the Faint Infrared Sky with Deep JWST NIRCam Imaging}",
      journal = {\apjs},
         year = 2023,
        month = nov,
       volume = {269},
       number = {1},
          eid = {16},
        pages = {16},
          doi = {10.3847/1538-4365/acf44d},
archivePrefix = {arXiv},
       eprint = {2306.02466},
 primaryClass = {astro-ph.GA},
       adsurl = {https://ui.adsabs.harvard.edu/abs/2023ApJS..269...16R}
}

@ARTICLE{CurtisLake2025,
       author = {{Curtis-Lake}, Emma and {Cameron}, Alex J. and {Bunker}, Andrew J. and {Scholtz}, Jan and {Carniani}, Stefano and {Parlanti}, Eleonora and {D'Eugenio}, Francesco and {Jakobsen}, Peter and {Willmer}, Christopher N.~A. and {Arribas}, Santiago and {Baker}, William M. and {Charlot}, St{\'e}phane and {Chevallard}, Jacopo and {Circosta}, Chiara and {Curti}, Mirko and {Duan}, Qiao and {Eisenstein}, Daniel J. and {Hainline}, Kevin and {Ji}, Zhiyuan and {Johnson}, Benjamin D. and {Jones}, Gareth C. and {Maiolino}, Roberto and {Maseda}, Michael V. and {Perna}, Michele and {P{\'e}rez-Gonz{\'a}lez}, Pablo G. and {Rawle}, Tim and {Rieke}, Marcia and {Rinaldi}, Pierluigi and {Robertson}, Brant and {Rodr{\'\i}guez Del Pino}, Bruno and {Saxena}, Aayush and {Shivaei}, Irene and {Smit}, Renske and {Tacchella}, Sandro and {{\"U}bler}, Hannah and {Venturi}, Giacomo and {Williams}, Christina C. and {Willott}, Chris},
        title = "{JADES data release 4 ─ Paper I. Sample selection, observing strategy and redshifts of the complete spectroscopic sample}",
      journal = {\mnras},
         year = 2026,
        month = jul,
       volume = {549},
       number = {4},
          eid = {stag836},
        pages = {stag836},
          doi = {10.1093/mnras/stag836},
archivePrefix = {arXiv},
       eprint = {2510.01033},
 primaryClass = {astro-ph.GA},
       adsurl = {https://ui.adsabs.harvard.edu/abs/2026MNRAS.549ag836C}
}

@ARTICLE{Deugenio2025,
       author = {{D'Eugenio}, Francesco and {Cameron}, Alex J. and {Scholtz}, Jan and {Carniani}, Stefano and {Willott}, Chris J. and {Curtis-Lake}, Emma and {Bunker}, Andrew J. and {Parlanti}, Eleonora and {Maiolino}, Roberto and {Willmer}, Christopher N.~A. and {Jakobsen}, Peter and {Robertson}, Brant E. and {Johnson}, Benjamin D. and {Tacchella}, Sandro and {Cargile}, Phillip A. and {Rawle}, Tim and {Arribas}, Santiago and {Chevallard}, Jacopo and {Curti}, Mirko and {Egami}, Eiichi and {Eisenstein}, Daniel J. and {Kumari}, Nimisha and {Looser}, Tobias J. and {Rieke}, Marcia J. and {Rodr{\'\i}guez Del Pino}, Bruno and {Saxena}, Aayush and {{\"U}bler}, Hannah and {Venturi}, Giacomo and {Witstok}, Joris and {Baker}, William M. and {Bhatawdekar}, Rachana and {Bonaventura}, Nina and {Boyett}, Kristan and {Charlot}, Stephane and {Danhaive}, A. Lola and {Hainline}, Kevin N. and {Hausen}, Ryan and {Helton}, Jakob M. and {Ji}, Xihan and {Ji}, Zhiyuan and {Jones}, Gareth C. and {Juod{\v{z}}balis}, Ignas and {Maseda}, Michael V. and {P{\'e}rez-Gonz{\'a}lez}, Pablo G. and {Perna}, Michele and {Pusk{\'a}s}, D{\'a}vid and {Shivaei}, Irene and {Silcock}, Maddie S. and {Simmonds}, Charlotte and {Smit}, Renske and {Sun}, Fengwu and {Villanueva}, Natalia C. and {Williams}, Christina C. and {Zhu}, Yongda},
        title = "{JADES Data Release 3: NIRSpec/Microshutter Assembly Spectroscopy for 4000 Galaxies in the GOODS Fields}",
      journal = {\apjs},
         year = 2025,
        month = mar,
       volume = {277},
       number = {1},
          eid = {4},
        pages = {4},
          doi = {10.3847/1538-4365/ada148},
archivePrefix = {arXiv},
       eprint = {2404.06531},
 primaryClass = {astro-ph.GA},
       adsurl = {https://ui.adsabs.harvard.edu/abs/2025ApJS..277....4D}
}

@ARTICLE{Salim_Nar2020,
       author = {{Salim}, Samir and {Narayanan}, Desika},
        title = "{The Dust Attenuation Law in Galaxies}",
      journal = {\araa},
         year = 2020,
        month = aug,
       volume = {58},
        pages = {529-575},
          doi = {10.1146/annurev-astro-032620-021933},
archivePrefix = {arXiv},
       eprint = {2001.03181},
 primaryClass = {astro-ph.GA},
       adsurl = {https://ui.adsabs.harvard.edu/abs/2020ARA&A..58..529S}
}

@ARTICLE{Salmon2016,
       author = {{Salmon}, Brett and {Papovich}, Casey and {Long}, James and {Willner}, S.~P. and {Finkelstein}, Steven L. and {Ferguson}, Henry C. and {Dickinson}, Mark and {Duncan}, Kenneth and {Faber}, S.~M. and {Hathi}, Nimish and {Koekemoer}, Anton and {Kurczynski}, Peter and {Newman}, Jeffery and {Pacifici}, Camilla and {P{\'e}rez-Gonz{\'a}lez}, Pablo G. and {Pforr}, Janine},
        title = "{Breaking the Curve with CANDELS: A Bayesian Approach to Reveal the Non-Universality of the Dust-Attenuation Law at High Redshift}",
      journal = {\apj},
         year = 2016,
        month = aug,
       volume = {827},
       number = {1},
          eid = {20},
        pages = {20},
          doi = {10.3847/0004-637X/827/1/20},
archivePrefix = {arXiv},
       eprint = {1512.05396},
 primaryClass = {astro-ph.GA},
       adsurl = {https://ui.adsabs.harvard.edu/abs/2016ApJ...827...20S}
}

@ARTICLE{Witt2000,
       author = {{Witt}, Adolf N. and {Gordon}, Karl D.},
        title = "{Multiple Scattering in Clumpy Media. II. Galactic Environments}",
      journal = {\apj},
         year = 2000,
        month = jan,
       volume = {528},
       number = {2},
        pages = {799-816},
          doi = {10.1086/308197},
archivePrefix = {arXiv},
       eprint = {astro-ph/9907342},
 primaryClass = {astro-ph},
       adsurl = {https://ui.adsabs.harvard.edu/abs/2000ApJ...528..799W}
}

@ARTICLE{Weingartner01,
       author = {{Weingartner}, Joseph C. and {Draine}, B.~T.},
        title = "{Dust Grain-Size Distributions and Extinction in the Milky Way, Large Magellanic Cloud, and Small Magellanic Cloud}",
      journal = {\apj},
         year = 2001,
        month = feb,
       volume = {548},
       number = {1},
        pages = {296-309},
          doi = {10.1086/318651},
archivePrefix = {arXiv},
       eprint = {astro-ph/0008146},
 primaryClass = {astro-ph},
       adsurl = {https://ui.adsabs.harvard.edu/abs/2001ApJ...548..296W}
}

@ARTICLE{Markov2025,
       author = {{Markov}, Vladan and {Gallerani}, Simona and {Ferrara}, Andrea and {Pallottini}, Andrea and {Parlanti}, Eleonora and {Mascia}, Fabio Di and {Sommovigo}, Laura and {Kohandel}, Mahsa},
        title = "{The evolution of dust attenuation in z {\ensuremath{\approx}} 2-12 galaxies observed by JWST}",
      journal = {Nature Astronomy},
         year = 2025,
        month = mar,
       volume = {9},
        pages = {458-468},
          doi = {10.1038/s41550-024-02426-1},
archivePrefix = {arXiv},
       eprint = {2402.05996},
 primaryClass = {astro-ph.GA},
       adsurl = {https://ui.adsabs.harvard.edu/abs/2025NatAs...9..458M}
}

@ARTICLE{Shivaei2025,
       author = {{Shivaei}, Irene and {Naidu}, Rohan P. and {Rodr{\'\i}guez Montero}, Francisco and {Matsumoto}, Kosei and {Leja}, Joel and {Matthee}, Jorryt and {Johnson}, Benjamin D. and {Oesch}, Pascal A. and {Chevallard}, Jacopo and {Adamo}, Angela and {Bodansky}, Sarah and {Bunker}, Andrew J. and {Covelo Paz}, Alba and {Di Cesare}, Claudia and {Egami}, Eiichi and {Furtak}, Lukas J. and {Heintz}, Kasper E. and {Kramarenko}, Ivan and {Meyer}, Romain A. and {Reddy}, Naveen A. and {Rinaldi}, Pierluigi and {Tacchella}, Sandro and {Torralba}, Alberto and {Witstok}, Joris and {Wozniak}, Michael A. and {Xiao}, Mengyuan},
        title = "{Diversity and evolution of dust-attenuation curves from redshift z {\ensuremath{\sim}} 1 to 9}",
      journal = {\aap},
         year = 2026,
        month = aug,
       volume = {712},
          eid = {A132},
        pages = {A132},
          doi = {10.1051/0004-6361/202557064},
archivePrefix = {arXiv},
       eprint = {2509.01795},
 primaryClass = {astro-ph.GA},
       adsurl = {https://ui.adsabs.harvard.edu/abs/2026A&A...712A.132S}
}

@ARTICLE{Gordon2003,
       author = {{Gordon}, Karl D. and {Clayton}, Geoffrey C. and {Misselt}, K.~A. and {Landolt}, Arlo U. and {Wolff}, Michael J.},
        title = "{A Quantitative Comparison of the Small Magellanic Cloud, Large Magellanic Cloud, and Milky Way Ultraviolet to Near-Infrared Extinction Curves}",
      journal = {\apj},
         year = 2003,
        month = sep,
       volume = {594},
       number = {1},
        pages = {279-293},
          doi = {10.1086/376774},
archivePrefix = {arXiv},
       eprint = {astro-ph/0305257},
 primaryClass = {astro-ph},
       adsurl = {https://ui.adsabs.harvard.edu/abs/2003ApJ...594..279G}
}

@ARTICLE{Kroupa,
       author = {{Kroupa}, Pavel},
        title = "{On the variation of the initial mass function}",
      journal = {\mnras},
         year = 2001,
        month = apr,
       volume = {322},
       number = {2},
        pages = {231-246},
          doi = {10.1046/j.1365-8711.2001.04022.x},
archivePrefix = {arXiv},
       eprint = {astro-ph/0009005},
 primaryClass = {astro-ph},
       adsurl = {https://ui.adsabs.harvard.edu/abs/2001MNRAS.322..231K}
}

@article{Maiolino_AGN,
	author = {{Maiolino} and {Scholtz, Jan} and {Curtis-Lake, Emma} and {Carniani, Stefano} and {Baker, William} and {de Graaff, Anna} and {Tacchella, Sandro} and {Übler, Hannah} and {D’Eugenio, Francesco} and {Witstok, Joris} and {Curti, Mirko} and {Arribas, Santiago} and {Bunker, Andrew J.}caand {Charlot, Stéphane} and {Chevallard, Jacopo} and {Eisenstein, Daniel J.} and {Egami, Eiichi} and {Ji, Zhiyuan} and {Jones, Gareth C.} and {Lyu, Jianwei} and {Rawle, Tim} and {Robertson, Brant} and {Rujopakarn, Wiphu} and {Perna, Michele} and {Sun, Fengwu} and {Venturi, Giacomo} and {Williams, Christina C.} and {Willott, Chris}},
	title = {JADES - The diverse population of infant black holes at 4 < z < 11: Merging, tiny, poor, but mighty},
	DOI= "10.1051/0004-6361/202347640",
	url= "https://doi.org/10.1051/0004-6361/202347640",
	journal = {\aap},
	year = 2024,
	volume = 691,
	pages = "A145",
}

@ARTICLE{Scholtz_AGN,
       author = {{Scholtz}, Jan and {Maiolino}, Roberto and {D'Eugenio}, Francesco and {Curtis-Lake}, Emma and {Carniani}, Stefano and {Charlot}, Stephane and {Curti}, Mirko and {Silcock}, Maddie S. and {Arribas}, Santiago and {Baker}, William and {Bhatawdekar}, Rachana and {Boyett}, Kristan and {Bunker}, Andrew J. and {Chevallard}, Jacopo and {Circosta}, Chiara and {Eisenstein}, Daniel J. and {Hainline}, Kevin and {Hausen}, Ryan and {Ji}, Xihan and {Ji}, Zhiyuan and {Johnson}, Benjamin D. and {Kumari}, Nimisha and {Looser}, Tobias J. and {Lyu}, Jianwei and {Maseda}, Michael V. and {Parlanti}, Eleonora and {Perna}, Michele and {Rieke}, Marcia and {Robertson}, Brant and {Del Pino}, Bruno Rodr{\'\i}guez and {Sun}, Fengwu and {Tacchella}, Sandro and {{\"U}bler}, Hannah and {Venturi}, Giacomo and {Williams}, Christina C. and {Willmer}, Christopher N.~A. and {Willott}, Chris and {Witstok}, Joris},
        title = "{JADES: A large population of obscured, narrow-line active galactic nuclei at high redshift}",
      journal = {\aap},
         year = 2025,
        month = may,
       volume = {697},
          eid = {A175},
        pages = {A175},
          doi = {10.1051/0004-6361/202348804},
archivePrefix = {arXiv},
       eprint = {2311.18731},
 primaryClass = {astro-ph.GA},
       adsurl = {https://ui.adsabs.harvard.edu/abs/2025A&A...697A.175S}
}

@article{Battisti_2016,
doi = {10.3847/0004-637X/818/1/13},
url = {https://dx.doi.org/10.3847/0004-637X/818/1/13},
year = {2016},
month = {feb},
publisher = {The American Astronomical Society},
volume = {818},
number = {1},
pages = {13},
author = {Battisti, A. J. and Calzetti, D. and Chary, R.-R.},
title = {CHARACTERIZING DUST ATTENUATION IN LOCAL STAR-FORMING GALAXIES: UV AND OPTICAL REDDENING},
journal = {\apj}
}

@ARTICLE{multinest_1,
       author = {{Feroz}, F. and {Hobson}, M.~P.},
        title = "{Multimodal nested sampling: an efficient and robust alternative to Markov Chain Monte Carlo methods for astronomical data analyses}",
      journal = {\mnras},
         year = 2008,
        month = feb,
       volume = {384},
       number = {2},
        pages = {449-463},
          doi = {10.1111/j.1365-2966.2007.12353.x},
archivePrefix = {arXiv},
       eprint = {0704.3704},
 primaryClass = {astro-ph},
       adsurl = {https://ui.adsabs.harvard.edu/abs/2008MNRAS.384..449F}
}

@ARTICLE{multinest_2,
       author = {{Feroz}, F. and {Hobson}, M.~P. and {Bridges}, M.},
        title = "{MULTINEST: an efficient and robust Bayesian inference tool for cosmology and particle physics}",
      journal = {\mnras},
         year = 2009,
        month = oct,
       volume = {398},
       number = {4},
        pages = {1601-1614},
          doi = {10.1111/j.1365-2966.2009.14548.x},
archivePrefix = {arXiv},
       eprint = {0809.3437},
 primaryClass = {astro-ph},
       adsurl = {https://ui.adsabs.harvard.edu/abs/2009MNRAS.398.1601F}
}

@article{Bruzual_Charlot,
    author = {Bruzual, G. and Charlot, S.},
    title = {Stellar population synthesis at the resolution of 2003},
    journal = {\mnras},
    volume = {344},
    number = {4},
    pages = {1000-1028},
    year = {2003},
    month = {10},
    issn = {0035-8711},
    doi = {10.1046/j.1365-8711.2003.06897.x},
    url = {https://doi.org/10.1046/j.1365-8711.2003.06897.x},
    eprint = {https://academic.oup.com/mnras/article-pdf/344/4/1000/2908334/344-4-1000.pdf},
}

@misc{ferland2023cloudy,
      title={The 2017 Release of Cloudy}, 
      author={G. J. Ferland and M. Chatzikos and F. Guzmán and M. L. Lykins and P. A. M. van Hoof and R. J. R. Williams and N. P. Abel and N. R. Badnell and F. P. Keenan and R. L. Porter and P. C. Stancil},
      year={2023},
      eprint={1705.10877},
      archivePrefix={arXiv},
      primaryClass={astro-ph.GA},
      url={https://arxiv.org/abs/1705.10877}, 
}

@ARTICLE{Calzetti_1994,
       author = {{Calzetti}, Daniela and {Kinney}, Anne L. and {Storchi-Bergmann}, Thaisa},
        title = "{Dust Extinction of the Stellar Continua in Starburst Galaxies: The Ultraviolet and Optical Extinction Law}",
      journal = {\apj},
         year = 1994,
        month = jul,
       volume = {429},
        pages = {582},
          doi = {10.1086/174346},
       adsurl = {https://ui.adsabs.harvard.edu/abs/1994ApJ...429..582C}
}

@ARTICLE{Calzetti2000,
       author = {{Calzetti}, Daniela and {Armus}, Lee and {Bohlin}, Ralph C. and {Kinney}, Anne L. and {Koornneef}, Jan and {Storchi-Bergmann}, Thaisa},
        title = "{The Dust Content and Opacity of Actively Star-forming Galaxies}",
      journal = {\apj},
         year = 2000,
        month = apr,
       volume = {533},
       number = {2},
        pages = {682-695},
          doi = {10.1086/308692},
archivePrefix = {arXiv},
       eprint = {astro-ph/9911459},
 primaryClass = {astro-ph},
       adsurl = {https://ui.adsabs.harvard.edu/abs/2000ApJ...533..682C}
}

@article{Merlin_2024,
   title={ASTRODEEP-JWST: NIRCam-HST multi-band photometry and redshifts for half a million sources in six extragalactic deep fields},
   volume={691},
   ISSN={1432-0746},
   url={http://dx.doi.org/10.1051/0004-6361/202451409},
   DOI={10.1051/0004-6361/202451409},
   journal={\aap},
   publisher={EDP Sciences},
   author={Merlin, E. and Santini, P. and Paris, D. and Castellano, M. and Fontana, A. and Treu, T. and Finkelstein, S. L. and Dunlop, J. S. and Arrabal Haro, P. and Bagley, M. and Boyett, K. and Calabrò, A. and Correnti, M. and Davis, K. and Dickinson, M. and Donnan, C. T. and Ferguson, H. C. and Fortuni, F. and Giavalisco, M. and Glazebrook, K. and Grazian, A. and Grogin, N. A. and Hathi, N. and Hirschmann, M. and Kartaltepe, J. S. and Kewley, L. J. and Kirkpatrick, A. and Kocevski, D. D. and Koekemoer, A. M. and Leung, G. and Lotz, J. M. and Lucas, R. A. and Magee, D. K. and Marchesini, D. and Mascia, S. and McLeod, D. J. and McLure, R. J. and Nanayakkara, T. and Napolitano, L. and Nonino, M. and Papovich, C. and Pentericci, L. and Pérez-González, P. G. and Pirzkal, N. and Ravindranath, S. and Roberts-Borsani, G. and Somerville, R. S. and Trenti, M. and Trump, J. R. and Vulcani, B. and Wang, X. and Watson, P. J. and Wilkins, S. M. and Yang, G. and Yung, L. Y. A.},
   year={2024},
   month=nov, pages={A240} }

@article{Markov_2024,
   title={The evolution of dust attenuation in z ≈ 2–12 galaxies observed by JWST},
   volume={9},
   ISSN={2397-3366},
   url={http://dx.doi.org/10.1038/s41550-024-02426-1},
   DOI={10.1038/s41550-024-02426-1},
   number={3},
   journal={Nature Astronomy},
   publisher={Springer Science and Business Media LLC},
   author={Markov, Vladan and Gallerani, Simona and Ferrara, Andrea and Pallottini, Andrea and Parlanti, Eleonora and Mascia, Fabio Di and Sommovigo, Laura and Kohandel, Mahsa},
   year={2024},
   month=nov, pages={458–468} }

@article{Reddy_2015,
doi = {10.1088/0004-637X/806/2/259},
url = {https://dx.doi.org/10.1088/0004-637X/806/2/259},
year = {2015},
month = {jun},
publisher = {The American Astronomical Society},
volume = {806},
number = {2},
pages = {259},
author = {Reddy, Naveen A. and Kriek, Mariska and Shapley, Alice E. and Freeman, William R. and Siana, Brian and Coil, Alison L. and Mobasher, Bahram and Price, Sedona H. and Sanders, Ryan L. and Shivaei, Irene},
title = {THE MOSDEF SURVEY: MEASUREMENTS OF BALMER DECREMENTS AND THE DUST ATTENUATION CURVE AT REDSHIFTS z ∼ 1.4–2.6*},
journal = {\apj}
}

@article{Battisti_2022,
   title={The average dust attenuation curve at z ∼ 1.3 based on HST grism surveys},
   volume={513},
   ISSN={1365-2966},
   url={http://dx.doi.org/10.1093/mnras/stac1052},
   DOI={10.1093/mnras/stac1052},
   number={3},
   journal={\mnras},
   publisher={Oxford University Press (OUP)},
   author={Battisti, A J and Bagley, M B and Baronchelli, I and Dai, Y S and Henry, A L and Malkan, M A and Alavi, A and Calzetti, D and Colbert, J and McCarthy, P J and Mehta, V and Rafelski, M and Scarlata, C and Shivaei, I and Wisnioski, E},
   year={2022},
   month=apr, pages={4431–4450} }

@article{Salim_2020,
   title={The Dust Attenuation Law in Galaxies},
   volume={58},
   ISSN={1545-4282},
   url={http://dx.doi.org/10.1146/annurev-astro-032620-021933},
   DOI={10.1146/annurev-astro-032620-021933},
   number={1},
   journal={\araa},
   publisher={Annual Reviews},
   author={Salim, Samir and Narayanan, Desika},
   year={2020},
   month=aug, pages={529–575} }

@article{Salim_2018,
doi = {10.3847/1538-4357/aabf3c},
url = {https://dx.doi.org/10.3847/1538-4357/aabf3c},
year = {2018},
month = {may},
publisher = {The American Astronomical Society},
volume = {859},
number = {1},
pages = {11},
author = {Salim, Samir and Boquien, Médéric and Lee, Janice C.},
title = {Dust Attenuation Curves in the Local Universe: Demographics and New Laws for Star-forming Galaxies and High-redshift Analogs},
journal = {\apj}
}

@ARTICLE{Shivaei_2020,
       author = {{Shivaei}, Irene and {Reddy}, Naveen and {Rieke}, George and {Shapley}, Alice and {Kriek}, Mariska and {Battisti}, Andrew and {Mobasher}, Bahram and {Sanders}, Ryan and {Fetherolf}, Tara and {Azadi}, Mojegan and {Coil}, Alison L. and {Freeman}, William R. and {de Groot}, Laura and {Leung}, Gene and {Price}, Sedona H. and {Siana}, Brian and {Zick}, Tom},
        title = "{The MOSDEF Survey: The Variation of the Dust Attenuation Curve with Metallicity}",
      journal = {\apj},
         year = 2020,
        month = aug,
       volume = {899},
       number = {2},
          eid = {117},
        pages = {117},
          doi = {10.3847/1538-4357/aba35e},
archivePrefix = {arXiv},
       eprint = {2005.01742},
 primaryClass = {astro-ph.GA},
       adsurl = {https://ui.adsabs.harvard.edu/abs/2020ApJ...899..117S}
}

@article{Battisti_2017,
   title={Characterizing Dust Attenuation in Local Star-forming Galaxies: Inclination Effects and the 2175 Å Feature},
   volume={851},
   ISSN={1538-4357},
   url={http://dx.doi.org/10.3847/1538-4357/aa9a43},
   DOI={10.3847/1538-4357/aa9a43},
   number={2},
   journal={\apj},
   publisher={American Astronomical Society},
   author={Battisti, A. J. and Calzetti, D. and Chary, R.-R.},
   year={2017},
   month=dec, pages={90} }

@ARTICLE{cardelli_1989,
       author = {{Cardelli}, Jason A. and {Clayton}, Geoffrey C. and {Mathis}, John S.},
        title = "{The Relationship between Infrared, Optical, and Ultraviolet Extinction}",
      journal = {\apj},
         year = 1989,
        month = oct,
       volume = {345},
        pages = {245},
          doi = {10.1086/167900},
       adsurl = {https://ui.adsabs.harvard.edu/abs/1989ApJ...345..245C}
}

@article{Shivaei_2015,
doi = {10.1088/0004-637X/804/2/149},
url = {https://dx.doi.org/10.1088/0004-637X/804/2/149},
year = {2015},
month = {may},
publisher = {The American Astronomical Society},
volume = {804},
number = {2},
pages = {149},
author = {Shivaei, Irene and Reddy, Naveen A. and Steidel, Charles C. and Shapley, Alice E.},
title = {INVESTIGATING Hα, UV, AND IR STAR-FORMATION RATE DIAGNOSTICS FOR A LARGE SAMPLE OF z ∼ 2 GALAXIES},
journal = {\apj}
}

@article{Price_2014,
doi = {10.1088/0004-637X/788/1/86},
url = {https://dx.doi.org/10.1088/0004-637X/788/1/86},
year = {2014},
month = {may},
publisher = {The American Astronomical Society},
volume = {788},
number = {1},
pages = {86},
author = {Price, Sedona H. and Kriek, Mariska and Brammer, Gabriel B. and Conroy, Charlie and Schreiber, Natascha M. Förster and Franx, Marijn and Fumagalli, Mattia and Lundgren, Britt and Momcheva, Ivelina and Nelson, Erica J. and Skelton, Rosalind E. and van Dokkum, Pieter G. and Whitaker, Katherine E. and Wuyts, Stijn},
title = {DIRECT MEASUREMENTS OF DUST ATTENUATION IN z ∼ 1.5 STAR-FORMING GALAXIES FROM 3D-HST: IMPLICATIONS FOR DUST GEOMETRY AND STAR FORMATION RATES},
journal = {\apj}
}

@article{Carnall_2018,
   title={Inferring the star formation histories of massive quiescent galaxies with bagpipes: evidence for multiple quenching mechanisms},
   volume={480},
   ISSN={1365-2966},
   url={http://dx.doi.org/10.1093/mnras/sty2169},
   DOI={10.1093/mnras/sty2169},
   number={4},
   journal={\mnras},
   publisher={Oxford University Press (OUP)},
   author={Carnall, A C and McLure, R J and Dunlop, J S and Davé, R},
   year={2018},
   month=aug, pages={4379–4401} }

@article{MS_Popesso,
    author = {Popesso, P and Concas, A and Cresci, G and Belli, S and Rodighiero, G and Inami, H and Dickinson, M and Ilbert, O and Pannella, M and Elbaz, D},
    title = {The main sequence of star-forming galaxies across cosmic times},
    journal = {\mnras},
    volume = {519},
    number = {1},
    pages = {1526-1544},
    year = {2022},
    month = {11},
    issn = {0035-8711},
    doi = {10.1093/mnras/stac3214},
    url = {https://doi.org/10.1093/mnras/stac3214},
    eprint = {https://academic.oup.com/mnras/article-pdf/519/1/1526/48447685/stac3214.pdf},
}

@ARTICLE{1983ApJ...266..713O,
       author = {{Oke}, J.~B. and {Gunn}, J.~E.},
        title = "{Secondary standard stars for absolute spectrophotometry.}",
      journal = {\apj},
         year = 1983,
        month = mar,
       volume = {266},
        pages = {713-717},
          doi = {10.1086/160817},
       adsurl = {https://ui.adsabs.harvard.edu/abs/1983ApJ...266..713O}
}

@ARTICLE{2019A&A...622A.169M,
       author = {{Merlin}, E. and {Pilo}, S. and {Fontana}, A. and {Castellano}, M. and {Paris}, D. and {Roscani}, V. and {Santini}, P. and {Torelli}, M.},
        title = "{A-PHOT: a new, versatile code for precision aperture photometry}",
      journal = {\aap},
         year = 2019,
        month = feb,
       volume = {622},
          eid = {A169},
        pages = {A169},
          doi = {10.1051/0004-6361/201833991},
archivePrefix = {arXiv},
       eprint = {1812.00727},
 primaryClass = {astro-ph.IM},
       adsurl = {https://ui.adsabs.harvard.edu/abs/2019A&A...622A.169M}
}

@ARTICLE{2024ApJ...975...83R,
       author = {{Rieke}, G.~H. and {Alberts}, Stacey and {Shivaei}, Irene and {Lyu}, Jianwei and {Willmer}, Christopher N.~A. and {P{\'e}rez-Gonz{\'a}lez}, Pablo and {Williams}, Christina C.},
        title = "{SMILES: A Prototype JWST Multiband Mid-infrared Survey}",
      journal = {\apj},
         year = 2024,
        month = nov,
       volume = {975},
       number = {1},
          eid = {83},
        pages = {83},
          doi = {10.3847/1538-4357/ad6cd2},
archivePrefix = {arXiv},
       eprint = {2406.03518},
 primaryClass = {astro-ph.GA},
       adsurl = {https://ui.adsabs.harvard.edu/abs/2024ApJ...975...83R}
}

@ARTICLE{2024ApJ...976..224A,
       author = {{Alberts}, Stacey and {Lyu}, Jianwei and {Shivaei}, Irene and {Rieke}, George H. and {P{\'e}rez-Gonz{\'a}lez}, Pablo G. and {Bonaventura}, Nina and {Zhu}, Yongda and {Helton}, Jakob M. and {Ji}, Zhiyuan and {Morrison}, Jane and {Robertson}, Brant E. and {Stone}, Meredith A. and {Sun}, Yang and {Williams}, Christina C. and {Willmer}, Christopher N.~A.},
        title = "{SMILES Initial Data Release: Unveiling the Obscured Universe with MIRI Multiband Imaging}",
      journal = {\apj},
         year = 2024,
        month = dec,
       volume = {976},
       number = {2},
          eid = {224},
        pages = {224},
          doi = {10.3847/1538-4357/ad7396},
archivePrefix = {arXiv},
       eprint = {2405.15972},
 primaryClass = {astro-ph.GA},
       adsurl = {https://ui.adsabs.harvard.edu/abs/2024ApJ...976..224A}
}

@ARTICLE{2002ASPC..281..228B,
       author = {{Bertin}, E. and {Mellier}, Y. and {Radovich}, M. and others},
        title = "{Astronomical Data Analysis Software and Systems XI}",
      journal = {ASPC},
         year = 2002,
       volume = {281},
        pages = {228},
}

@ARTICLE{2016A&A...595A..97M,
       author = {{Merlin}, E. and {Bourne}, N. and {Castellano}, M. and {Ferguson}, H.~C. and {Wang}, T. and {Derriere}, S. and {Dunlop}, J.~S. and {Elbaz}, D. and {Fontana}, A.},
        title = "{T-PHOT version 2.0: Improved algorithms for background subtraction, local convolution, kernel registration, and new options}",
      journal = {\aap},
         year = 2016,
        month = nov,
       volume = {595},
          eid = {A97},
        pages = {A97},
          doi = {10.1051/0004-6361/201628751},
archivePrefix = {arXiv},
       eprint = {1609.00146},
 primaryClass = {astro-ph.IM},
       adsurl = {https://ui.adsabs.harvard.edu/abs/2016A&A...595A..97M}
}

@ARTICLE{Storey1995,
       author = {{Storey}, P.~J. and {Hummer}, D.~G.},
        title = "{Recombination line intensities for hydrogenic ions-IV. Total recombination coefficients and machine-readable tables for Z=1 to 8}",
      journal = {\mnras},
         year = 1995,
        month = jan,
       volume = {272},
       number = {1},
        pages = {41-48},
          doi = {10.1093/mnras/272.1.41},
       adsurl = {https://ui.adsabs.harvard.edu/abs/1995MNRAS.272...41S}
}

@ARTICLE{Ferrarotti2006,
       author = {{Ferrarotti}, A.~S. and {Gail}, H.-P.},
        title = "{Composition and quantities of dust produced by AGB-stars and returned to the interstellar medium}",
      journal = {\aap},
         year = 2006,
        month = feb,
       volume = {447},
       number = {2},
        pages = {553-576},
          doi = {10.1051/0004-6361:20041198},
       adsurl = {https://ui.adsabs.harvard.edu/abs/2006A&A...447..553F}
}

\appendix
\section{NIRSpec artefacts}
\begin{figure}[h!]
    \centering
    \includegraphics[width=1\linewidth]{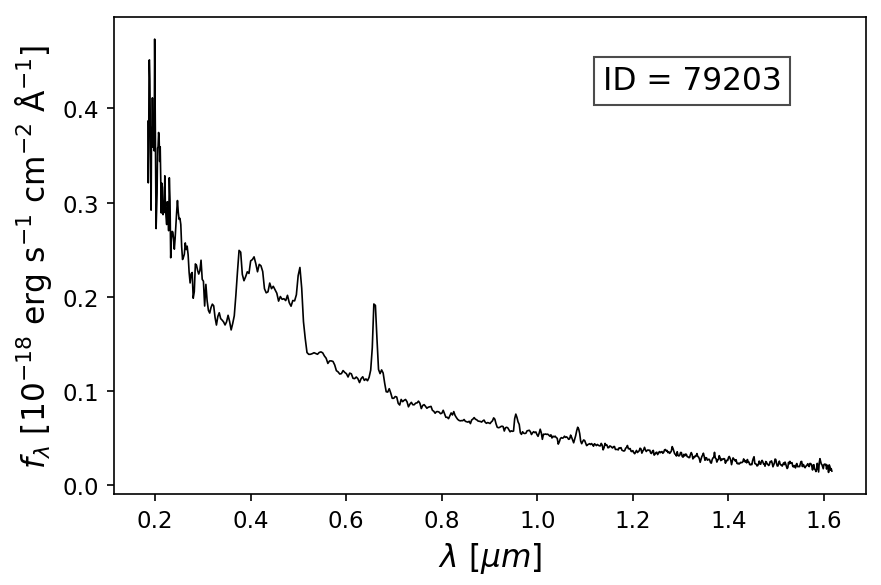}
    \caption{Example of a shutter bump feature in the rest-frame spectrum of an object. Sources with a similar unphysical box like structure around the Balmer break have been discarded from the final sample.}
    \label{fig:bump_plot}
\end{figure}

\section{Physical properties of galaxies as a function of Balmer optical depth bins}
\label{appB}
\begin{figure}[]
    \centering
    \includegraphics[width=1.\linewidth]{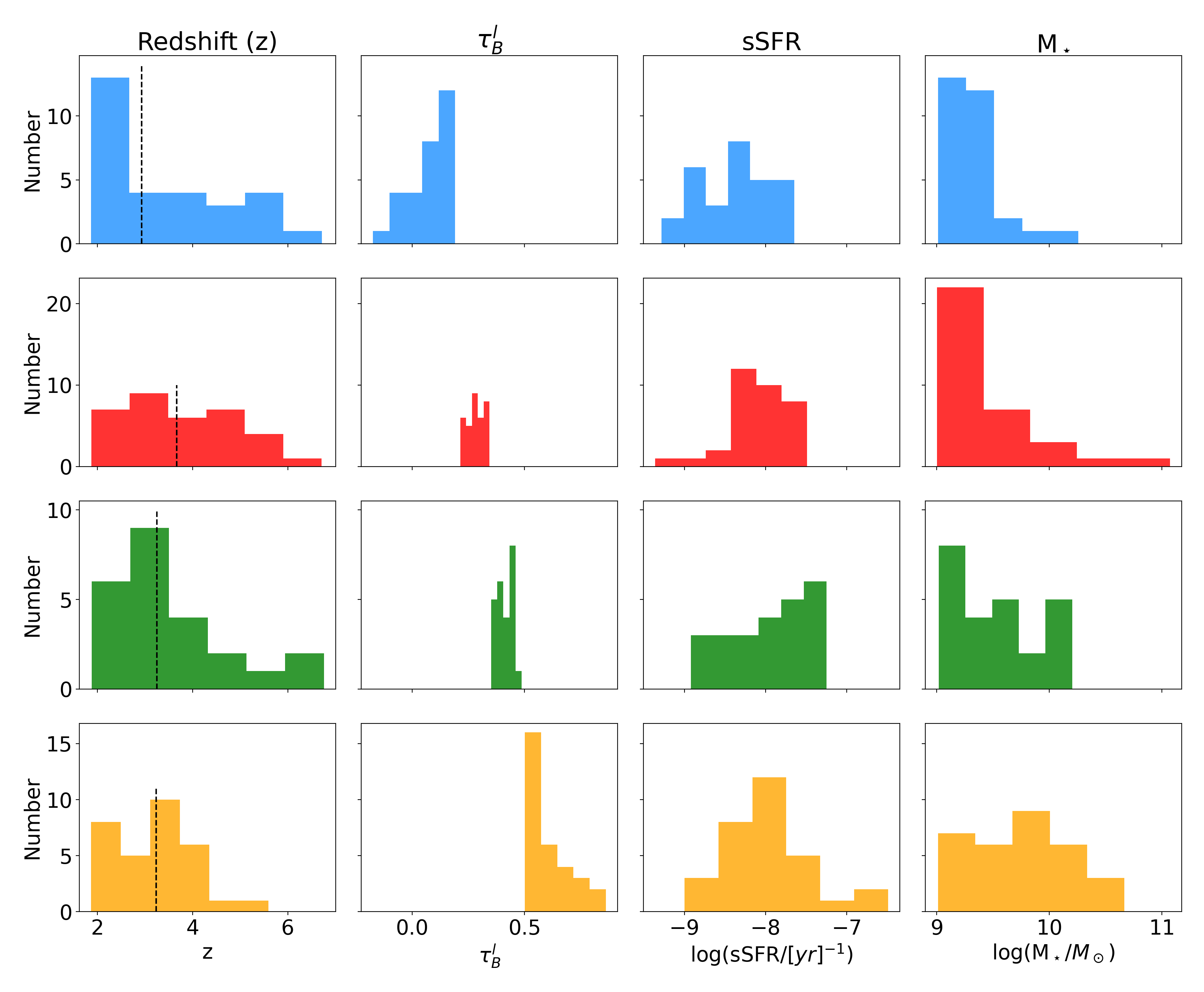}
    \caption{From left to right: Distributions of redshift ($z$), Balmer optical depth ($\tau_B^l$), specific star-formation rate (sSFR), and stellar mass ($M_\star$) for the galaxies in each $\tau_B^l$ bin. The colour coding matches that adopted for the average spectral templates in the main text. In the redshift panels, the vertical dashed lines indicate the mean redshift of each subsample.}
    \label{z_tau_sSFR}
\end{figure}
In Fig.\,\ref{z_tau_sSFR} we show the redshift, Balmer optical depth, specific star-formation rate, and stellar-mass distributions for each subsample defined by the $\tau_B^l$ binning. The redshift distribution is well sampled up to $z \lesssim 4$ in all bins. At higher redshifts, the sampling becomes less uniform; however, as discussed in Sect.\,\ref{beta-tau-relation}, this does not bias our analysis, since the relation $\beta\mbox{--}\tau^l_B$ shows no significant dependence on the redshift. The $\tau^l_B$ distributions appear fairly uniform across the bins, with the exception of the $0.5 < \tau^l_B < 0.9$ bin, which is less populated. However, this does not affect the construction of the dust attenuation curve, since each SED ratio is normalised using the median $\tau^l_B$ value of the corresponding bin, following the procedure described in Sect.\,\ref{sec:selective_curve}. The sSFR distributions indicate that the targets exhibit, on average, similar star formation properties, with all $\tau^l_B$ bins peaking around $\log_{10}(\mathrm{sSFR}/\mathrm{yr}^{-1}) \sim -8.0$.

The stellar-mass distributions show some dependence on $\tau_B^l$, with galaxies in the higher-$\tau_B^l$ bins tending towards larger stellar masses. This behaviour is consistent with the positive correlation between $M_\star$ and $\tau_B^l$ discussed in Sect.~5.2. Nevertheless, the stellar-mass distributions span substantially overlapping ranges across the different $\tau_B^l$ bins, indicating that the stacked spectra are not constructed from completely disjoint stellar-mass populations. We note, however, that this comparison does not exclude residual differences in the intrinsic stellar populations among the $\tau_B^l$ bins. In particular, properties such as stellar age and metallicity may also vary across the bins and affect the intrinsic continuum shape. These dependencies are not explicitly constrained in the present analysis and therefore represent a potential source of systematic uncertainty in the differential derivation of the attenuation curve.

\end{document}